\documentclass[11pt,twoside,b5paper]{book}
\usepackage[english]{babel}
\usepackage{amsmath,amsfonts,amssymb,bm}\interdisplaylinepenalty=2500
\ifx\pdfoutput\undefined
  \usepackage{graphicx,color}   
  \usepackage[nativepdf,backref,bookmarks]{hyperref} 
  \usepackage{rotating}
\else
  \usepackage[backref,bookmarks]{hyperref}  
  \usepackage[pdftex]{graphicx,color}     
  \usepackage[pdftex]{rotating}              
\fi
\graphicspath{{figs/}}
\mathchardef\varphi="011E        \mathchardef\phi="0127
\ifx\undefined\degrees\def\degrees{\ensuremath{^{\circ}}}\fi
\ifx\undefined\celsius\def\celsius{\ensuremath{^{\circ}\mathrm{C}}}\fi
\ifx\undefined\unit\def\unit#1{\ensuremath{\mathrm{\,#1}}}\fi
\ifx\undefined\micro\def\micro{\ensuremath{\mu}}\fi
\ifx\undefined\sups\def\sups#1{\ensuremath{^{\mathrm{#1}}}}\fi
\ifx\undefined\subs\def\subs#1{\ensuremath{_{\mathrm{#1}}}}\fi
\ifx\undefined\ohm\def\ohm{\ensuremath{\mathrm{\Omega}}}\fi
\def\req#1{(\ref{#1})}

\newcommand{\namedgraphics}[2]{
 \parbox{\textwidth}{\rotatebox{90}{~\ttfamily\scriptsize#2}%
 \hspace*{\fill}\includegraphics[scale=#1]{#2}\hspace*{\fill}}}

\makeatletter
\def\@makechapterhead#1{%
  \vspace*{50\p@}{\parindent \z@ \raggedright \normalfont
    \ifnum \c@secnumdepth >\m@ne \if@mainmatter 
        \LARGE\sffamily\@chapapp\space \thechapter\par\nobreak\vskip 20\p@
      \fi\fi
    \interlinepenalty\@M
    \huge\sffamily#1\par\nobreak
    \vskip 40\p@}}
\def\@makeschapterhead#1{%
  \vspace*{50\p@}{\parindent \z@ \raggedright\normalfont
    \interlinepenalty\@M
    \huge\sffamily#1\par\nobreak
    \vskip 40\p@}}
\makeatother

\newtheorem{rremark}{Remark}[chapter]
\newenvironment{remark}{%
  \begin{rremark}\normalfont}{\mbox{}\hfill$\blacktriangle$\end{rremark}}
\newenvironment{proof}[1][Proof.~~]{\vspace{1ex}\noindent\textsc{#1}}{\mbox{}\hfill$\blacktriangle$}
\newenvironment{subproof}[1]{\noindent\emph{#1}}{}
\newtheorem{eexample}{Example}[chapter]
\newenvironment{example}{%
	\begin{eexample}\normalfont}{\mbox{}\hfill$\blacktriangle$\end{eexample}}
\newtheorem{pproperty}{Property}[chapter]

\newenvironment{statement}[1]{\vspace{1ex}\noindent\textbf{#1}\itshape}{}

\title{The Leeson Effect\\[0em]---\\[0em]\Large Phase Noise in Quasilinear Oscillators}
\author{\\[1em]Enrico Rubiola\\[3em]
Universit\'e Henri Poincar\'e, Nancy, France\\
{\normalsize\textsc{esstin} and \textsc{lpmia}}\\[0.8em]
\small web page \texttt{www.rubiola.org}\\[0ex]
\small e-mail~\texttt{enrico@rubiola.org}\\[8.5em]\mbox{~}}
\date{Rev.~$1.0$,~February 23, 2005}

\begin{document}
\frontmatter
\maketitle
\thispagestyle{empty}\mbox{}
\setcounter{tocdepth}{2}        
\tableofcontents

\chapter{Most used symbols}
\begin{center}
\begin{tabular}[t]{ll}
$b_i$		& coefficients of the power-law approximation of $S_\phi(f)$,\\
$\phantom{QQQQQQQ}$		
		& Eq. \req{eqn:le-power-law-sphi},  Fig.\ \ref{fig:le-power-law}\\
$b(t)$		& resonator phase response\\
$f$		& Fourier frequency, Hz\\
$f_c$		& amplifier corner frequency, Hz, Fig.\ \ref{fig:le-ampli-noise}\\
$f_L$		& Leeson frequency, Hz, 
		Eq.\ \req{eqn:le-leeson-heuristic-fl-def} \\
$h(t)$		& impulse response\\
$h_i$		& coefficients of the power-law approximation of $S_y(f)$,\\
		& Eq. \req{eqn:le-power-law-sy},  Fig.\ \ref{fig:le-power-law}\\
$j$		& imaginary unit, $j^2=-1$ \\
$k$		& Boltzmann constant, $k=1.38{\times}10{-23}$ J/K\\
$m$		& harmonic order (Chapter \ref{chap:le-delayline})\\
$n$		& harmonic order (Chapter \ref{chap:le-delayline})\\
$n(t)$		& random noise, as a function of time\\
$v(t)$		& voltage, as a function of time\\
$x$		& a generic variable\\
$x(t)$		& phase time fluctuation, Eq.\ \req{eqn:le-phase-time-def}\\
$y(t)$		& fractional frequency fluctuation, 
		Eq.\ \req{eqn:le-fractional-frequency-def}\\
$A$		& amplifier voltage gain\\
$B(s)$	& resonator phase response, $B(s)=\mathcal{L}\{b(t)\}$\\
$\mathcal{D}$	& denominator of a transfer function\\
$F$		& amplifier noise figure, Eq.\ \req{eqn:le-noise-fig-def}\\
$H(s)$	& transfer function, Eq.\ \req{eqn:le-def-hvolt}\\
$\mathcal{H}(s)$	& phase transfer function
\end{tabular}
\end{center}
\clearpage
\begin{center}
\begin{tabular}[t]{ll}
$\mathcal{L}(f)$	& single-sideband noise spectrum, 
		Eq.\ \req{eqn:le-script-ell-def},  dBc/Hz\\
$\mathcal{L}(\,\cdot\,)$	& Laplace transform operator\\
$N$		& noise spectrum density, esp.\ RF/microwave, W/Hz\\
$P$		& power, W, esp.\ carrier power\\
$Q$		& resonator merit factor\\
$S_a(f)$, $S_a(\omega)$	& power spectral density of the quantity $a$\\
$T$		& period, $T=1/\nu$, s; also, absolute temperature\\
$U(t)$	& Heaviside (step) function, $U(t)=\int\delta(t')\,dt'$\\
$V$		& either dc voltage (constant) or phasor\\
$V(s)$	& Laplace transform of $v(t)$\\
$\alpha(t)$	& (normalized) amplitude noise, 
		Eq.\ \req{eqn:le-oscillator-signal-polar}\\
$\beta(s)$	& transfer function of the feedback path, 
		Fig.\ \ref{fig:le-loop}\\
$\delta(t)$	& Dirac delta function\\
$\theta$	& argument of the resonator transfer function 
		$\rho e^{j\theta}$\\
$\mu$	& harmonic order in the phase space, 
		(Chapter \ref{chap:le-delayline})\\
$\nu$		& frequency (Hz), used for carriers,
		Eq.\ \req{eqn:le-oscillator-signal-polar}\\
$\rho$	& modulus of the resonator transfer function 
		$\rho e^{j\theta}$\\
$\sigma$	& real part of the complex variable $s=\sigma+j\omega$\\
$\sigma_y(\tau)$	& Allan deviation, 
		square root of the Allan variance $\sigma^2_y(\tau)$\\
		& (used only with the fractional frequency fluctuation $y$\\ 
$\tau$	& measurement time, in $\sigma(\tau)$\\
$\tau$	& resonator relaxation time, or delay of a delay line\\
$\phi(t)$	& phase noise, Eq.\ \req{eqn:le-oscillator-signal-polar}\\
$\chi$		& dissonance, Eq.\ \req{eqn:le-app-disson}\\
$\psi(t)$	& amplifier phase noise; also, a constant phase\\
$\omega$	& angular frequency (both carrier and Fourier)\\  
$\Phi(t)$	& phase noise, $\Phi(s)=\mathcal{L}\{\phi(t)\}$\\
$\Psi(t)$	& amplifier phase noise, $\Psi(s)=\mathcal{L}\{\psi(t)\}$\\
$\Omega$	& detuning angular frequency\\  
\multicolumn{2}{l}{\textbf{Note: 
		$\bm{\omega}$ is used as a shorthand for $\bm{2\pi f}$
		or $\bm{2\pi f}$, and viceversa}}
\end{tabular}
\end{center}
\vfill~

\mainmatter
\addtocounter{chapter}{-1}

\chapter{Preface}\label{chap:preface}

Time, and equivalently frequency, is the most precisely measured physical quantity.  It is therefore inevitable that  virtually all domains of engineering and physics need reference oscillators.   The oscillator noise can be decomposed into amplitude noise and phase noise.  The latter, far more important,  affects timing, for it is related to precision and accuracy of measurements. 

The oscillator, inherently, turns the phase noise of the internal parts into frequency noise.  This is a necessary consequence of the Barkhausen condition, which states that the loop gain must be of one, with zero phase, for stationary oscillation.  There follows that oscillator phase noise, which is the integral of the frequency noise, diverges in the long run.  
This phenomenon is often referred to as the ``Leeson model'' after a short article published in 1966 by David B. Leeson \cite{leeson66pieee}, and called \emph{Leeson effect} here, in order to emphasize that it is far more general  than model.  In 2001, David B. Leeson received the W. G. Cady award of the 
IEEE International Frequency Control Symposium
``For  clear physical insight and model of the effects of noise on oscillators''.

Since spring 2004, I had the opportunity to give some seminars on noise in oscillators at the Jet Propulsion Laboratory, at the IEEE Frequency Control Symposium, at the FEMTO-ST Laboratory, and at the Universit\'e Henri Poincar\'e.
These seminars had the purpose to provide a \emph{tutorial}, as opposed to a report on advanced science, addressed to a variety of people including technicians, PhD students, and senior scientists.
This monograph derives from these seminars, and from numerous discussions with colleagues.

The topics covered can be divided into three parts.  Chapter 1 addresses 
language and general physical mechanisms.  Chapter 2 aims at understanding the inside of commercial oscillators through the analysis of the specifications.  Chapter 3 and 4 focus on the use of the Laplace transform to describe the oscillator and its phase noise.

\def\scratch{%
I wish to thank some manufacturer and their representatives for kindness and prompt help:  Jean-Pierre Aubry from Oscilloquartz,  Art Faverio and Charif Nasrallah from Miteq, and Mark Henderson from Oewaves.

of colleagues disproportioned with the small ambition of this booklet.  A non exhaustive list is: 
Roger Bourquin,
R\'emi Brendel,
Giorgio Brida,
Vincent Candelier,
G. John Dick,
Michele Elia,
Vincent Giordano,
Serge Galliou, 
Charles (Chuck) Greenhall,
Jacques Groslambert,
Lute Maleki,
Mark Oxborrow,
Fran\c{o}is Vernotte,
Nan Yu,

I have been discussing for eight years about phase noise with Vincent Giordano, 
}

\vspace{1em}\noindent{Nancy, Feb 23, 2005}\\[2em]
\hspace*{50ex}Enrico Rubiola

\chapter{Heuristic approach to the Leeson effect}\label{chap:le-fundamentals}
\section{Phase noise fundamentals}
%
This introductory section provides a summary about phase noise and of
its properties.  The material is available in many classical references, 
such \cite{rutman78pieee,ccir90rep580-3,vanier:frequency-standards,ieee99std1139}.

The quasi-perfect sinusoidal signal of oscillators is modeled as
\begin{equation}
v(t)=V_0[1+\alpha(t)]\cos[2\pi\nu_0t+\phi(t)]~,
\label{eqn:le-oscillator-signal-polar}
\end{equation}
where $\nu_0$ is the carrier frequency; the random variables
$\alpha(t)$ and $\phi(t)$ are the fractional amplitude noise and the
phase noise, respectively.  The physical dimension of $\phi(t)$ is
rad, $\alpha(t)$ is dimensionless.  

It is sometimes convenient to rewrite the signal
\req{eqn:le-oscillator-signal-polar} in the equivalent Cartesian form
\begin{equation}
v(t)=V_0\cos(2\pi\nu_0t)+v_c(t)\cos(2\pi\nu_0t)-v_s(t)\cos(2\pi\nu_0t)~.
\label{eqn:le-oscillator-signal-cartesian}
\end{equation}
In low noise conditions ($|\alpha|\ll1$ and $|\phi|\ll1$), it holds that
\begin{equation}
\alpha(t)=\frac{v_c(t)}{V_0} \qquad\text{and}\qquad
\phi(t)=\frac{v_s(t)}{V_0}~.
\end{equation}
In the absence of noise, the spectrum of $v(t)$ is a Dirac
$V_0^2\,\smash{\frac12}\delta(\nu-\nu_0)$ function.  Noise broadens the spectrum.  Most of
the art of measuring the oscillator noise is related to the ability to
measure extremely narrow-band signals, for the radiofrequency spectrum
turns out to be a poor tool.  The oscillator noise is better described
in terms of the power spectrum density $S(f)$ of the amplitude and
phase noise, thus $S_\alpha(f)$ and $S_\phi(f)$, as a function of the Fourier frequency $f$.
Only phase noise is analyzed here.  Nonetheless, one should be aware that
the effect of amplitude noise may not be negligible, and that the resonant 
frequency of some resonators may be affected by the amplitude. 

The physical unit of $S_\phi(f)$ is \unit{rad^2/Hz}.  Phase noise
spectra are (almost) always plotted on a log-log scale.  The technical
unit ``decibel'', $S_\mathrm{dB}=10\log_{10}(S)$, is commonly used.
Manufacturers prefer the quantity $\mathcal{L}(f)$ (pronounce
`script-ell') to $S_\phi(f)$.  In physics and mathematics
$S_\phi(f)$ is preferred.  $\mathcal{L}(f)$ and $S_\phi(f)$ are
equivalent since $\mathcal{L}(f)$ is now%
\footnote{Formerly, $\mathcal{L}(f)$ was defined as the
	single-sideband noise power in 1~Hz bandwidth divided by the carrier
	power.  This definition has been superseded by
	\req{eqn:le-script-ell-def} because it was ambiguous when amplitude
	noise and phase noise have not the same spectrum.} 
defined as
\begin{equation}
\mathcal{L}(f)=\frac{1}{2}S_\phi(f)~.
\label{eqn:le-script-ell-def}
\end{equation}
$\mathcal{L}(f)$ is always given in dBc/Hz, which stands for dB below
the carrier in 1-Hz bandwidth.  In decibels,
$\mathcal{L}(f)=S_\phi(f)-3\unit{dB}$.

A model that has been found useful in describing the oscillator noise
spectra is the power-law
\begin{equation}
S_\phi(f)=\sum_{i=0}^{-4}b_if^i~.
\label{eqn:le-power-law-sphi}
\end{equation}
Table~\ref{tab:le-noise-conversion} shows the phase noise terms of
\req{eqn:le-power-law-sphi}.  If needed, the sum
\req{eqn:le-power-law-sphi} may be extended adding additional negative
terms.

\begin{table}[t]
\begin{center}
\caption{Noise types, power spectral densities, and Allan variance.%
  \vrule width 0pt height 1.5ex depth 1.5ex}
\label{tab:le-noise-conversion}
\begin{tabular}[c]{lccccc}\hline
noise type           &$S_\phi(f)$   &$S_y(f)$      &$S_\phi\leftrightarrow S_y$
	&$\sigma_y^2(\tau)$ & $\mathrm{mod}\,\sigma_y^2(\tau)$\rule[-1.5ex]{0ex}{4ex}\\\hline
white $\phi$          &$b_0$         &$h_{2}f^{2}$  &$h_{2}=\frac{b_0}{\nu_0^2}$
	&$\propto\tau^{-2}$ &$\propto\tau^{-3}$\rule[-1.5ex]{0ex}{4ex}\\
flicker $\phi$        &$b_{-1}f^{-1}$&$h_{1}f$      &$h_{1}=\frac{b_{-1}}{\nu_0^2}$
	&$\propto\tau^{-2}$ &$\propto\tau^{-2}$\rule[-1.5ex]{0ex}{4ex}\\
white $f$      &$b_{-2}f^{-2}$&$h_0$         &$h_{0}=\frac{b_{-2}}{\nu_0^2}$
	&$\frac{1}{2}h_0\tau^{-1}$ &$\frac{1}{4}h_0\tau^{-1}$\rule[-1.5ex]{0ex}{4ex}\\
flicker $f$    &$b_{-3}f^{-3}$&$h_{-1}f^{-1}$&$h_{-1}=\frac{b_{-3}}{\nu_0^2}$
	&$2\ln(2)\;h_{-1}$ &$\frac{27}{20}\ln(2)\;h_{-1}$\rule[-1.5ex]{0ex}{4ex} \\
rand.\,walk $f$ &$b_{-4}f^{-4}$&$h_{-2}f^{-2}$&$h_{-2}=\frac{b_{-4}}{\nu_0^2}$
	&$\frac{4\pi^2}{6}h_{-2}\tau$ &$\frac{4\pi^2}{6}h_{-2}\tau$\rule[-1.5ex]{0ex}{4ex}
\\\hline
\end{tabular}
\end{center}
\end{table}

Two other quantities are often used to characterize the oscillator
noise,
\begin{align}
  x(t)&=\frac{\phi(t)}{2\pi\nu_0} &&\text{phase time}
  \label{eqn:le-phase-time-def}\\
  y(t)&=\frac{\Dot{\phi}(t)}{2\pi\nu_0}
  &&\text{fractional frequency fluctuation}~
  \label{eqn:le-fractional-frequency-def}.
\end{align}
The phase time (fluctuation) $x(t)$ is the phase fluctuation $\phi(t)$
converted into time, and measured in seconds.  The fractional frequency
fluctuation $y(t)$ is the instantaneous frequency fluctuation
normalized to the nominal carrier frequency $\nu_0$.  $y(t)$ is
dimensionless.  The power spectral densities are
\begin{align}
  S_x(f)&=\frac{1}{\nu_0^2}S_\phi(f)
  \label{eqn:le-phase-time-spectrum}\\
  S_y(f)&=\frac{f^2}{\nu_0^2}S_\phi(f)~.
  \label{eqn:le-fractional-frequency-spectrum}
\end{align}
$S_y(f)$ [Eq.\req{eqn:le-fractional-frequency-spectrum}] is obtained
from the definition \req{eqn:le-fractional-frequency-def} using the
property that the time-domain derivative maps into a multiplication by
$j\omega=j 2\pi f$ in the Fourier transform domain, thus by 
$\omega^2=4\pi^2f^{2}$ in the spectrum.

The power-law model applies to $S_x(f)$ and $S_y(f)$.  The coefficients
of $S_y(f)$ are denoted by $h_i$ in the literature, hence
\begin{equation}
S_y(f)=\sum_{i=2}^{-2}h_if^i~.
\label{eqn:le-power-law-sy}
\end{equation}
Table~\ref{tab:le-noise-conversion} helps in conversions between
$S\phi(f)$ and $S_y(f)$.

\begin{figure}[t]
\namedgraphics{0.8}{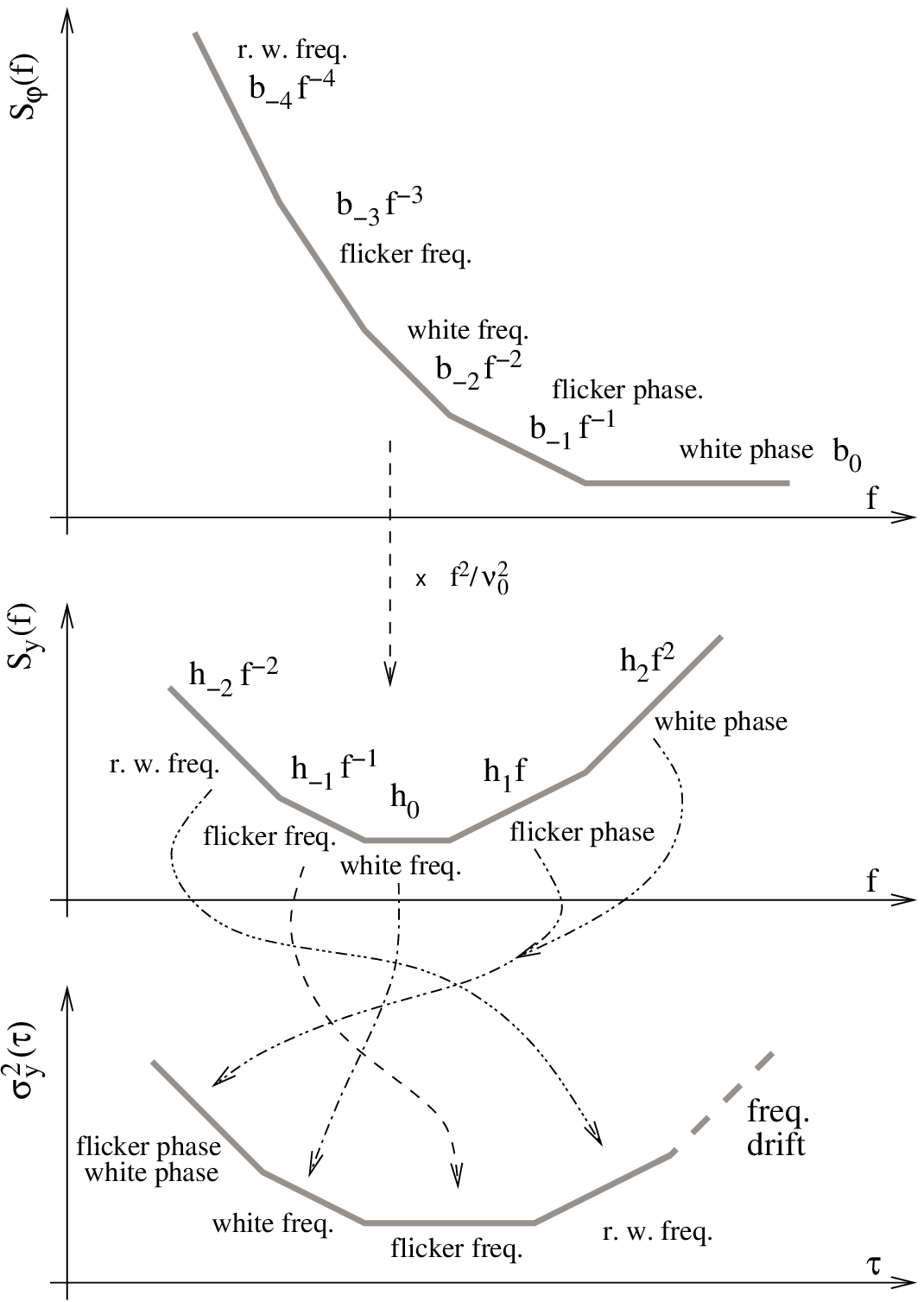}
\caption{Power-law, spectra, and Allan variance.}
\label{fig:le-power-law}
\end{figure}

Another tool often used in the oscillator characterization is the
Allan variance $\sigma^2_y(\tau)$, as a function of the measurement
time $\tau$.  The Allan variance is always estimated by averaging.
Given a stream of $M$ data $\overline{y}$, each representing a measure
of the quantity $y(t)$ averaged over a duration $\tau$ ending at the
time $t=k\tau$, the estimated Allan variance is
\begin{equation}
\sigma^2_y(\tau)=\frac{1}{2(M-1)}\sum_{k=1}^{M-1}
  \left(\overline{y}_{k+1}-\overline{y}_k\right)^2~.
\label{eqn:le-allan-variance-estimated}
\end{equation}
Table~\ref{tab:le-noise-conversion} provides some conversion formulae
to calculate $\sigma^2_y(\tau)$ from $S_y(f)$.  It is important to
understand that $\sigma^2_y(\tau)$ can always be calculated from
$S_y(f)$, but the inverse is not free from errors \cite{greenhall98im}
in the general case.
The modified Allan variance, not analyzed here, is also commonly used.
Figure~\ref{fig:le-power-law} provides a summary of the power-law spectra and Allan variance.
%

\section{Oscillator fundamentals}
%
\begin{figure}
\namedgraphics{0.8}{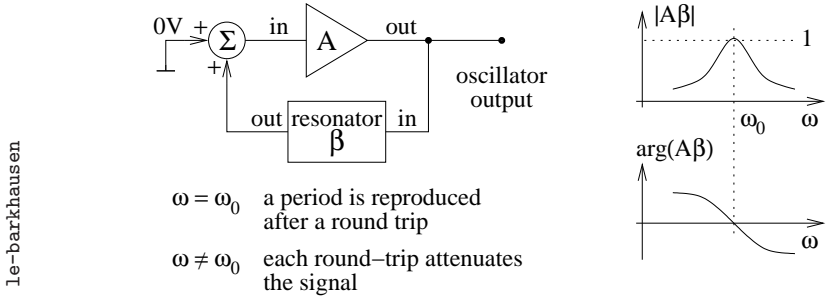}
\caption{Basic feedback oscillator.}
\label{fig:le-barkhausen}
\end{figure}
The basic feedback oscillator (Fig.\ \ref{fig:le-barkhausen}) is a loop
in which the gain $A$ of the sustaining amplifier compensates for the
loss [gain $\beta(\omega)$ in the figure] of the resonator at a given
angular frequency frequency $\omega_0$.  The condition for the
oscillation to be stationary, known as the \emph{Barkhausen}
condition, is
\begin{align}
&A\beta(\omega)=1 &\text{Barkhausen} \label{eqn:le-barkhausen-complex}\\ 
&|A\beta(\omega)|=1        \label{eqn:le-barkhausen-modulus}\\
&\arg[A\beta(\omega)]=0	\label{eqn:le-barkhausen-phase}\
\end{align}
at $\omega=\omega_0$.   

The unused input (0\,V) in Fig.\ \ref{fig:le-barkhausen} serves to set the initial condition that triggers the oscillation, and to  introduce noise in the loop.

It is often convenient use a constant-gain amplifier ($A$ is independent of frequency), 
and a bandpass filter as $\beta=\beta(\omega)$ in the feedback path.
Some small frequency dependence of the amplifier gain, which
is always present in real-world amplifier, can be moved from $A$ to
$\beta=\beta(\omega)$.  The function $\beta(\omega)$, still 
unspecified, is described graphically in Fig.\ \ref{fig:le-barkhausen}.

\begin{figure}
\namedgraphics{0.8}{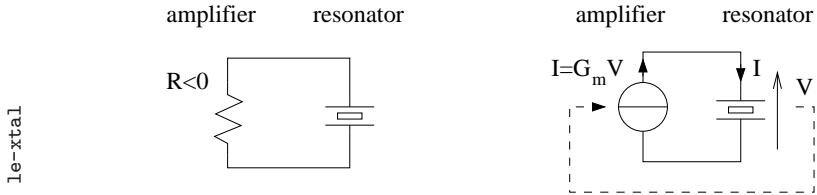}
\caption{Negative-resistance oscillator.}
\label{fig:le-xtal}
\end{figure}

The model of Fig.\ \ref{fig:le-barkhausen} is general.  It applies to a
variety of systems, electrical, mechanical, and others.  A little
effort may be necessary to identify $A$ and $\beta$.  If for example
the resonator is a two-port microwave cavity connected to an amplifier
in closed loop, matching it to Fig.\ \ref{fig:le-barkhausen} is
trivial.  A less trivial example is the negative-resistance oscillator
shown in Fig.\ \ref{fig:le-xtal}.  The feedback function
$\beta(\omega)$ is the resonator impedance
$Z(\omega)=\frac{V(\omega)}{I(\omega)}$, thus $I(\omega)$ is the input and $V(\omega)$
the output.  The resonator impedance is a complex function of
frequency that takes a real value (a resistance) at $\omega=\omega_0$.
A negative conductance $G$ plays the role of the amplifier.  We match
the oscillator of Fig.\ \ref{fig:le-xtal} to the general scheme
(Fig.\ \ref{fig:le-barkhausen}) by observing that the controlled
current generator is a transimpedance amplifier that senses the voltage
$V$ across the resonator and delivers a current $I=G_mV$.  The game of
signs deserves some attention.  The condition $|A\beta|=1$ requires
that $G_m>0$.  The sign of the current can follow two conventions, in
a generator the current is positive when it exits, in a load the the
current is positive when it enters.  Interpreting the controlled
generator as a resistor, the sign of the current is to be changed.
Thus $G_m=-G$.

\begin{figure}[t]
\namedgraphics{0.8}{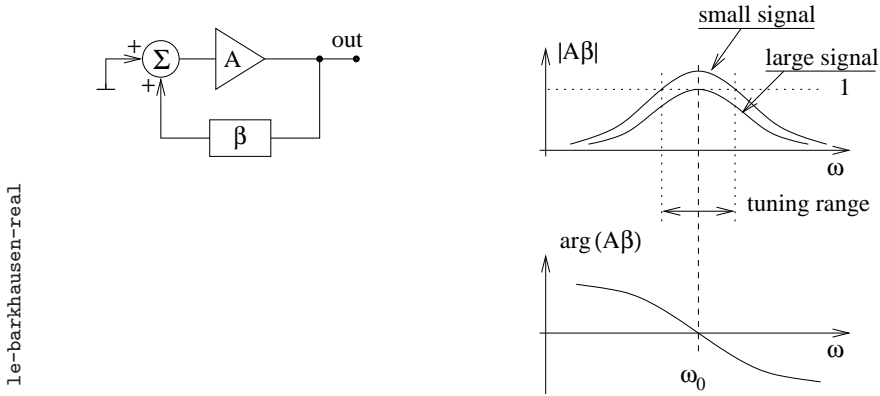}
\caption{Starting the oscillator.}
\label{fig:le-barkhausen-real}
\end{figure}
\begin{figure}[t]
\namedgraphics{0.8}{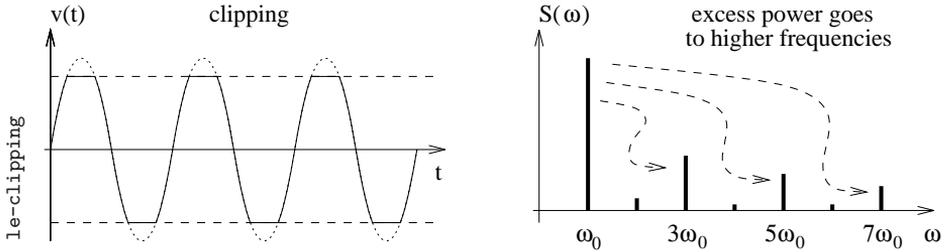}
\caption{Time-domain clipping results in power squeezed to higher harmonics.  In the example shown clipping is almost symmetric, for the odd harmonics are privileged, but the even harmonics are still present.}
\label{fig:le-clipping}
\end{figure}

Oscillation starts from noise or from the switch-on transient.
In the spectrum of such random signal, only a small energy is initially 
contained at $\omega_0$.  For the oscillation to grow up to 
a desired amplitude, it is necessary that $|A\beta(\omega)|>1$ 
at $\omega=\omega_0$ for small signals (Fig.\ \ref{fig:le-barkhausen-real}).
In such condition, oscillation at the frequency $\omega_0$ that derives from $\arg[A\beta(\omega)]=0$ rises exponentially.  
As the oscillation amplitude approaches the desired value, an amplitude
control (not shown Fig.\ \ref{fig:le-barkhausen-real}) reduces the loop gain, so that  the loop reaches the stationary condition $A\beta(\omega)=1$.  
The amplitude can be stabilized by an external AGC (automatic 
gain control), or by the large-signal saturation of the amplifier.  
Figure \ref{fig:le-clipping} shows the effect of saturation.  When the input amplitude exceeds the saturation level, the output signal is clipped.  Further increasing the input level, the gain decreases at the fundamental frequency $\omega_0$, and the excess power is squeezed into the harmonics at frequencies multiple of $\omega_0$. 

In summary, it is important to understand that in real-world
oscillators
\begin{enumerate} 
\item it is necessary that $|A\beta(\omega)|>1$ for small signals,
\item the condition $|A\beta(\omega)|=1$ results from large-signal gain saturation, 
\item the oscillation frequency is determined only by the phase condition
   $\arg[A\beta(\omega)]=0$.
\end{enumerate}
If a static phase $\psi$ is inserted in the loop (Fig.\ \ref{fig:le-tuning} right), 
the Barkhausen phase condition becomes $\arg\beta(\omega)+\psi=0$.
Hence the loop oscillates at the frequency 
\begin{equation} 
\omega_0+\Delta\omega
\qquad\text{at which}\qquad
\arg\beta(\omega)=-\psi~.  
\label{eqn:le-barkhausen-tuning}
\end{equation}
The effect of $\psi$ on the oscillation frequency is obtained by 
inverting Eq.~\ref{eqn:le-barkhausen-tuning}.  Within the accuracy of
linearization, it holds that 
\begin{equation} 
\Delta\omega=-\dfrac{\psi}{\frac{d}{d\omega}\arg\beta(\omega)}  ~.
\label{eqn:le-barkhausen-derivative}
\end{equation}
If the resonator is a simple circuit governed by a second-order differential equation
with low damping factor (i.e., large merit factor $Q$), in the vicinity of the resonant frequency $\omega_0$ it holds that $\frac{d}{d\omega}\arg\beta(\omega)=\smash{-\frac{2Q}{\omega_0}}$.  Thus
\begin{align} 
\frac{\Delta\omega}{\omega_0}=\frac{\Delta\nu}{\nu_0}
     &=\frac{\psi}{2Q} 
     &\text{resonator, for~} \frac{\Delta\omega}{\omega_0}\ll\frac{1}{2Q}~.
\end{align}
For reference, $\psi>0$ means that the loop leads in the time domain, consequently the 
oscillator is pulled to a frequency higher than the exact resonance.

\subsection{Pulling the oscillator frequency}\label{ssec:le-heur-tuning}
There exist (at least) two ways to tune an oscillator to the desired frequency.

\begin{figure}[t]
\namedgraphics{0.8}{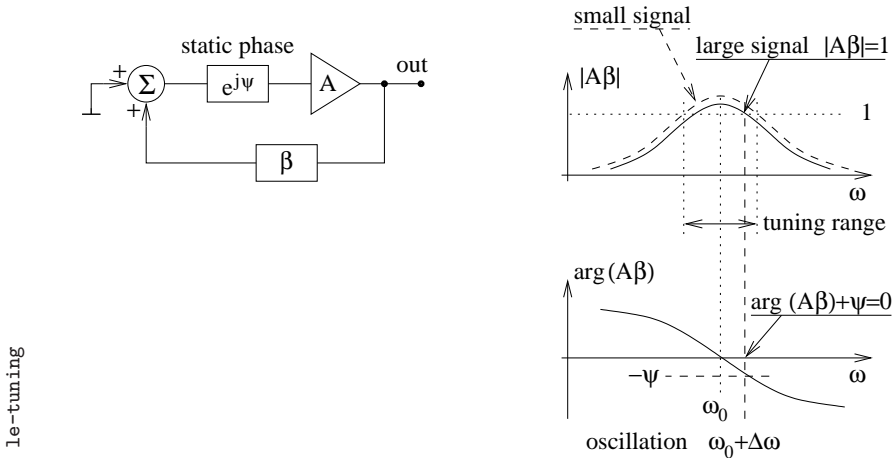}
\caption{Tuning the oscillation frequency by insertion of a static phase.}
\label{fig:le-tuning}
\end{figure}

\paragraph{Introduce a static phase shift in the loop}
The first method consists of introducing a static phase lead or lag $\psi$ in the loop, as in Fig.\ \ref{fig:le-tuning}.  Oscillation is ruled by the Barkhausen condition \req{eqn:le-barkhausen-complex}, with saturated amplitude.  Therefore, the oscillator tuning range the frequency range in which
\begin{align} 
|A\beta(\omega)|&>1\qquad\text{(small signal)}~.
\end{align}
In this region, the gain can be reduced by saturation and the phase determines the oscillation frequency.  Out of this range, the response to a perturbation decays exponentially, hence no oscillation is possible.  This method is used in microwave oscillators, where a phase shifter is used to set the static phase $\psi$.

\begin{figure}[t]
\namedgraphics{0.8}{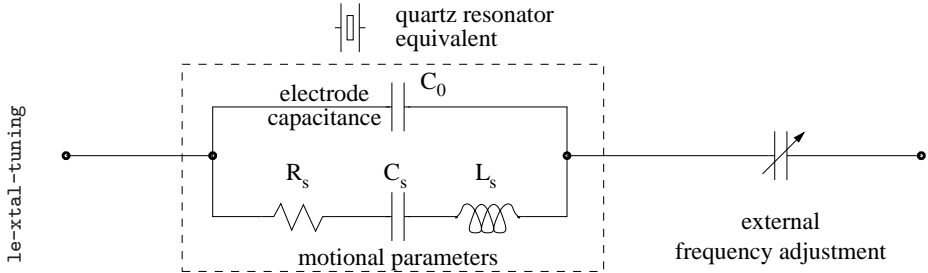}
\caption{Typical tuning scheme for quartz oscillators.}
\label{fig:le-xtal-tuning}
\end{figure}

\paragraph{Change the natural frequency of the resonator}
The~second~method consists of pulling the natural frequency of the resonator by modifying the parameters of the resonator differential equation. The adjustment circuit is no longer distinct from the resonator, and there is no reason to introduce the static phase $\psi$.  
This method is often used in quartz oscillators, where a variable capacitor is used to alter on the resonator natural frequency (Fig.\ \ref{fig:le-xtal-tuning}).

\paragraph{The effect on phase noise}
Analyzing noise, the two methods are quite different.  
The static phase $\psi$, inherently, increases the noise bandwidth of the resonator.
Conversely, the reactance used to pull the natural frequency of the resonator has not such intrinsic effect.  Nonetheless, in practice the resonator noise bandwidth still increases because the additional loss introduced by the external reactance reduces the merit factor.
More details are given in Section \ref{ssec:le-theory-tuning}.

\section{The Leeson formula}
%
Let us consider an oscillator in which the feedback circuit $\beta$ is
an ideal resonator, free from frequency fluctuations, with a 
large\footnote{Strictly, only $Q\gg1$ is necessary.} merit factor $Q$.  The resonator relaxation time is
\begin{align} 
\tau = \frac{Q}{\pi}T_0 = \frac{Q}{\pi\nu_0} = \frac{2Q}{\omega_0} 
\label{eqn:le-relaxation-time-heuristic}
\end{align}
Let us then replace the static phase $\psi$ of Eq.~\req{eqn:le-barkhausen-derivative}
with a random phase fluctuation $\psi(t)$ (Fig.\ \ref{fig:le-tuning}) that accounts 
for all the phase noise sources in the loop.
There follows an oscillator output signal of the form
\begin{align} 
v(t) = V_0\cos[\omega_0t+\phi(t)]~, 
\end{align}
where $\phi(t)$ is the effect of $\psi(t)$.  We analyze the mechanism with which the power spectrum density of $\psi$ is transferred into $\phi$.
\begin{figure}
\namedgraphics{1}{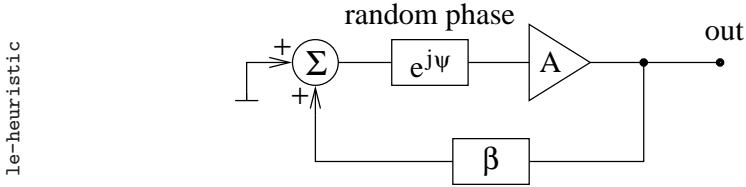}
\caption{The phase noise of the amplifier and of all other
  components of the loop is modeled as a random phase $\psi$ at the
  input of the amplifier.}
\label{fig:le-heuristic}
\end{figure}

For the \emph{slow components} of $\psi(t)$, slower than the inverse of the
relaxation time, $\psi$ can be treated as quasi-static perturbation.
Hence 
\begin{align} 
\Delta\nu&=\frac{\nu_0}{2Q}\psi(t)
\intertext{and}
S_{\Delta\nu}(f)&=\left(\frac{\nu_0}{2Q}\right)^2S_\psi(f)~.
\end{align}
The instantaneous output phase is 
\begin{align} 
\phi(t)&= 2\pi\int(\Delta\nu)\:dt~.
\end{align}
The time-domain integration maps into a multiplication
by $\smash{\frac{1}{j\omega}}$ in the Fourier transform domain, thus 
into a multiplication by $\smash{\frac{1}{(2\pi f)^2}}$ in the spectrum.
Consequently, the oscillator spectrum density is
\begin{align} 
S_\phi(f)&=\frac{1}{f^2}\left(\frac{\nu_0}{2Q}\right)^2S_\psi(f)~.
\label{eqn:le-leeson-slow}
\end{align}

For the \emph{fast fluctuations} of $\psi$, faster than the inverse of the
relaxation time, the resonator is flywheel that steers the signal.
Loosely speaking, it is open circuit for the phase fluctuation.  The 
fluctuation $\psi(t)$ crosses the amplifier and shows up at the output, without being fed back at the amplifier input.  No noise regeneration takes place in this conditions, thus
$\phi(t)=\psi(t)$, and
\begin{align}
S_\phi(f)&=S_\psi(f)~.
\label{eqn:le-leeson-fast}
\end{align}
Under the assumption that there is no correlation between fast
and slow fluctuations, we can add the effects stated by 
Equations \req{eqn:le-leeson-slow} and \req{eqn:le-leeson-fast}
\begin{align} 
S_\phi(f)&=\left[1+\frac{1}{f^2}\left(\frac{\nu_0}{2Q}\right)^2\right]S_\psi(f)
	\qquad\begin{array}{c}\text{Leeson}\\[-0.4ex]\text{formula}\end{array}.
\label{eqn:le-leeson-heuristic}
\end{align}
The above can be rewritten as
\begin{align} 
S_\phi(f)&=\left[1+\frac{f^2_L}{f^2}\right]S_\psi(f)~,
\label{eqn:le-leeson-heuristic-with-fl}
\intertext{where}
f_L&=\frac{\nu_0}{2Q}=\frac{1}{2\pi\tau}
  &\hspace*{-30ex}\text{Leeson frequency}
\label{eqn:le-leeson-heuristic-fl-def}
\end{align}
is the Leeson frequency.  
\begin{figure}[t]
\namedgraphics{0.8}{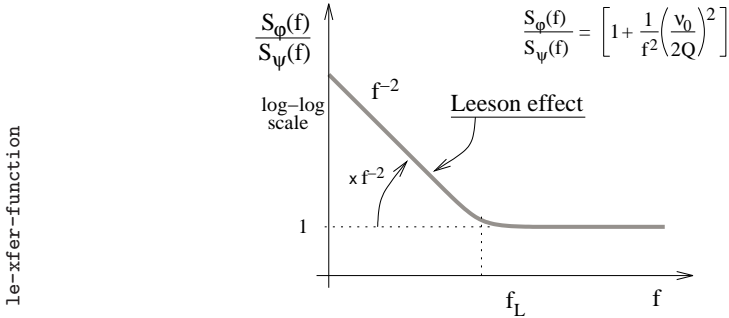}
\caption{The Leeson-effect transfer function.}
\label{fig:le-xfer-function}
\end{figure}%
By inspection on Eq.~\req{eqn:le-leeson-heuristic-with-fl}, the oscillator
behavior is that of a \emph{first-order filter} with a perfect
integrator (a pole in the origin in the Laplace transform domain) and
a cutoff frequency $f_L$ (a zero on the real left-axis), as 
shown in Fig.\ \ref{fig:le-xfer-function}.  The filter
time constant is the relaxation time $\tau$ of the resonator. 

It is to be made clear that Eq.~\req{eqn:le-leeson-heuristic}, and
equivalently Eq.~\req{eqn:le-leeson-heuristic-with-fl}, accounts only for
the phase-to-frequency conversion inherent in the loop.  The
\emph{resonator noise} is still to be added for the noise spectrum to
be correct.

\begin{figure}[t]
\namedgraphics{0.8}{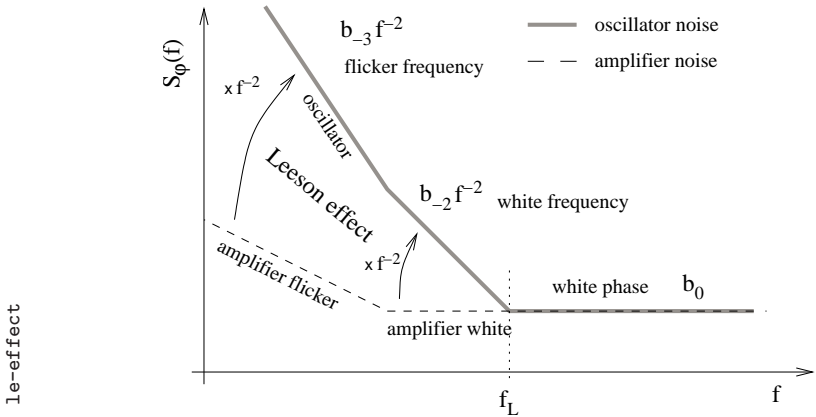}
\caption{The Leeson effect, shown for a typical microwave oscillator.}
\label{fig:le-effect}
\end{figure}

To summarize, the Leeson effect [Eq.~\req{eqn:le-leeson-heuristic}--\req{eqn:le-leeson-heuristic-with-fl}]
consists of the multiplication by $f^{-2}$ of the amplifier phase noise
spectrum below the Leeson frequency $f_L=\frac{\nu_0}{2Q}$.  This
behavior is quite general since the amplifier noise is still
unspecified.  Figure~\ref{fig:le-effect} shows the Leeson effect in a
typical case (microwave oscillator), in which the amplifier shows white
and flicker phase noise.

The formula \req{eqn:le-leeson-heuristic} was originally proposed by David
B. Leeson \cite{leeson66pieee} as a model for short-term frequency fluctuations, 
inspired to the magnetron for radar applications.  It was perfectly sound from this standpoint to
consider the cavity ideally stable in the short term (1 ms corresponds to a round-trip of 300 km), 
and to ascribe all the noise to the amplifier, which relies on an electron beam.

\subsection{Delay-line oscillator}
\begin{figure}[t]
\namedgraphics{0.8}{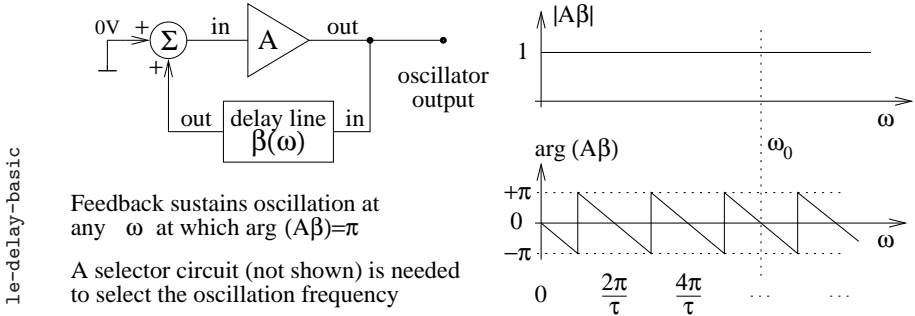}
\caption{Basic delay-line oscillator.}
\label{fig:le-delay-basic}
\end{figure}

The frequency reference can be a delay line instead of the resonator, as shown in 
Fig.\ \ref{fig:le-delay-basic}.   In the frequency domain, the delay line is described by $\beta(\omega)=e^{-j\omega\tau}$.  Thus the loop can sustain any oscillation frequency for which $\arg\beta(\omega)=0$. 
A selector circuit, not shown in Fig.\ \ref{fig:le-delay-basic}, is therefore necessary to select a specific oscillation frequency $\omega_0$.

The Leeson effect is derived in quasistatic conditions from Eq.~\req{eqn:le-barkhausen-derivative} 
\begin{equation} 
\Delta\omega=-\dfrac{\psi}{\frac{d}{d\omega}\arg\beta(\omega)} ~.
\end{equation}
In the case of the delay line it holds that $\smash{\frac{d}{d\omega}}\arg\beta(\omega)=-\tau$.  
Interestingly, the delay line is equivalent to a resonator of resonant frequency 
$\nu_0$ and merit factor
\begin{align} 
Q&=\pi\nu_0\tau~,
\intertext{thus}
f_L=\frac{1}{\pi\tau}
\end{align}
to the extent that it has the same slope $\smash{\frac{d}{d\omega}}\arg\beta(\omega)$
of the resonator.

For \emph{slow} fluctuations it holds that 
\begin{equation} 
\Delta\nu=\dfrac{\psi}{2\pi\tau}\qquad f\ll f_L~,
\end{equation}
and therefore 
\begin{equation} 
S_\phi(f)=\frac{1}{f^2}\:\frac{1}{4\pi^2\tau^2}\:S_\psi(f)\qquad f\ll f_L~.
\end{equation}
The noise propagation of \emph{fast} phase fluctuations ($f\gtrsim f_L$) from the 
amplifier input to the oscillator output is far more complex.  
Chapter \ref{chap:le-delayline} is devoted to this topic.

\section{Amplifier noise}\label{sec:le-ampli-noise}
%
\subsection{Additive white noise}  
The amplifier noise is described in terms of the \emph{noise temperature} $T_a$ defined as follows.  When the amplifier is input-terminated to a resistor at the temperature $T_0$, the equivalent spectrum density at the amplifier input is $N=k(T_a+T_0)$. The amplifier noise is therefore
\begin{align}
N_a&=kT_a~.
\label{eqn:le-noise-temp-def}
\end{align}
The spectrum density of the equivalent input noise can also be written as 
\begin{align}
N_a&=FkT_0~,
\label{eqn:le-noise-fig-def}
\end{align}
The above Eq.\ \req{eqn:le-noise-fig-def} defines the \emph{noise figure} $F$.     
By equating \req{eqn:le-noise-temp-def} and \req{eqn:le-noise-fig-def}, we find  
\begin{align}
N_a&=(F-1)kT_0~,
\intertext{and}
F&=\frac{T_a+T_0}{T_0}~.
\end{align}
The unambiguous definition of $F$ requires that the temperature $T_0$
is specified.  The standard value is $290$ K (17 \celsius).  Accordingly, 
it holds that $kT_0=4{\times}10^{-21}$ J, that is, $-174$ dBm in 1 Hz bandwidth.

When amplifiers are cascaded, the input noise $F_n-1$ of the $n$-th amplifier is divided by the power gain $\prod_{m=1}^{n-1}A_m^2$ of the preceding amplifiers.  The noise figure of the chain is given by the Friis formula \cite{friis44ire}
\begin{align}
F &= F_1 + \frac{F_2-1}{A_1^2} + \frac{F_3-1}{A_1^2A_2^2} + \ldots
\qquad\begin{array}{c}\text{Friis}\\[-0.4ex]\text{formula}\end{array}
\end{align}

The typical noise figure of low-noise amplifiers is of 1--2 dB,
depending on technology, on frequency, and on bandwidth.  
The effect of bandwidth shows up clearly in microwave amplifiers, where the active devices have low noise temperature and capacitive input.
The loss of the impedance-matching network necessary to match the resistive amplifier input (50~\ohm) to the capacitive transistor gate turns into increased noise figure.
Needless to say, larger bandwidth design turns into higher input loss, thus into higher noise figure.

The definitions of $T_a$ and $F$ implicitly assume that the
amplifier noise is a random process \emph{added} to the useful signal,
and not correlated to it.  In the presence of a sinusoidal carrier of
power $P_0$, the phase noise is
\begin{align}
S_\phi(f)=b_0=\frac{FkT_0}{P_0}~&& \text{constant}.
\label{eqn:le-sphi0-noise-figure}
\end{align}
%

\subsection{Flicker noise}  
It has been experimentally observed \cite{halford68fcs,walls97uffc,hati03fcs}
that phase flickering of different amplifier types falls in a relatively narrow range, and that
for a given amplifier the phase flickering is about independent of the
carrier power.  Consequently the flicker noise of $m$ cascaded amplifiers of the same type is about $m$ times the noise of one amplifier.  This is radically different from the case white noise, where the noise of a stage referred to the input is divided by the gain of all the preceding stages.  

Table~\ref{tab:le-phase-flickering} shows the typical phase flickering of commercial amplifiers.
\begin{table}
\caption{Typical phase flickering of amplifiers.}
\label{tab:le-phase-flickering}
\begin{center}
\begin{tabular}{lccl}
~~RATE~~ & MOS          &  bipolar HF\\
                  & microwave &  HF/UHF \\ \hline     
~~fair &    $-100$      &  $-120$    & \unit{dBrad^2/Hz}   \\ 
~~good &    $-110$      &  $-130$       \\ 
~~best &    $-120$      &  $-140$ 
\end{tabular}\end{center}
\end{table}

\begin{figure}[t]
\namedgraphics{0.8}{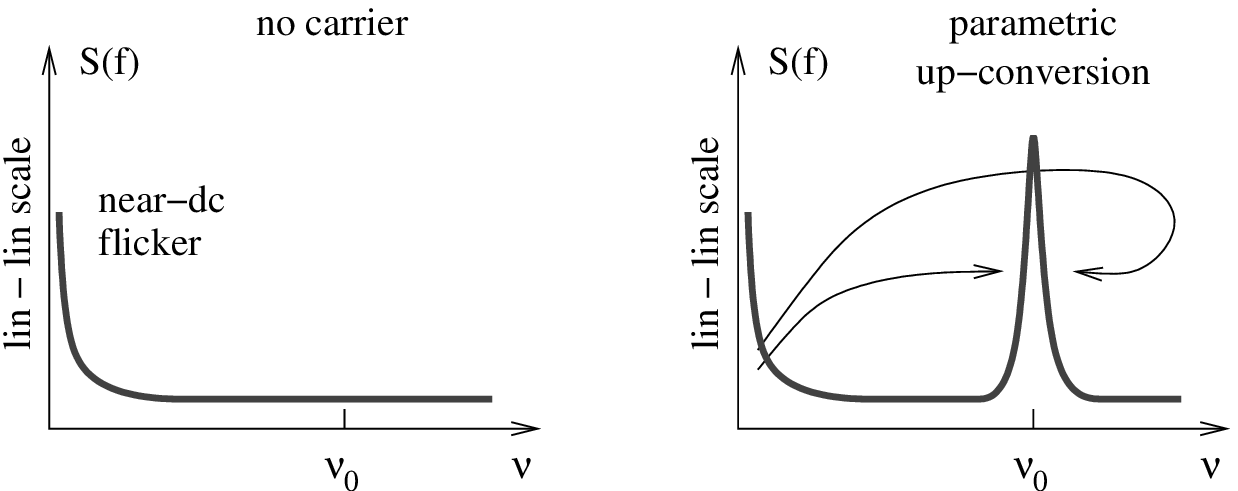}
\caption{Parametric up-conversion of near-dc flicker in amplifiers.}
\label{fig:le-param-noise}
\end{figure}

The mechanism that originates phase flickering is a low-frequency (close to dc) random process with spectrum of the flicker type that modulates the carrier (Fig.\ \ref{fig:le-param-noise}).  This mechanism is often called \emph{parametric noise}\footnote{Of course, the term \emph{parametric noise} is more general than the phase flickering.} because the near-dc flickering modulates a parameter of the device high-frequency model.

A simplified nonlinear model provides insight in the counterintuitive behavior of phase noise.  Let us consider a signal $x(t)=V_0\cos(\omega_0t)+n(t)$ inside a device, where the sinusoid is the input signal, and $n(t)$ is the near-dc flickering of the dc bias.  Additionally, let us assume that the device is slightly nonlinear, and that the nonlinearity can be expanded as the 2nd-degree polynomial $P(x)=a_0+a_1x+a_2x^2$.  Feeding $x(t)$ into the polynomial, we get a carrier term $a_1V_0\cos(\omega_0t)$, plus a close-in noise term $a_2V_0n(t)\cos(\omega_t)$.  Inspecting on the radio-frequency spectrum around $\omega_0$, we find that the power of the noise sidebands is proportional to $V_0^2$, thus to the power of the input signal.  This is in agreement with the simple observation that there can not be close-in noise in the absence of a carrier.  As a consequence, the fractional amplitude fluctuation, i.e., the close-in noise divided by the carrier, is $\alpha(t)=\frac{2a_2}{a_1}n(t)$, independent of the carrier power.  Similarly, a 2nd-degree nonlinear model that involves a variable reactance mechanism (varactor effect in transistors and MOS) leads to close-in phase noise independent of the carrier power. 

Phase flickering depends on the physical size of the amplifier active region.  This can be proved through a gedankenexperiment in which we split the input signal into $n$ equal branches, amplify and recombine.  The power gain is that of a single branch.
Conversely, the flicker noise is 
$(b_{-1})_\mathrm{tot}=\smash{\frac1n}(b_{-1})_\mathrm{branch}$ because the branch amplifiers are independent and the phase flickering of each is not affected by having reduced the power by $n$.  If we join the $n$ amplifiers in a single one that employs an active region (base in the case of bipolar transistor, or channel in the case of field-effect transistors) $n$ times larger, the phase noise reduction is kept.  The additional hypothesis is required, that the near-dc flickering takes place at microscopic scale, for there is no correlation between the different regions of the $n$-volume device.  This hypothesis is consistent with the two most accredited models for the flicker noise \cite{hooge69pla,mcwhorter57}.

Finally, phase flickering is related to the amplifier gain.  This is a side effect of the number of stages needed for a given gain, rather than a gain effect in a single stage.

\subsection{Other noise types}  
Noise phenomena with a slope significantly steeper than $-1$ in the
spectrum, say $f^{-2}$, do not exist in amplifiers.  If such 
phenomenon was present, the delay of the amplifier would diverge
rapidly.  This never happens in practice.  This general statement does
not exclude some bumps in the spectrum, for example due to the
environment temperature, which in some frequency range yield a spectrum locally steeper than $f^{-1}$.

\begin{figure}
\namedgraphics{0.8}{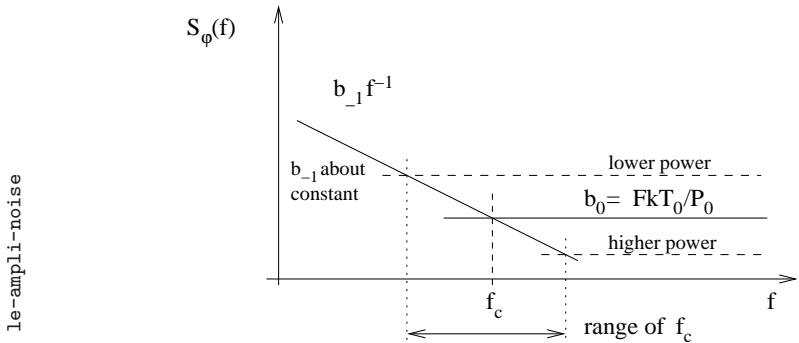}
\caption{Typical phase noise of an amplifier.}
\label{fig:le-ampli-noise}
\end{figure}
%
\subsection{Phase noise spectrum}
The total phase noise spectrum results from adding the white and the
flicker noise spectra, as in Fig.\ \ref{fig:le-ampli-noise}.  This
relies on the assumption that white and flicker phenomena are
independent, which is true for actual amplifiers.

It is important to understand that $b_0$ (white)
is proportional to the inverse of the carrier power $P_0$, while
$b_{-1}$ (flicker) is about independent of $P_0$.  The
corner frequency $f_c$ depends on the input power.  
The belief that $f_c$ is a noise parameter of the amplifier 
is a common mistake.

\subsection{Noise-corrected amplifiers}
%
The amplifier flicker noise is of paramount importance to the
oscillator frequency stability.  Reducing the flicker, even if this is
done at expense of higher white noise, results in improved oscillator
stability.  A new generation of oscillators make use of a
noise-corrected amplifier in the loop.  This technique is based on a bridge 
scheme that takes the difference between the input and output of the amplifier, 
compensating for gain and delay.  Balancing the bridge, the differential
signal is the amplifier distortion and noise.  The latter is amplified, detected, and used
to compensate for the amplifier noise in closed loop.

In a noise-corrected amplifier the input is split into two branches, that is, the main amplifier and the noise-correction amplifier.  A directional coupler may be necessary, which introduce a loss of 3 dB, plus the dissipative loss.  Such loss turns into an increase in noise figure of the same amount.  If the noise figure of the internal amplifier is of 1--2 dB and the total loss of the coupler is of 4 dB, the noise figure of the corrected amplifier is of 5--6 dB\@.  On the other hand, the correction results in a reduction of 20--40 dB in the phase flickering.

\section{The phase noise spectrum of real oscillators}
%
\begin{figure}[t]
\namedgraphics{0.8}{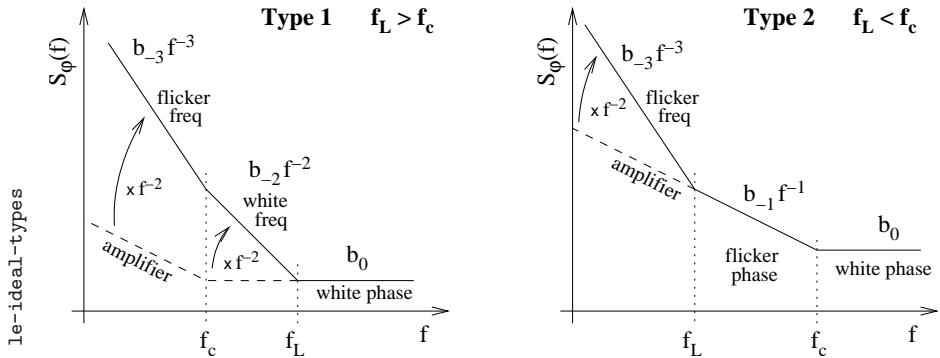}
\caption{With a flickering amplifier, 
  the Leeson effect yields two types of spectrum, type 1 for $f_c<f_L$
  and type 2 for $f_c>f_L$.  A noise-free resonator is assumed.}
\label{fig:le-ideal-types}
\end{figure}
For a given amplifier, the phase noise (Fig.\ \ref{fig:le-ampli-noise})
is white at high $f$, and of the flicker type below the cutoff frequency
$f_c$ that results from the carrier power $P_0$.  When such amplifier
is inserted in an oscillator, it interacts with the resonator in the way shown 
in Fig.\ \ref{fig:le-ideal-types}.
Two basic types of interaction are possible.

Type 1 is the most frequently encountered.  It is typical of microwave
oscillator and high-frequency ($\ge100$ MHz) piezoelectric oscillators,
in which $f_L$ is made high by the high resonant frequency and by the
low merit factor $Q$.  Looking at Fig.\ \ref{fig:le-ampli-noise} from
right to left, the amplifier phase noise is white and the Leeson
effect originates white frequency noise ($b_{-2}f^{-2}$).  At lower
frequencies the amplifier phase noise turns into flicker, hence the
oscillator noise turns into frequency flickering ($b_{-3}f^{-3}$).  No
flicker is present in the output phase spectrum.

Type 2 is found in low-frequency (5--10 MHz) high-stability quartz
oscillator, in which the merit factor may exceed $10^6$.  Looking at
Fig.\ \ref{fig:le-ampli-noise} from right to left, the amplifier phase
noise turns from white to flicker at $f=f_c$.  Accordingly, phase
flickering ($b_{-1}f^{-1}$) is visible at the oscillator output.  At
lower frequencies the Leeson effect takes place, hence the oscillator
noise turns into frequency flickering ($b_{-3}f^{-3}$).  No white
frequency noise ($b_{-2}f^{-2}$) is present in the output phase
spectrum.

By inspection on Fig.\ \ref{fig:le-ampli-noise}, there can be either the $f^{-1}$
or the $f^{-2}$ noise types, not both.

\subsection{The effect of the resonator noise}
\begin{figure}[t]
\namedgraphics{0.8}{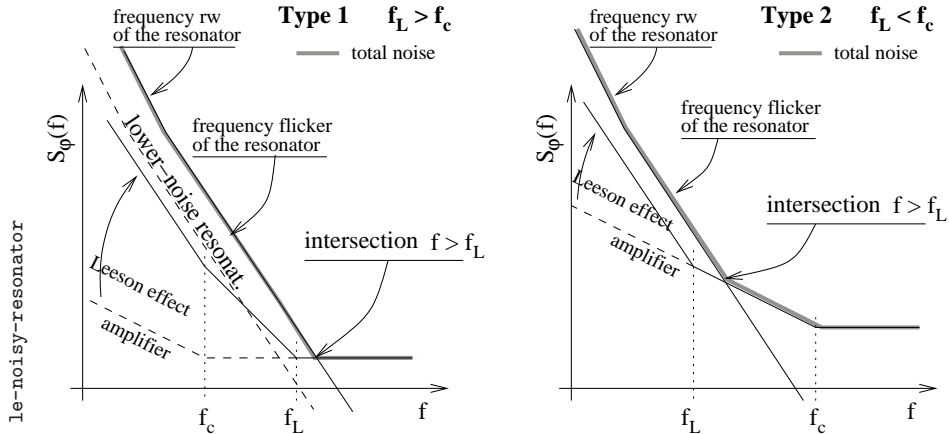}
\caption{Effect of the resonator frequency 
  fluctuations on the oscillator noise.}
\label{fig:le-noisy-resonator}
\end{figure}
The dissipative loss of the resonator, inherently, originates white noise.  Yet, 
the noise phenomena most relevant to the oscillator stability are the flicker and the random
walk of the resonant frequency $\nu_0$. Thus, the spectrum $S_y(f)$ of
the fractional frequency fluctuation $y=\Delta\nu/\nu_0$ shows a term
$h_{-1}f^{-1}$ for the frequency flicker, and $h_{-2}f^{-2}$ for the
frequency random walk.  The relationship\footnote{Start from the
  fractional frequency fluctuation
  $y=\smash{\frac{1}{2\pi\nu_0}\frac{d\phi(t)}{dt}}$.  The time-domain
  derivative maps into a multiplication by $\omega^2$, thus by $(2\pi f)^2$, in the
  spectrum.} between $S_\phi(f)$ and $S_y(f)$ is
\begin{equation}
S_y(f)=\frac{f^2}{\nu_0^2}~S_\phi(f)~.
\end{equation}
Accordingly, the term $h_{-1}f^{-1}$ of the resonator fluctuation
yields a term proportional to $f^{-3}$ in the phase noise, and the
term $h_{-2}f^{-2}$ yields a term $f^{-4}$.  The resonator fluctuation
is independent of the amplifier noise, for it adds to the oscillator noise.  

Figure ~\ref{fig:le-noisy-resonator} shows the two basic spectra
of Fig.\ \ref{fig:le-ideal-types}, after introducing the resonator frequency fluctuation.  
The resonator fluctuation may hide the amplifier corner frequency $f_c$, the Leeson effect, or both.  Three behavior types deserve attention.
\begin{enumerate}
\item The resonator noise hides $f_c$ but not $f_L$.  This is found in type-1 spectra.
The cross point of the $f^{-3}$ resonator noise and the $f^{-2}$ noise due to the Leeson effect has the same graphical signature of $f_c$, but is not.

\item The resonator noise hides $f_c$ and $f_L$.  This is found in type-1 and type-2 spectra.  Only one corner is visible on the plot, where the $f^{-3}$ resonator noise crosses the  $f^{0}$ amplifier noise.  This behavior is found for example in VHF quartz oscillators. 

\item The resonator noise hides $f_L$, but not $f_c$.  This is expected\footnote{This case needs some more theoretical analysis.} in type-2 spectra.  
\end{enumerate}

\subsection{The effect of the output buffer}\label{ssec:le-out-buffer}
%
\begin{figure}[t]
\namedgraphics{0.8}{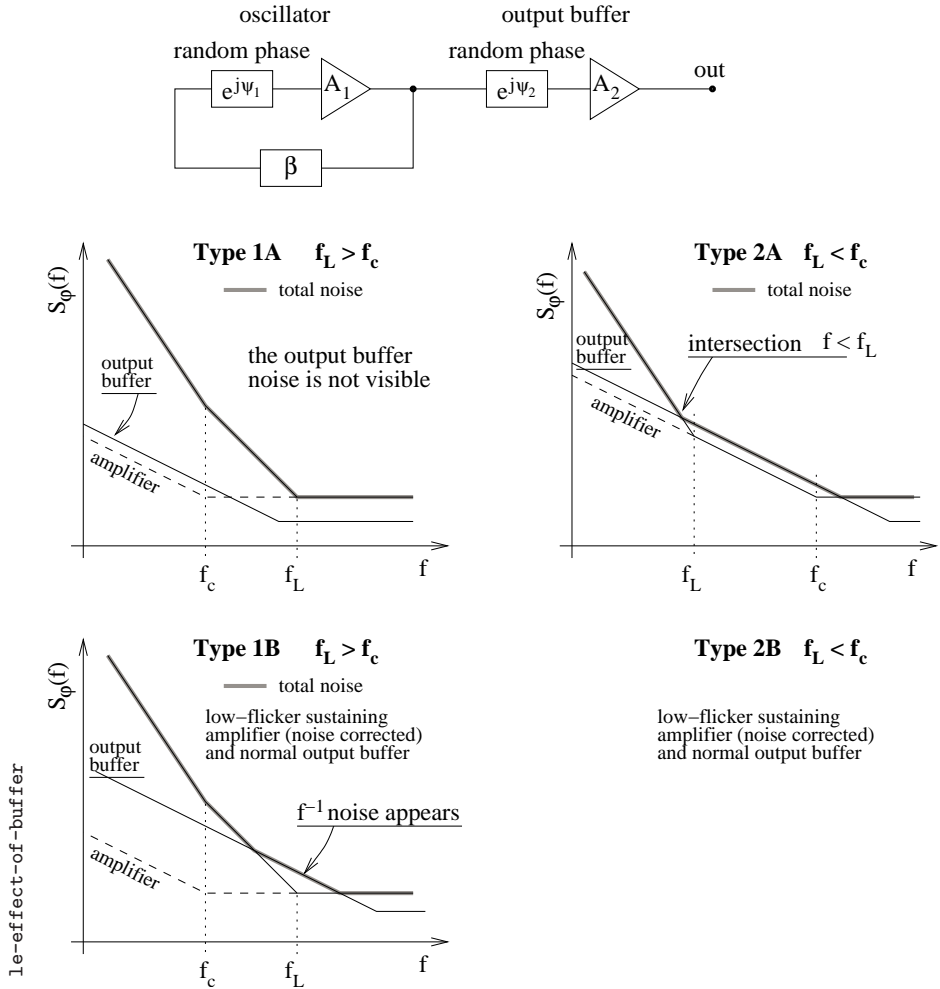}
\caption{Effect of the output buffer on the oscillator noise.}
\label{fig:le-effect-of-buffer}
\end{figure}
It is interesting to analyze the case of an oscillator with ideal
resonator and noisy amplifier, followed by a noisy output buffer
(Fig.\ \ref{fig:le-effect-of-buffer}).  Of course the output buffer is
independent of the oscillator, for the two noise spectra barely add.

The \emph{white noise} comes from the additive white noise $N$
referred to the carrier power $P_0$.  The noise $FkT_0$ of the
sustaining amplifier is amplified at the input of the buffer.
Therefore, the white phase noise of the buffer ends up to be
negligible in virtually all practical cases.  On the other hand,
the \emph{flicker phase
  noise} of amplifiers is about independent of the carrier power,
hence the phase noise of the buffer is not expected to be negligible.
Accounting for the buffer noise, the spectra of Fig.\ \ref{fig:le-ideal-types} 
are to be modified as shown in Fig.\ \ref{fig:le-effect-of-buffer}.

In type 1A and 2A spectra, the sustaining amplifier and the buffer have similar flicker
characteristics.
In type 1A ($f_L>f_c$), the phase flickering of the buffer is hidden by
the Leeson effect.  The insertion of the output buffer lets the
spectrum is substantially unchanged.

In type 2A ($f_L<f_c$), the phase flickering of the buffer adds to the
phase flickering of the oscillator.  As a consequence, the
corner point at which the $f^{-1}$ noise turns into $f^{-3}$ is
pushed towards lower frequencies.  This corner point can be easily
mistaken for the Leeson frequency because it has the same graphical
signature.   

In type 1B spectrum, the sustaining amplifier exhibits low phase flickering, 
significantly lower than that of the buffer. 
As a result, $f_c$ is a low frequency.  This is the case of
some low-noise microwave oscillators, thus we assume $f_L>f_c$ even in
the case of large merit-factor oscillators.  A simple amplifier is
used as the buffer.  If the merit factor of the resonator is large
enough (low $f_L$), the $f^{-1}$ phase noise of the buffer shows up,
hiding the $f^0$ to $f^{-2}$ slope transition characteristic 
of the Leeson effect. Yet, the Leeson frequency can be estimated
extrapolating the $f^{-2}$  segment still visible.

If we assume that $f_L<f_c$, the case of the noise corrected amplifier
and simple buffer is similar to the type 2A, but for a larger buffer noise.
Consequently, the $f^{-1}\rightarrow f^{-3}$ corner slides leftwards, farther from the true Leeson frequency.

\chapter{Oscillator hacking}
The combined knowledge of the oscillator noise theory, of general
physics, and of electronic technology, enables to understand the
inside of an oscillator from the data sheet, and to guess some relevant
internal parameters like $P_0$, $Q$, $f_L$, amplifier $1/f$ noise, etc.
We describe the guidelines of this process and 
show some examples.  The need of guessing the internal technology 
is a source of difficulties and of inconsistencies which may make the interpretation 
only partially reliable.  Coping with this is a part of the message addressed to the
reader.

\paragraph{Inspection on the data sheet.}
The first step consists of reading carefully the data sheet focusing
on the resonator and on the amplifier technology, and bringing up to
the mind as many related facts as possible.  For example, a 5 MHz
quartz can have a merit factor in excess of $10^{6}$, but it must be
driven at very low power, say 10--20 $\mu$W, for best
long-term stability.  The merit factor of a dielectric resonator can
be of 1000 or more, depending on size and frequency.  So on for the
other resonator types, and for the amplifiers.  Similar oscillators
encountered in the past may have a similar spectrum, or be
surprisingly different.

\paragraph{Parametric estimation of the spectrum.}
This part of the process consists of matching the phase noise spectrum
with the polynomial $S_\phi(f)=\sum_{i=-4}^{0}b_if^i$ in order to
identify the coefficients $b_i$.  A term $b_if^i$ on a log-log plot appears as a straight line
\begin{center}
\begin{tabular}{llc}
noise type        & term           & slope \\\hline
white phase       & $b_0$          & 0 \\
flicker phase     & $b_{-1}f^{-1}$ & $-10$ dB/dec \\
white frequency   & $b_{-2}f^{-2}$ & $-20$ dB/dec \\
flicker frequency & $b_{-3}f^{-3}$ & $-30$ dB/dec \\
frequency r.\,w.\ & $b_{-4}f^{-4}$ & $-40$ dB/dec
\end{tabular}
\end{center}
The actual spectra are of the form $S_\phi(f)=\sum_{i=-4}^{0}b_if^i+\sum_js_j(f)$, where the terms $s_j(f)$ account for the residuals of the mains (50 Hz or 60 Hz and multiples), for bumps due to feedback and for other stray phenomena.  Figures~\ref{fig:le-miteq-dro-mod} to \ref{fig:le-oew-oeo-mod} provide some examples of actual phase noise spectra.
The mathematical process of matching the spectrum to a model is called 
\emph{parametric estimation} \cite{percival:spectrum-analysis,jenkins:spectral-analysis}.   
Some a-priori knowledge of the nature of the stray signals may be necessary to match  the complete model $\sum_{i=-4}^{0}b_if^i+\sum_js_j(f)$ to the observed spectrum.
Although (almost) only in the power-law coefficients $b_i$ are relevant in the end, the $s_j(f)$ are essential in that they reduce the bias and residuals of the estimation.

Whereas computers provide accuracy, a general parametric estimator is not easy to implement.  Conversely, the human eyes do well in filtering out the stray signals and getting a good straight-line (polynomial) approximation. The inspection on a log-log plot by sliding old-fashion squares and rulers proves to be surprisingly useful.
Spectra are often shown as $\mathcal{L}(f)$, don't forget that $\mathcal{L}(f)=\frac{1}{2}S_\phi(f)$.  Pick up the reference slope $f^i$ using the largest possible area on the coordinate frame, slide the square until its side coincides to the corresponding portion of the spectrum ($b_if^i$), and get the coefficients $b_i$.  Proceed from right to left, thus from $b_0$ to $b_{-4}$.  

Generally, at the corner between two straight lines the true spectrum is 3~dB above the corner point.  This is due to either one of the following reasons.  
In the first case, the difference in slope is 1 at the corner point.  
When this occurs, there are two independent random processes whose spectrum takes the same value ($b_if^i=b_{i+1}f^{i+1}$) at the corner point.
In the second case the difference in slope is of 2, for at the corner point it
holds that $b_if^i=b_{i+2}f^{i+2}$.  This occurs when a single noise process is filtered, due to the Leeson effect.  The factor 2 (3~dB) at the corner point results from a single real zero of the complex transfer function.

In some cases the difference between the spectrum and the straight-line approximation at a corner is not of 3 dB\@.  When this occurs, one should work two estimations, one based on the straight-line fitting, and the other based on the 3 dB difference between straight lines and true spectrum at the corner frequencies.  The best estimate is a weighted average of the two worked-out spectra.
Physical judgment should be used to assign unequal weights.

\paragraph{Interpretation.}  
This part of the process starts from the identification of the spectrum type, among those analyzed in the previous chapter.  Then we get into the learning process, in which each oscillator is a unique case.  As a general rule, one should proceed from the right-hand side of the spectrum to left, thus from white phase noise to frequency flicker or to random walk.

Starting from the white phase noise, we evaluate the power $P_0$ at the
input of the sustaining amplifier using $S_\phi(f)=b_0=FkT_0/P_0$
[Eq.~\req{eqn:le-sphi0-noise-figure}]. Thus, 
\begin{align}
P_0&=\frac{FkT}{b_0} 
\end{align}
One may admit a noise figure $F=1$ dB for conventional amplifiers, and of $F=5$ dB for noise-corrected amplifiers, due to the input power splitter.  Thanks to the gain of the  sustaining amplifier, the white noise of the output buffer can generally be neglected.

The next step is to evaluate $f_c$ (flicker of the sustaining amplifier) and $f_L$, in order of occurrence from right to left. It is then necessary to guess the oscillator sub-type (Fig.~\ref{fig:le-effect-of-buffer}).  A major difficulty is to understand whether the oscillator stability derives from the Leeson effect or from the resonator fluctuation.  
Inverting
Eq.~\req{eqn:le-leeson-heuristic-fl-def}, the Leeson frequency gives the merit factor
\begin{align}
Q&=\frac{\nu_0}{2f_L}~.
\end{align}
The corner frequency $f_c$ reveals the phase flickering of the amplifier
\begin{align}
(b_{-1})_\mathrm{ampli}&=b_0f_c~
\end{align}
If the the spectrum is of the type 1B (noise-corrected sustaining amplifier),
the $1/f$ noise is the phase noise of the output buffer.

Finally, the Allan variance of the fractional frequency fluctuation (Tab.~\ref{tab:le-noise-conversion}), i.e., the oscillator stability, is
\begin{align}
\sigma^2_y(\tau)
&= \ldots + \frac{1}{2}\frac{b_{-2}}{\nu_0^2}\:\frac1\tau +
	2\ln(2)\frac{b_{-3}}{\nu_0^2} +
	\frac{4\pi^2}{6}\frac{b_{-4}}{\nu_0^2}\:\tau +
	\ldots
\end{align}
%

%

\section{Miteq DRO mod.\ D-210B}
\begin{figure}[t]
\namedgraphics{0.8}{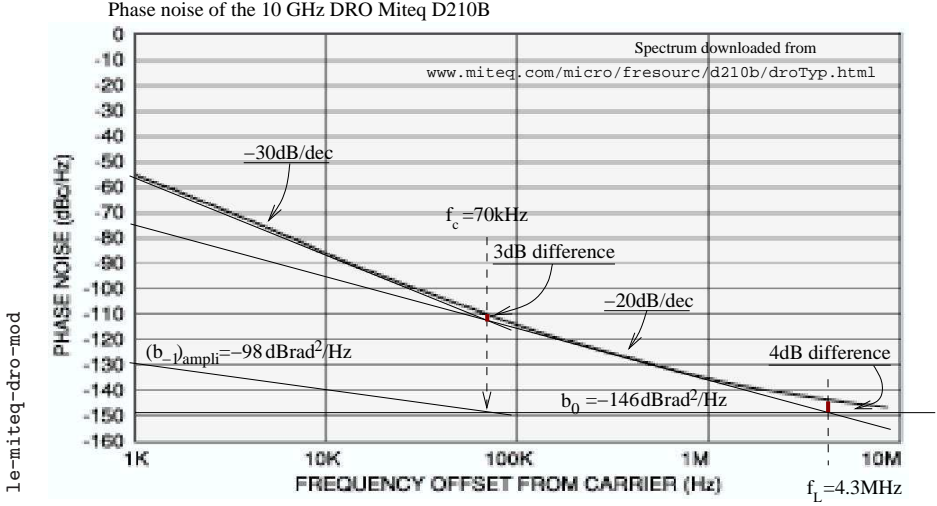}
\caption{Phase noise of the 10 GHz DRO Miteq D210B\@. Courtesy of Miteq Inc.  
Interpretation, comments and mistakes are of the author.}
\label{fig:le-miteq-dro-mod}
\end{figure}
Figure~\ref{fig:le-miteq-dro-mod} shows the phase noise spectrum of the dielectric-resonator oscillator (DRO) Miteq D-210B, taken from the device data sheet.
The plot is fitted by the polynomial $\sum_{i=-3}^0b_if^i$,  with 
\begin{center}
\begin{tabular}{lccl}
$b_0$    & $-146$ dB & $2.5{\times}10^{-15}$  & \unit{rad^2/Hz} \\
$b_{-1}$ &           & (not visible)          &  \\
$b_{-2}$ & $-11$ dB  & $7.9{\times}10^{-2}$   & \unit{rad^2/Hz}  \\
$b_{-3}$ & $+37$ dB  & $5.0{\times}10^{3}$    & \unit{rad^2/Hz}
\end{tabular}
\end{center}
This indicates that the spectrum is of the type 1A of Fig.~\ref{fig:le-effect-of-buffer}. 
 
One might be tempted to fit the spectrum with a smaller $b_0$ (say, $-147$ \unit{dBrad^2/Hz}) and to add a term $b_{-1}f^{-1}$ tangent to the curve at $f\approx2$~MHz.  In this case the spectrum would be of the type 1B, which contains the signature of the output buffer. 
We discard this alternate interpretation because the noise of the output buffer would be $b_{-1}\approx10^{-8}$ \unit{rad^2/Hz} ($-80$~dB), which is too high for a microwave amplifier (Table~\ref{tab:le-phase-flickering}).

The spectrum gives the following indications.
\begin{enumerate}
\item The coefficient $b_0$ derives from the amplifier noise $FkT$ referred to the carrier power $P_0$ at the input of the sustaining amplifier, that is, $\smash{b_0=\frac{FkT}{P_0}}$. 
Assuming that the noise figure is $F=1$ dB, thus 
$FkT=5.1{\times}10^{-21}$~\unit{rad^2/Hz} ($-173$ \unit{dBrad^2/Hz}), it follows that $\smash{P_0=\frac{FkT}{b_0}=2}$~$\mu$W ($-27$ dBm).  

\item However arbitrary the assumption $F=1$ dB may seem, it is representative of actual microwave amplifiers.  Depending on bandwidth and technology, the noise figure of a ``good'' amplifier is between 0.5 dB and 2 dB\@.   In this range, we find $P_0$ between 1.8 $\mu$W and 2.5 $\mu$W\@.

\item The spectrum changes slope from $f^0$ to $f^{-2}$ at the Leeson frequency $f_L\simeq4.3$~MHz.  At this frequency, the asymptotic approximation is some 4~dB lower than the measured spectrum, instead of the expected 3~dB\@. This discrepancy is tolerable. From $f_L\simeq4.3$~MHz, it follows that $\smash{Q=\frac{\nu_0}{2f_L}\simeq1160}$, quite plausible for a dielectric resonator.

\item The white frequency coefficient is $b_{-2}=7.9{\times}10^{-2}$~\unit{rad^2/Hz} ($-11$~\unit{dBrad^2/Hz}).

\item The corner point at which the slope changes from $-2$ to $-3$ is $70$~kHz.  This is the corner frequency $f_c$ of the amplifier, at which it holds that $(b_{-1})_\mathrm{ampli}f^{-1}=(b_0)_\mathrm{ampli}$.  Hence $(b_{-1})_\mathrm{ampli}=1.8{\times}10^{-10}$ \unit{rad^2/Hz} ($-98$~\unit{dBrad^2/Hz}).

\item The flicker frequency coefficient is $b_{-3}=5{\times}10^{3}$~\unit{rad^2/Hz} ($+37$~\unit{dBrad^2/Hz}).

\item The white and flicker frequency noise, transformed into Allan variance (Table~\ref{tab:le-noise-conversion}), is
\begin{align}
\sigma^2_y(\tau)
&=\frac{h_0}{2\tau} + 2\ln(2)\:h_{-1}\nonumber\\
&=\frac{b_{-2}}{\nu_0^2} \frac{1}{2\tau} + 2\ln(2)\:\frac{b_{-3}}{\nu_0^2}\nonumber\\
&\simeq\frac{7.9{\times}10^{-2}}{2\times(10^{10})^2}\:\frac{1}{\tau} + 
	1.39{\times}\frac{5{\times}10^3}{(10^{10})^2}~,\nonumber
\intertext{thus}
\sigma^2_y(\tau)&\simeq\frac{4{\times}10^{-22}}{\tau} + 6.9{\times}10^{-17}\nonumber\\
\sigma_y(\tau)&\simeq\frac{2{\times}10^{-11}}{\sqrt{\tau}}+8.3{\times}10^{-9}~.\nonumber
\end{align}
\end{enumerate}
Finally, one should note that the oscillator flicker shows up in the 1--100 kHz region. Common sense suggests that temperature and other environmental fluctuations have no effect at this time scale, and that the flickering of the dielectric constant in the resonator will not exceed the amplifier noise.  Consequently, in this region the oscillator flicker is due to the amplifier through the Leeson effect, rather than to the resonator.

\section{Poseidon DRO-10.4-FR (10.4 GHz)}
\begin{figure}[t]
\namedgraphics{0.8}{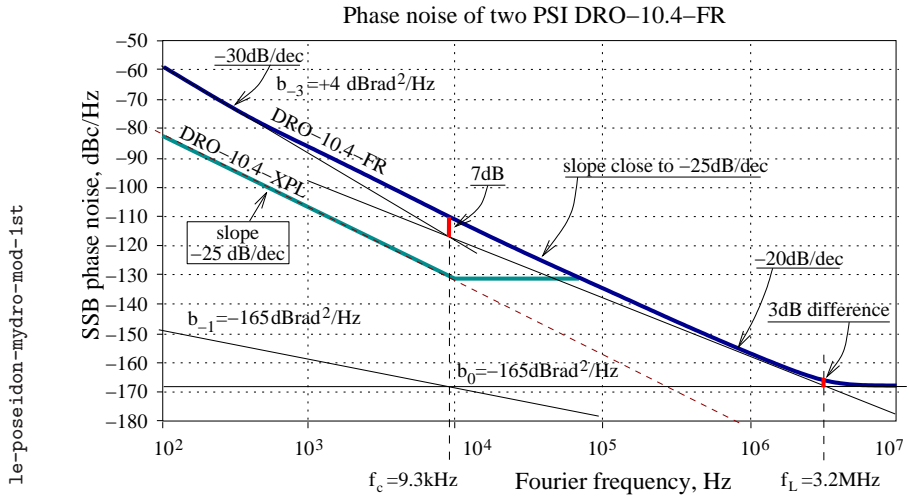}
\caption{Poseidon DRO 10.4-FR\@.  The spectra are plotted using the preliminary data available on the manufacturer web site.  Interpretation, comments and mistakes are of the author.}
\label{fig:le-poseidon-mydro-mod-1st}
\end{figure}
The Poseidon DRO-10.4-FR is another example of oscillator based on a dielectric resonator. 
Figure~\ref{fig:le-poseidon-mydro-mod-1st} shows the phase noise spectrum, from a preliminary data sheet.  $S_\phi(f)$ is fitted by the polynomial $\sum_{i=-3}^0b_if^i$, with
\begin{center}
\begin{tabular}{lccl}
$b_0$       & $-165$ dB & $3.2{\times}10^{-17}$  & \unit{rad^2/Hz} \\
$b_{-1}$ &                    & (not visible)          &  \\
$b_{-2}$ & $-35$ dB &  $3.2{\times}10^{-4}$ & \unit{rad^2/Hz}\\
$b_{-3}$ & $+4$ dB   & $2.5$                    & \unit{rad^2/Hz}
\end{tabular}
\end{center}
Once again, the spectrum is of the type 1A (Fig.~\ref{fig:le-effect-of-buffer}), typical of microwave oscillators. Yet, the discrepancy with respect to the theoretical model is larger than in the case of the Miteq oscillator.  The spectrum gives the following indications.

\begin{enumerate}
\item The spectrum results from the comparison of two DRO-10.4 oscillators.  In the absence of other indications, we believe that the numerical data refer to a single oscillator, after taking away 3 dB from the raw data.

\item The white phase noise is $b_0=3.2{\times}10^{-17}$  \unit{rad^2/Hz} ($-165$~\unit{dBrad^2/Hz}).  Thus $\smash{P_0=\frac{FkT}{b_0}\simeq160}$~$\mu$W ($-8$ dBm), assuming that $F=1$~dB\@.

\item The Leeson frequency is $f_L=3.2$ MHz.   Accordingly, the merit factor is $\smash{Q=\frac{\nu_0}{2f_L}\simeq1625}$, which is reasonable for a dielectric resonator.

\item In a type-1A spectrum it holds that $b_{-2}f^{-2}=b_0$ at $f=f_L$.  Thus, the white frequency noise is $b_{-2}\simeq3.2{\times}10^{-4}$~\unit{rad^2/Hz} ($-35$~\unit{dBrad^2/Hz}).

\item There is some discrepancy between the Leeson model and the true spectrum.  In the region from 2~kHz to 200~kHz, the spectrum seems to be close to a line of slope $f^{-5/2}$, rather than $f^{-3}$ or $f^{-2}$.   At the present time this discrepancy, (up to 4 dB at $f\approx10$~kHz) is unexplained.

\item The corner frequency of the amplifier (i.e., the frequency at which the oscillator spectrum changes from $f^{-2}$ to $f^{-3}$) is $f_c=9.3$~kHz.  Accordingly, the phase noise spectrum of the amplifier, on the left hand of $f=f_c$, is $(b_{-1})_\mathrm{ampli}=b_0f_c=2.9{\times}10^{-13}$~\unit{rad^2/Hz} ($-125$~\unit{dBrad^2/Hz}).  

\item The amplifier flickering, 5 dB lower than the best in Table~\ref{tab:le-phase-flickering} is surprisingly low for a microwave amplifier contained a commercial product.  Such a low noise could be obtained with SiGe technology, with a single-stage
amplifier employing a large-volume transistor, or with some feedback or feedforward noise degeneration scheme. A noise degeneration scheme seems incompatible with the size of the packaged oscillator.  Yet, nothing can be taken for sure on the basis of the available information.

\item The flicker frequency coefficient is $b_{-3}=2.5$~\unit{rad^2/Hz} ($+4$~\unit{dBrad^2/Hz}).

\item The white and flicker frequency noise, transformed into Allan variance (Table~\ref{tab:le-noise-conversion}), is
\begin{align}
\sigma^2_y(\tau)&=\frac{h_0}{2\tau} + 2\ln(2)\:h_{-1}\nonumber\\
&=\frac{b_{-2}}{\nu_0^2} \frac{1}{2\tau} + 2\ln(2)\:\frac{b_{-3}}{\nu_0^2}\nonumber\\
	&\simeq\frac{2.5}{2\times(10.4{\times}10^{9})^2}\:\frac{1}{\tau} + 
	1.39{\times}\frac{5{\times}10^3}{(10.4{\times}10^{9})^2}~,\nonumber
\intertext{thus}
\sigma^2_y(\tau)&\simeq\frac{1.5{\times}10^{-24}}{\tau} + 3.2{\times}10^{-20}\nonumber\\
\sigma_y(\tau)&\simeq\frac{1.2{\times}10^{-12}}{\sqrt{\tau}}+1.8{\times}10^{-10}~.\nonumber
\end{align}
\end{enumerate}
Figure~\ref{fig:le-poseidon-mydro-mod-1st} also reports the phase noise spectrum of the DRO-10.4-XPL oscillator, which is a different version of the same base design, intended for phase-locked loops.  Below a cutoff frequency of about 70 kHz, this oscillator is locked to an external reference, for the spectrum gives no additional information in this region.  Nonetheless, it is to be noted that the spectrum is proportional to $f^{-5/2}$, i.e., $-25$~dB/decade below the loop cutoff frequency of 10 kHz.  This is the signature of a fractional-order control system, like that proposed in \cite{cretin85im}.

\section{Poseidon Shoebox (10 GHz sapphire resonator)}
%
The Poseidon Shebox integrates a sapphire whispering gallery (WG) resonator and an interferometric noise degeneration scheme.  There results a low-noise oscillator intended for high short-term stability applications.
\begin{figure}[t]
\namedgraphics{0.785}{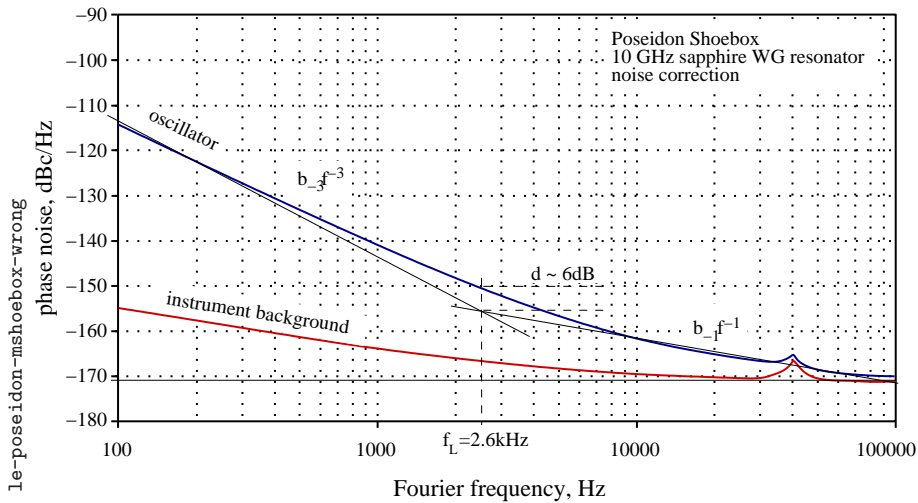}
\caption{Poseidon Shoebox, first attempt to interpret the spectrum.  Data are taken from the manufacturer data sheet.  Interpretation, comments and mistakes are of the author.}
\label{fig:le-poseidon-mshoebox-wrong}
\end{figure}
Figure~\ref{fig:le-poseidon-mshoebox-wrong} shows the phase noise spectrum with a tentative interpretation.  The spectrum seems to be of the type 2 (Fig.~\ref{fig:le-ideal-types}), with $f_L<f_c$.  Qualitatively, this is consistent with the fact that the WG resonator features high $Q$.  Yet this interpretation suffers from three problems.
\begin{enumerate}
\item From $\smash{Q=\frac{\nu_0}{2f_L}}$, we get $Q\approx1.9{\times}10^6$ at $\nu_0=10$~GHz.  This value is incompatible with the dielectric loss of the sapphire.  For comparison, the typical merit factor of a 10 GHz WG resonator is in the range of
\begin{center}\begin{tabular}{ccl}
$2{\times}10^5$ & 295 K & room temperature\\
$3{\times}10^7$ & 77~K  & liquid N\\
$5{\times}10^9$ & 4~K    & liquid He
\end{tabular}\end{center}
The dielectric loss of the sapphire is a reproducible function of temperature.  
Thus, the loss of the resonator depends on the space distribution of the electric field, 
i.e., on the mode, in a narrow range.  A merit factor of $1.9{\times}10^6$ can not be obtained by moderate cooling (Peltier cells).  Finally, the size and weigh (3~\unit{dm^3} and 6.5~kg) indicate that the oscillator works at room temperature, at most with temperature stabilization.

\item The phase flicker of the sustaining amplifier is $b_{-1}f^{-1}$ with $b_{-1}\approx10^{-12}$~\unit{rad^2/Hz}.  This is seen on the plot, $-160$~\unit{dBrad^2/Hz} at $f=10$~kHz with slope $f^{-1}$, thus $-120$~\unit{dBrad^2/Hz} at 1~Hz. This value is too high for a sophisticated amplifier that makes use of the interferometric noise correction technique.

\item At $f_L$, where $b_{-3}f^{-3}=b_{-1}f^{-1}$, the true spectrum differs from the asymptotic approximation by 6 dB instead of 3 dB\@.  This discrepancy is an additional reason to reject the interpretation of Fig.~\ref{fig:le-poseidon-mshoebox-wrong}.  
\end{enumerate}
\begin{figure}[t]
\namedgraphics{0.8}{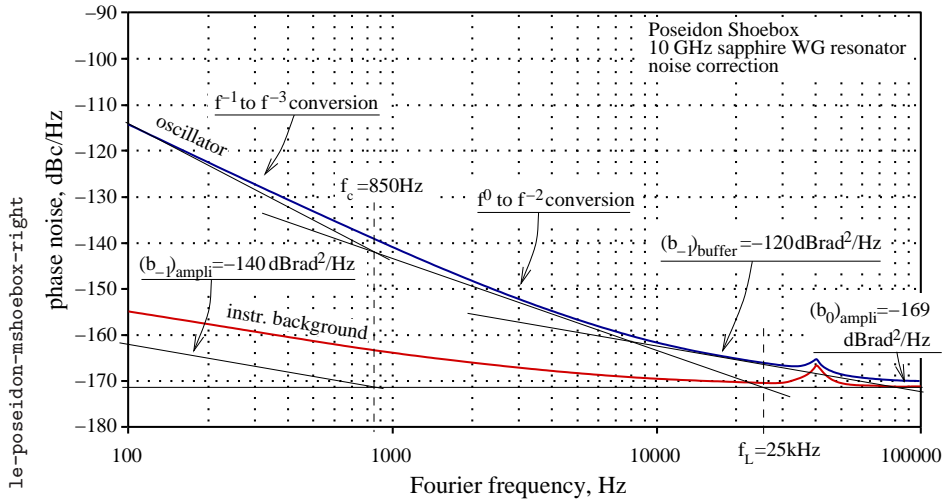}
\caption{Poseidon Shoebox, revised interpretation.  Interpretation, comments and mistakes are of the author.}
\label{fig:le-poseidon-mshoebox-right}
\end{figure}
The above difficulties make us understand that the spectrum is of the type 2A (Fig.~\ref{fig:le-effect-of-buffer}), in which the flicker noise of the output buffer 
shows up in the region around $f_L$. There follows new interpretation, shown in Fig.~\ref{fig:le-poseidon-mshoebox-right}.
\begin{enumerate}
\item As usual, we start from the white noise floor.  On the Figure, we observe that $b_0=1.3{\times}10^{-17}$~\unit{rad^2/Hz} ($-169$~\unit{dBrad^2/Hz}).  This is ascribed to the sustaining amplifier.

\item The sustaining amplifier makes use of an interferometric noise degeneration circuit to reduce the flicker noise.  In this circuit there are two amplifiers, the first amplifies the input signal, and the second amplifies the noise of the first, at the output of a carrier suppression circuit. The second amplifier corrects for the noise of the first by means of a feedback circuit.   We guess a noise figure $F=5$~dB, which results from the intrinsic loss of the power splitter at the input of the noise corrected amplifier (3~dB), from the resistive loss of the power splitter and of the lines (1~dB), and from the noise figure of the second amplifier (1~dB).  

\item From $\smash{b_0=\frac{FkT}{P_0}}$, we get $P_0=1$~mW (0~dBm). 

\item The phase flickering of the output buffer shows up on the right-hand part of the spectrum, at $10^5$--$10^6$ Hz.  The noise coefficient is $(b_{-1})_\mathrm{buf}\simeq10^{-12}$~\unit{rad^2/Hz} ($-120$~\unit{dBrad^2/Hz}).  The output buffer is a good microwave amplifier.

\item After removing the buffer phase noise, the white frequency noise $b_{-2}f^{-2}$ of the oscillator is clearly identified.  The decade centered at 4 kHz is used to find $b_{-2}=7.9{\times}10^{-17}$~\unit{rad^2/Hz} ($-81$~\unit{dBrad^2/Hz}) on the plot.

\item From $b_{-2}f^{-2}=b_0$, found on the plot, it follows that $f_L\simeq25$~kHz.
Hence, $\smash{Q=\frac{\nu_0}{2f_L}}\simeq2{\times}10^5$.  This is the typical value for a room-temperature WG sapphire resonator.
  
\item It is seen on the plot that  $b_{-3}\simeq7.9{\times}10^{-6}$~\unit{rad^2/Hz} ($-51$~\unit{dBrad^2/Hz}).

\item In the spectrum of the type 2A, the corner frequency $f_c$ of the sustaining amplifier is the frequency at which the oscillator noise changes slope from $f^{-3}$ to $f^{-2}$.   Thus, $f_c\simeq850$~Hz.  The flicker noise of the sustaining amplifier is $(b_{-1})_\mathrm{ampli}f^{-1}$ with $(b_{-1})_\mathrm{ampli}=b_0f_c$. 
Thus, $(b_{-1})_\mathrm{ampli}\simeq10^{-14}$~\unit{rad^2/Hz} ($-140$~\unit{dBrad^2/Hz}).

\item The white and flicker frequency noise, transformed into Allan variance (Table~\ref{tab:le-noise-conversion}), is
\begin{align}
\sigma^2_y(\tau)&=\frac{h_0}{2\tau} + 2\ln(2)\:h_{-1}\nonumber\\
&=\frac{b_{-2}}{\nu_0^2} \frac{1}{2\tau} + 2\ln(2)\:\frac{b_{-3}}{\nu_0^2}\nonumber\\
&\simeq\frac{7.9{\times}10^{-9}}{2\times(10{\times}10^{9})^2}\:\frac{1}{\tau} + 
	1.39{\times}\frac{7.9{\times}10^{-6}}{(10{\times}10^{9})^2}~,\nonumber
\intertext{thus}
\sigma^2_y(\tau)&\simeq\frac{4{\times}10^{-29}}{\tau} + 1.1{\times}10^{-25}\nonumber\\
\sigma_y(\tau)&\simeq\frac{6.3{\times}10^{-15}}{\sqrt{\tau}}+3.3{\times}10^{-13}~.\nonumber
\end{align}

\item The bump at $f=40$~kHz is ascribed to the feedback control of the noise degeneration circuit.

\end{enumerate}

\section{Oscilloquartz OCXO 8600 (5 MHz AT-cut BVA)}
\begin{figure}[t]
\namedgraphics{0.8}{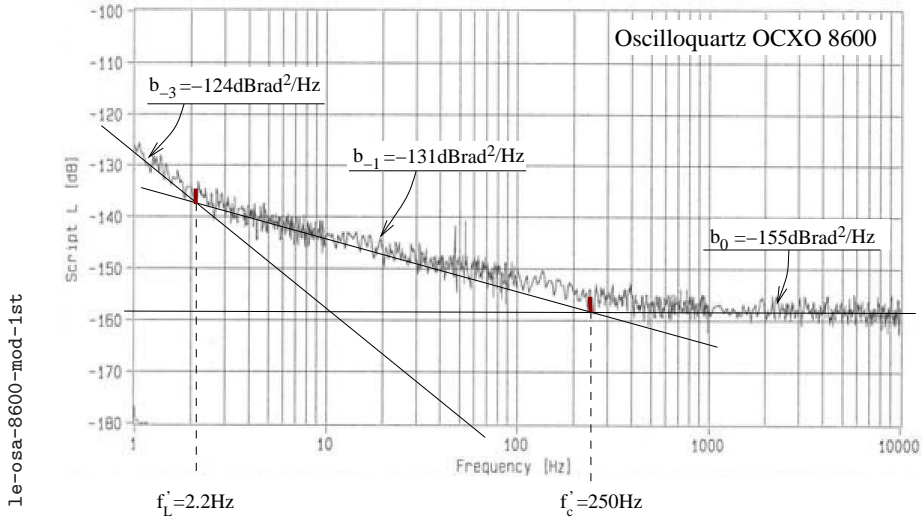}
\caption{Phase noise of the oven-controlled quartz oscillator Oscilloquartz OCX0 8600, with a preliminary (wrong) interpretation.  Courtesy of Oscilloquartz S.A\@.  Interpretation, comments and mistakes are of the author.}
\label{fig:le-osa-8600-mod-1st}
\end{figure}

\begin{figure}[t]
\namedgraphics{0.8}{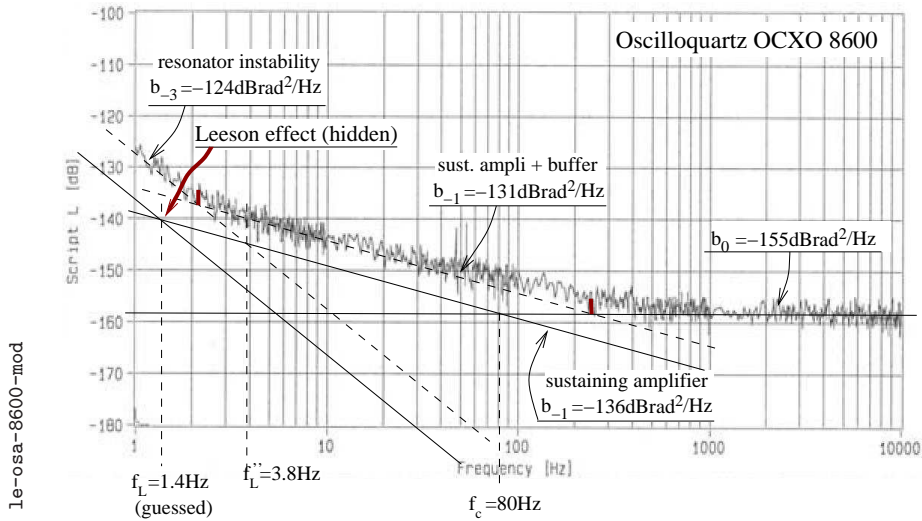}
\caption{Phase noise of the oven-controlled quartz oscillator Oscilloquartz OCX0 8600, with revised interpretation.  Courtesy of Oscilloquartz S.A\@. Interpretation, comments and mistakes are of the author.}
\label{fig:le-osa-8600-mod}
\end{figure}

Figure~\ref{fig:le-osa-8600-mod-1st} shows the phase noise of the oven-controlled quartz oscillator Oscilloquartz OCX0 8600.   This oscillator has been chosen as an example because of its outstanding stability in the 0.1--10~s region.  The spectrum is of the type 2 ($f_L<f_c$), typical of low-frequency oscillators with high-$Q$ resonators.
The plot is fitted by the polynomial $\sum_{i=-3}^0b_if^i$,  with 
\begin{center}
\begin{tabular}{lccl}
$b_0$    & $-155$ dB & $3.2{\times}10^{-16}$  & \unit{rad^2/Hz} \\
$b_{-1}$ & $-131$ dB &  $7.9{\times}10^{-14}$ & \unit{rad^2/Hz} \\
$b_{-2}$ &           & (not visible) & \\
$b_{-3}$ & $-124$ dB  & $4{\times}10^{-13}$    & \unit{rad^2/Hz}
\end{tabular}
\end{center}
A preliminary interpretation  the spectrum as follows.
\begin{enumerate}
\item The white phase noise is $b_0=3.2{\times}10^{-16}$~\unit{rad^2/Hz} ($-155$~\unit{dBrad^2/Hz}).  Thus 
\begin{displaymath}
P_0=\frac{FkT}{b_0}\simeq16~\unit{\mu W}\quad\text{($-18$ dBm)}~,
\end{displaymath}
assuming that $F=1$~dB\@.  This power level is consistent with the general experience on 5--10 MHz oscillators, in which the power is kept low for best stability.

\item $f_c\simeq250$~Hz (from the plot), hence $(b_{-1})_\mathrm{ampli}=7.9{\times}10^{-14}$. 

\item On the plot we observe that $f_L\simeq2.2$~Hz ($f'_L$ in Fig.~\ref{fig:le-osa-8600-mod-1st}), and  $b_{-3}\simeq4{\times}10^{-13}$.

\item  $\smash{Q=\frac{\nu_0}{2F_L}}$, hence $Q\simeq1.1{\times}10^6$.

\end{enumerate}
The item 1 is correct because the effect of the output buffer on the white noise is divided by the gain of the sustaining amplifier, thus it is negligible. The white noise originates in the series resistance of the resonator and in the sustaining amplifier input.  
Conversely, the flicker noise of the output buffer is not negligible because it results from parametric modulation.  Thus we change the interpretation (Fig.~\ref{fig:le-osa-8600-mod}) as follows.
\begin{enumerate}
\item Section \ref{ssec:le-out-buffer} indicates that the effect of the output buffer can not be neglected if phase flickering shows up in the phase-noise spectrum.  We guess that there are two stages between the sustaining amplifier and the output and that the $1/f$ noise of each stage is at least equal to that of the sustaining amplifier.  Thus the $1/f$ noise of the sustaining amplifier is not higher than $1/3$ of the total noise.  Let us take away 5 dB\@.  In reality, the difference could be more then 5 dB because the output amplifier can be more complex, and because superior technology can be used in the sustaining amplifier, rather than in the buffers.  

\item The new estimate of the flickering of the sustaining amplifier is   $(b_{-1})_\mathrm{ampli}=2.5{\times}10^{-14}$ \unit{rad^2/Hz} ($-136$ \unit{dBrad^2/Hz}).  This affects the corner frequency and in turn the Leeson frequency.  The new values are
\begin{displaymath}f_c\simeq80\unit{Hz}\end{displaymath}
and $f_L\simeq3.8$~Hz ($f''_L$ in Fig.~\ref{fig:le-osa-8600-mod}).  The latter is still provisional.

\item Using $f_L=3.8$ Hz in $\smash{f_L=\frac{\nu_0}{2Q}}$, we get $Q\simeq6.6{\times}10^5$. 

\item $Q\simeq6.6{\times}10^5$ is a too low value for a top-technology oscillator.  The literature suggests that the product $\nu_0Q$ is between $10^{13}$ and $2{\times}10^{13}$. 

\item $Q$ is higher than value of $6.6{\times}10^5$.  This means that the Leeson effect is hidden, and that we are not able to calculate $Q$.  
Henceforth, our knowledge about $Q$ relies only upon experience and on the literature.

\item If we guess  $Q=2.5{\times}10^{6}$, reduced to $Q=1.8{\times}10^{6}$ (loaded) by the dissipative effect of the amplifier input, we get
\begin{displaymath}
f_L=\frac{\nu_0}{2Q}\approx1.4\unit{Hz}
\end{displaymath}
This indicates that the oscillator frequency flicker $b_{-3}\simeq3.2 {\times}10^{13}$ is due to frequency flickering in the resonator.  The Leeson effect is hidden by the resonator instability and by the phase noise of the output buffer.  Thus, the corner frequency $f''_L\simeq3.8$~Hz of Fig.~\ref{fig:le-osa-8600-mod} is not the Leeson frequency.

\item The flicker noise of the sustaining amplifier combined with the Leeson frequency gives the stability limit of the electronics of the oscillator.  This is the solid line $f^{-3}$ in Fig.~\ref{fig:le-osa-8600-mod}, some 10 dB below the phase noise.  

\item The frequency flicker turned into Allan variance (Table~\ref{tab:le-noise-conversion}), is
\begin{align}
\sigma^2_y(\tau)&=2\ln(2)\:h_{-1}
= 2\ln(2)\:\frac{b_{-3}}{\nu_0^2}\nonumber
=1.39{\times}\frac{4{\times}10^{-13}}{(5{\times}10^{6})^2}~,\nonumber
\end{align}
thus
\begin{align}
\sigma^2_y(\tau)&\simeq2.2{\times}10^{-26} &
\sigma_y(\tau)&\simeq1.5{\times}10^{-13}~.\nonumber
\end{align}
This is lower than the value 
$\sigma_y(\tau)<3{\times}10^{-13}$ for $0.2\unit{s}\le\tau\le30\unit{s}$.
given in the specifications.  On the other hand, specifications are conservative,
while a sample can be better.
\end{enumerate}

\section{FEMTO-ST prototype (10 MHz LD-cut quartz)}
The interest of the LD cut is the low isochronism defect $\frac{\Delta\nu}{\nu_0\Delta P}$, i.e., the low dependence of frequency on the drive power \cite{gufflet02uffc}.  For comparison, the isochronism defect can be of $1.2{\times}10^{-11}$ / $\mu$W (typical) for the popular SC-cut crystals, and as low as $10^{-11}$ / $\mu$W for the LD cut.  
This oscillator is a laboratory prototype.  It has been included in this Section otherwise devoted to commercial oscillators because we know some internal parameters, for the analysis process is somewhat different. 

\begin{figure}[t]
\namedgraphics{0.80}{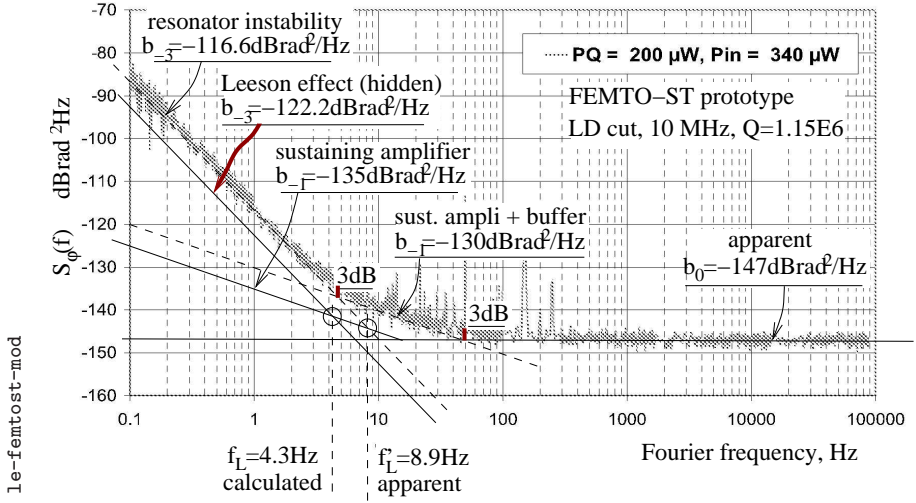}
\caption{Phase noise of a FEMTO-ST oscillator prototype based on a LD-cut resonator. Courtesy of the FEMTO-ST laboratory.  Interpretation, comments and mistakes are mine.}
\label{fig:le-femtost-mod}
\end{figure}
Figure~\ref{fig:le-femtost-mod} shows the phase noise spectrum.  The spectrum is the total noise obtained by comparing two equal devices, and divided by two.
The plot is fitted by the polynomial $\sum_{i=-3}^0b_if^i$,  with 
\begin{center}
\begin{tabular}{lccl}
$b_0$    & $-147$ dB & $2{\times}10^{-15}$  & \unit{rad^2/Hz} \\
$b_{-1}$ & $-130$ dB &  $1{\times}10^{-13}$ & \unit{rad^2/Hz} \\
$b_{-2}$ &           & (not visible) & \\
$b_{-3}$ & $-116.6$ dB  & $2.2{\times}10^{-12}$    & \unit{rad^2/Hz}
\end{tabular}
\end{center}
We interpret the spectrum as follows.
\begin{enumerate}

\item The white phase noise is $b_0=2{\times}10^{-115}$~\unit{rad^2/Hz} ($-147$~\unit{dBrad^2/Hz}).  As we know the power at the amplifier input, $P_0=340$ $\mu$W, we can calculate the noise figure $F=\smash{\frac{b_0}{kT}}\simeq167$ ($+22$ dB).  This value is far too high for the noise figure of a HF-VHF amplifier, thus the observed floor can not be the noise floor $b_0$ of the oscillator.  Therefore we ascribe it to the white noise floor of the instrument, used at low input power.

\item As $b_0$ is hidden, the corner frequency $f_c$ of the amplifier is also hidden.

\item The phase flickering is clearly visible on a span of one decade.  At the corner frequencies, 5 Hz and 50 Hz, the difference between experimental data and asymptotic approximation is of 3 dB as expected.  The residual noise of the instrument is not shown on the plot.  Yet experience suggests that the value $(b_{-1})_\mathrm{osc}=10^{-13}$ is sufficiently larger than the noise of a HF-VHF mixer.  This elements indicate that the value $(b_{-1})_\mathrm{osc}=10^{-13}$ can be trusted.

\item From the plot, the frequency flicker is $b_{-3}=2.2{\times}10^{-12}$.  We use it to estimate the Leeson frequency, at which $b_{-3}f^{-3}=b_{-1}f^{-1}$.  There results the value $f'_L=8.9$ Hz.

\item Let us assume that the sustaining amplifier contributes $1/3$ of the total phase flickering, guessing that there are two buffer stages  (Section \ref{ssec:le-out-buffer}) based on the same technology of the sustaining amplifier.  There follows that $(b_{-1})_\mathrm{ampli}=3.16{\times}10^{-13}$.

\item The merit factor of the resonator in the circuit load conditions is known, $Q=1.15{\times}10^6$.  The calculated Leeson frequency is  
\begin{displaymath}
f_L=\frac{\nu_0}{2Q}=4.35 \unit{Hz}~.
\end{displaymath}
This value suffers from the uncertainty with which we guess the phase flickering of the sustaining amplifier as a fraction of the total phase flickering.  Nonetheless, the buffer is necessary, which ensures that only a fraction of the phase flickering is due to the sustaining amplifier. 

\item The calculated Leeson frequency (4.35 Hz) is sufficiently lower than the apparent  value (8.9 Hz) to provide evidence that the Leeson effect is not visible on the spectrum, and that the $1/f^3$ line is the frequency flickering of the resonator.

\item The frequency flicker turned into Allan variance (Table~\ref{tab:le-noise-conversion}), is
\begin{displaymath}
\sigma^2_y(\tau)
=2\ln(2)\:h_{-1}
= 2\ln(2)\:\frac{b_{-3}}{\nu_0^2}
=2{\times}\frac{2.2{\times}10^{-12}}{(10{\times}10^{6})^2}~,
\end{displaymath}
hence
\begin{align}
\sigma^2_y(\tau)&\simeq3{\times}10^{-26}
&\sigma_y(\tau)&\simeq1.74{\times}10^{-13}~.\nonumber
\end{align}

\item   The total $1/f^3$ noise results from the amplifier noise through the Leeson effect and from resonator frequency flickering
\begin{align}
(b_{-3})_\mathrm{osc}&=(b_{-3})_\mathrm{Le}+(b_{-3})_\mathrm{reson}~.
	\label{eqn:le-xtal-f-noise}
\end{align}
Thus we can infer and the resonator stability
\begin{displaymath}
\sigma^2_y(\tau)=2\ln(2) \left(\frac{(b_{-3})_\mathrm{osc}}{\nu_0^2}
		-\frac{(b_{-3})_\mathrm{Le}}{\nu_0^2}\right)
\end{displaymath}
The numerical result
\begin{align}
\sigma^2_y(\tau)&\simeq2.2{\times}10^{-26}
&\sigma_y(\tau)&\simeq1.5{\times}10^{-13}~.\nonumber
\end{align}
should be taken as an indication only, rather than as a measure, because it suffers from uncertainty enhancement of the difference, and from the lack of reliable measurement of the sustaining amplifier. 

\end{enumerate}

\begin{remark}
This example provides a method for the independent measurement of the resonator stability.  The accurate measurement of the amplifier flicker $(b_{-1})_\mathrm{ampli}$ and of the merit factor $Q$ in the appropriate load conditions turns into a reliable measure of $f_L$ and of $(b_{-3})_\mathrm{Le}$ to be fed in Eq.\ \ref{eqn:le-xtal-f-noise}.
\end{remark}

\section{Wenzel 501-04623 (100 MHz SC-cut quartz)}
\begin{figure}[t]
\namedgraphics{0.8}{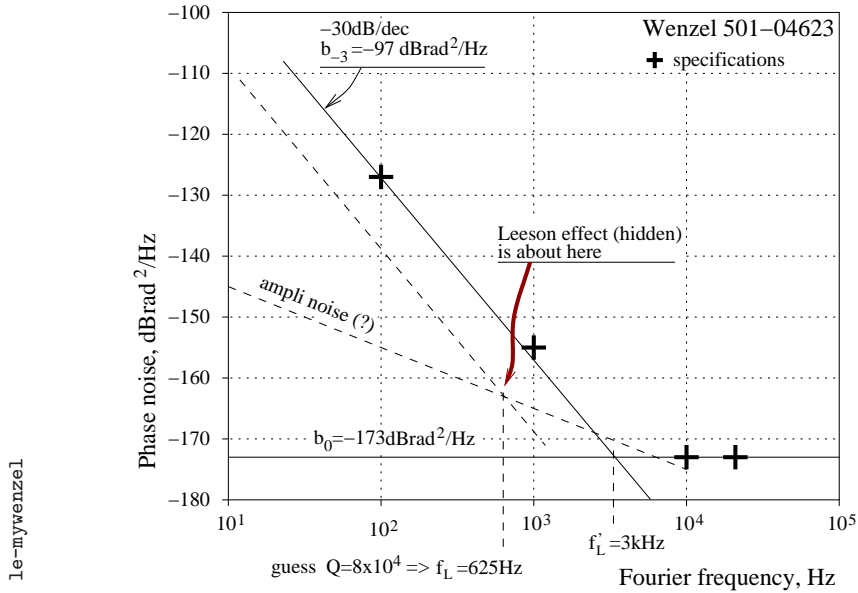}
\caption{Phase noise of the 100 MHz oscillator Wenzel 501-04623.  Data are from
the manufacturer web site.  Interpretation, comments and mistakes are mine.}
\label{fig:le-mywenzel}
\end{figure}

%
The Wenzel 501-04623 oscillator is chosen as an example because of its outstanding low phase noise floor, and because of its high carrier frequency (100 MHz).  These two features are necessary for low noise after frequency multiplication.
A table of the values of phase noise is given by the manufacturer, instead of the complete spectrum. These values are used to plot the spectrum (Fig.~\ref{fig:le-mywenzel})
\begin{center}
\begin{tabular}{lccl}
$b_0$    & $-173$ dB & $5{\times}10^{-18}$  & \unit{rad^2/Hz} \\
$b_{-1}$ &                & (not visible) & \\
$b_{-2}$ &           & (not visible) & \\
$b_{-3}$ & $-97$ dB  & $2{\times}10^{-10}$    & \unit{rad^2/Hz}
\end{tabular}
\end{center}
and to interpret it as follows.
\begin{enumerate}
\item The white phase noise is $b_0=5{\times}10^{-17}$~\unit{rad^2/Hz} ($-173$~\unit{dBrad^2/Hz}).  Thus 
\begin{displaymath}
P_0=\frac{Fk_BT}{b_0}\simeq1~\unit{mW}\quad\text{($0$ dBm)}~,
\end{displaymath}
assuming that $F=1$~dB\@.  This relatively large power is necessary for low short-term noise, although it may be detrimental to the medium-term and long-term stability.  

\item The terms $b_{-2}f^{-2}$ and $b_{-1}f^{-1}$ are virtually absent, as there is no room for them in the region between the two specified points, 1 kHz and 10 kHz.  A confirmation comes from the measurement of some samples.  Consequently, the point at which the slope changes by 2 (either $f^{0}\rightarrow f^{-2}$ or $f^{-1}\rightarrow f^{-3}$) is not visible.  This indicates that the Leeson effect is hidden by the frequency flicker of the resonator.
 
\item Of course, the corner point between the flicker of frequency and the white noise, $b_{-3}f^{-3}=b_0$, which occurs at $f=3$ kHz ($f'_L$ in Fig.~\ref{fig:le-mywenzel}), is not the Leeson frequency.  

\item As the Leeson effect is hidden, we are not able to calculate $Q$.  Relying upon the literature, we can guess that a high-stability 100 MHz quartz resonator can have a merit factor of $1.2{\times}10^5$, reduced to $Q=8{\times}10^4$ (loaded) by the power dissipation at the amplifier input.  Accordingly, the Leeson frequency is 
\begin{displaymath}
F_L=\frac{\nu_0}{2Q}\simeq625~\unit{Hz}~.
\end{displaymath}

\item The flicker noise of a  state-of-the-art HF/VHF amplifier is $(b_{-1})_\mathrm{ampli}\approx3.2{\times}10^{-14}$  \unit{rad^2/Hz} ($-135$ \unit{dBrad^2/Hz}).  This value, combined with the Leeson frequency, gives the stability limit of the electronics of the oscillator.  This is the dashed line $f^{-3}$ in Fig.~\ref{fig:le-mywenzel}, some 10 dB below the phase noise.  

\item The frequency flicker turned into Allan variance (Table~\ref{tab:le-noise-conversion}), is
\begin{align}
\sigma^2_y(\tau)
&=2\ln(2)\:h_{-1}
= 2\ln(2)\:\frac{b_{-3}}{\nu_0^2}
=1.39{\times}\frac{2{\times}10^{-10}}{(100{\times}10^{6})^2}~,\nonumber
\end{align}
thus
\begin{align}
\sigma^2_y(\tau)&\simeq2.8{\times}10^{-26}
& \sigma_y(\tau)&\simeq1.7{\times}10^{-13}~.\nonumber
\end{align}

\end{enumerate}
%

\section{Oewaves Tidalwave (10 GHz OEO)}
%
The unique feature of the opto-electronic oscillator (OEO) is that the frequency 
reference is an optical delay line instead of a traditional resonator.  The basic structure
consists of a loop in which the microwave sinusoid modulates a laser beam. 
After traveling through a long optical fiber, the microwave signal is reconstructed 
by a high-speed photodetector, and fed back into the amplifier.  
Yet the actual structure may be more complex.
Beside laboratory prototypes,  Oewaves is up to now the only one manufacturer of 
this type of oscillator.  Studying the noise in oscillators, this is a challenging example 
because the general experience earned with traditional oscillators is of scarce usefulness.
\begin{figure}[t]
\namedgraphics{0.8}{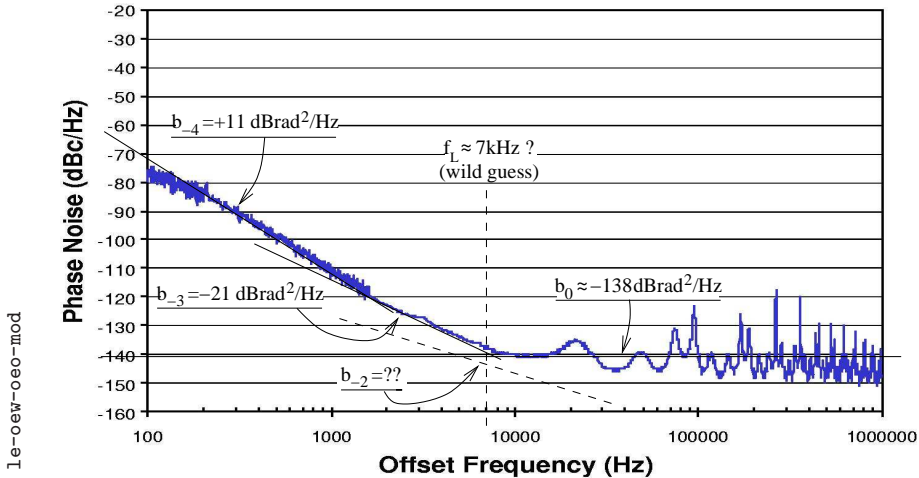}
\caption{Phase noise spectrum of the Oewaves Tidalwave photonic oscillator 
Courtesy of Oewaves Inc. Interpretation, comments and mistakes are of the author.}
\label{fig:le-oew-oeo-mod}
\end{figure}
The phase noise spectrum (Fig;~\ref{fig:le-oew-oeo-mod}) is fitted by the polynomial $\sum_{i=-3}^0b_if^i$,  with 
\begin{center}
\begin{tabular}{lccl}
$b_0$    & $-138$ dB & $1.6{\times}10^{-14}$  & \unit{rad^2/Hz} \\
$b_{-1}$ &                & (not visible) & \\
$b_{-2}$ &                & (not visible) & \\
$b_{-3}$ & $-21$ dB  & $7.9{\times}10^{-3}$    & \unit{rad^2/Hz}\\
$b_{-4}$ & $+11$ dB & $1.26{\times}10^{1}$   & \unit{rad^2/Hz}
\end{tabular}
\end{center}
\begin{enumerate}
\item In a delay-line oscillator with a delay $\tau$ we expect (Chapter \ref{chap:le-delayline}) a ``clean'' comb of spectral lines at $f_i=i/\tau$, integer $i\ge1$.  This structure is not present in the spectrum of Fig;~\ref{fig:le-oew-oeo-mod}.  Instead, there is a series of smaller bumps and spectral lines. This indicates that the Tidalwave is not a simple delay-line oscillator, and that it contains some additional circuits that reduce or almost eliminate the spectral lines at $f_\mu=\mu/\tau$, integer $\mu$. The remedy suggested by the literature is the dual-loop oscillator, in which a second delay line, significantly, is used to remove the largest peaks.  One conference article \cite{eliyahu03fcs} provides some interesting details, yet insufficient to identify clearly the inside of this specific oscillator.  

\item The right-hand side of the spectrum, in the white-noise region, an horizontal $b_0$ is only a poor approximation of the spectrum.  This is inherent in the delay-line oscillator.  Discarding the peaks, we take $b_0\approx1.6{\times}10^{-14}$  \unit{rad^2/Hz}
($-138$ \unit{dBrad^2/Hz}).  

\item The equivalent noise figure used to estimate the microwave power must be referred to the lowest-power point of the circuit, which is the output of the photodetector.  Noise includes at least the intensity fluctuation of the laser, the shot noise, and the noise of the microwave amplifier.  Let us guess that $F=6$ dB\@.  Consequently, the microwave power $P_0=\smash{\frac{FkT}{b_0}}$ at the phodetector output is $P_0=1$ $\mu$W ($-30$ dBm).

\item One may think that the transition $f^{-3}\rightarrow f^0$ around $f=7$ kHz is the signature of the frequency flicker of the reference, as in the Wenzel quartz oscillator.  Yet an alternate interpretation is possible in this case because $f_c$ and $f_L$ might be close to one another and fall in this region.  

\item A corner frequency $f_c\approx7$ kHz with $b_0\approx1.6{\times}10^{-14}$  
\unit{rad^2/Hz} requires that $b_{-1}=10^{-10}$  \unit{rad^2/Hz} ($-100$ \unit{dBrad^2/Hz}).  The latter value can be ascribed to the microwave amplifier (Tab.~\ref{tab:le-phase-flickering}), or to the optical system.

\item The delay $\tau$ is related to the Leeson frequency by $f_L=\smash{\frac{1}{2\pi\tau}}$
Thus $f_L\approx7$ kHz requires that $\tau\approx23$ $\mu$s, hence a fiber length 
$l=\smash{\frac{c}{n}}\tau\approx4.7$ km, which is likely. 

\item The $f^{-3}\rightarrow f^0$ transition at $f\approx7$ kHz can be ascribed either to the fact that $f_c\approx f_L$, or to the flickering of the delay.  The analysis of the spectrum is not sufficient to say more.

\item The frequency flicker and random walk,  turned into Allan variance (Table~\ref{tab:le-noise-conversion}), is
\begin{align}
\sigma^2_y(\tau)
&=2\ln(2)\:h_{-1} + \frac{4\pi^2}{6}\:h_{-2}\tau\nonumber\\
&= 2\ln(2)\:\frac{b_{-3}}{\nu_0^2}+\frac{4\pi^2}{6}\:\frac{b_{-4}}{\nu_0^2}\;\tau\nonumber\\
& =1.39\times\frac{7.9{\times}10^{-3}}{(10^{10})^2}+
   \frac{4\pi^2}{6}\times\frac{12.6}{(10^{10})^2}\:\tau~,\nonumber
\intertext{thus}
\sigma^2_y(\tau)&\simeq1.1{\times}10^{-22}+8.3{\times}10^{-18}\:\tau\nonumber\\
\sigma_y(\tau)&\simeq1.05{\times}10^{-11}+9{\times}10^{-9}\:\sqrt{\tau}\nonumber
\end{align}

\end{enumerate}

\chapter{Phase noise and linear feedback theory}\label{chap:le-theory}
\begin{figure}[t]
\namedgraphics{0.8}{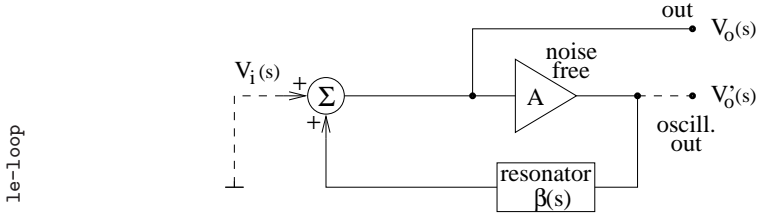}
\caption{Oscillator loop.}
\label{fig:le-loop}
\end{figure}

Figure~\ref{fig:le-loop} proposes a model for the oscillator loop in the domain
of the Laplace transforms.  The amplifier of gain $A$ (constant) and feedback path $\beta(s)$ are the blocks we are familiar with. 
The signal $V_i(s)$ at the input of the summer $\Sigma$ allows initial 
conditions and noise to be introduced in the loop.  Interestingly, if $V_i(s)$ is a driving signal, Fig.~\ref{fig:le-loop} models for the injection-locked oscillator. 
Figure \ref{fig:le-loop} differs form the classical control theory in that the at the feedback input of $\Sigma$ positive, and in that the output is taken at the amplifier input instead of at the output.  This choice simplifies some equations without affecting the phase noise. 
The elementary feedback theory tells us that the transfer function is
\begin{align}
H(s)&=\frac{V_o(s)}{V_i(s)}\qquad\text{def.\,of $H(s)$}
\label{eqn:le-def-hvolt}
\intertext{is}
H(s)&=\frac{1}{1-A\beta(s)}\qquad\text{Fig.\ \ref{fig:le-loop}}~.
\label{eqn:le-closed-loop}
\end{align}
From this standpoint, an oscillator is a system that has (at least) a pair of imaginary conjugate poles, excited with suitable initial conditions. 
One may also interpret frequency stability as the stability of the poles.
We first study the properties of Eq.\ \req{eqn:le-closed-loop} when $\beta(s)$ is the transfer function of a simple resonator, ruled by a second-order differential equation.  
Then, we extend the analysis to the phase space in the Laplace domain, which describes phase noise. 

\section{Oscillator and Laplace transforms}
\begin{figure}[t]
\namedgraphics{0.8}{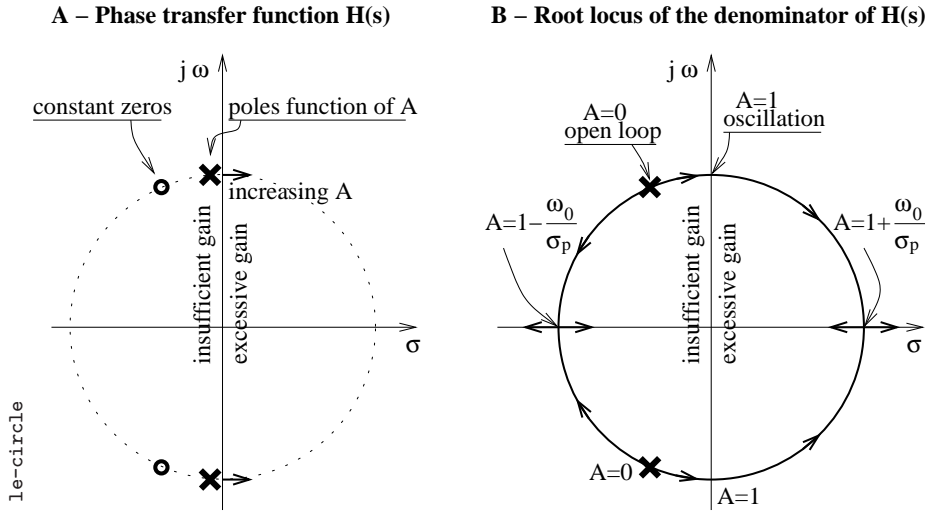}
\caption{A: Noise transfer function $H(s)$ [Eq.~\req{eqn:le-reson-closed-loop}], and B: root locus of the denominator of $H(s)$, as a function of the gain $A$.}
\label{fig:le-circle}
\end{figure}

The normalized transfer function of a resonator (Appendix \ref{app:le-resonator}) is
\begin{align}
\beta(s) &= \frac{\omega_0}{Q}\:\frac{s}{(s-s_p)(s-s_p^*)}
\qquad\begin{array}{c}\text{resonator}\\[-0.5ex]\text{poles}\end{array}
\left\{\!\!\!\begin{array}{l}s_p=\sigma_p+j\omega_p\\
	s_p^*=\sigma_p-j\omega_p\\ \sigma_p^2+\omega_p^2=\omega_0^2
	\end{array}\right.
\end{align}
By replacing $\beta(s)$ in Eq.~\req{eqn:le-closed-loop}, we get
\begin{align}
H(s) &= \frac{(s-s_p)(s-s_p^*)}{(s-s_p)(s-s_p^*)-A\frac{\omega_0}{Q}\:s}
\label{eqn:le-reson-closed-loop}
\end{align}
The above $H(s)=\smash{\frac{\mathcal{N}(s)}{\mathcal{D}(s)}}$ is a second-order rational function with real coefficients.  It has two poles, either real or complex conjugates, depending
on the gain $A$.  The root locus, shown in Figure \ref{fig:le-circle}, has the following properties.
\begin{enumerate}
\item The poles are imaginary conjugates for $A=1$.
\item The poles are complex conjugates for 
	$1-\frac{\omega_0}{\sigma_p}<A<1+\smash{\frac{\omega_0}{\sigma_p}}$,
	with $\sigma_p=\smash{-\frac{\omega_0}{2Q}}$; real elsewhere.
\item The complex conjugate poles are on a circle of radius $\omega_0$ centered in the origin.
\end{enumerate}
The proof is given underneath.

\begin{statement}{Property}
The poles are imaginary conjugates for $A=1$. 
\end{statement}

\begin{proof}
\begin{subproof}
We first rewrite the denominator $\mathcal{D}(s)$ in a more convenient form
\begin{align}
\mathcal{D}(s)
&=(s-s_p)(s-s^*_p)-A\frac{\omega_0}{Q}s\nonumber\\
&=(s-\sigma_p-j\omega_p)(s-\sigma_p+j\omega_p)-A\frac{\omega_0}{Q}s\nonumber\\
&=(s-\sigma_p^2)+\omega_p^2-A\frac{\omega_0}{Q}s\nonumber\\
&=s^2-2\sigma_ps+\sigma_p^2+\omega^2_p+2A\sigma_ps\nonumber
\intertext{and finally}
\mathcal{D}(s)&=s^2+2\sigma_p(A-1)s+\omega_0^2
\label{eqn:le-reson-loop-poles}
\end{align}
To obtain the above, we have used the properties $\sigma_p^2+\omega_p^2=\omega_0^2$, and $\sigma_p=-\frac{\omega_0}{2Q}$.
The poles of $H(s)$ are the solutions of $\mathcal{D}(s)=0$, thus
\begin{equation}
s_1,\;s_2=-\sigma_p(A-1)\pm\sqrt{\sigma^2_p(A-1)^2-\omega_0^2}
\label{eqn:le-reson-loop-poles-sol}
\end{equation}
\end{subproof}

\begin{subproof}{Condition for stationary oscillation.}
Stationary oscillation requires that the poles are imaginary conjugates  
\begin{align}
A&=1  &\text{which yields}\quad s&=\pm j\omega_0\nonumber
\end{align}
This is obtained by replacing $A=1$ in Eq.~\ref{eqn:le-reson-loop-poles}.
The discriminator
\begin{align}
\Delta\mathcal{D}&=\sigma^2_p(A-1)^2-\omega_0^2\nonumber
\end{align}
reduces to $-\omega_0^2$, and the real part $-\sigma_p(A-1)$ of the solutions $s_1$, $s_2$ vanishes. 
\end{subproof}
\end{proof}

\begin{statement}{Property}
The poles are complex conjugates for
$1-\frac{\omega_0}{\sigma_p}<A<1+\smash{\frac{\omega_0}{\sigma_p}}$,
with $\sigma_p=\smash{-\frac{\omega_0}{2Q}}$; real elsewhere.
\end{statement}

\begin{proof}
The poles of \req{eqn:le-reson-closed-loop} are complex conjugates 
when the discriminator $\Delta\mathcal{D}<0$, thus
\begin{align}
\sigma_p^2(A-1)^2-\omega_0^2<0~.\nonumber
\end{align}
We first solve $\Delta\mathcal{D}=0$, which yields 
\begin{align}
A_1,\:A_2=1\pm\frac{\omega_0}{\sigma_p}~.\nonumber
\end{align}
As the coefficient of $A$ in $\Delta\mathcal{D}$ is positive, we conclude that 
\begin{align}
\Delta\mathcal{D}&<0\qquad\text{for}~A_1<A<A_2~.\nonumber
\end{align}

The poles of $H(s)$ are real coincident when $\Delta\mathcal{D}=0$.
We replace $A=A_1$ and $A=A_2$ in $\mathcal{D}$
\begin{equation}
s^2+2\sigma_p\left(1\pm\frac{\omega_0}{\sigma_p}-1\right)+\omega_0^2=0~.
\nonumber
\end{equation}
Hence the poles are real coincident for $s^2+2\omega_0s+\omega_0^2=0$, thus 
\begin{equation}
s_3,\:s_4=\pm\omega_0~.\nonumber
\end{equation}
\end{proof}

\begin{statement}{Property}
The complex conjugate poles are on a circle of radius $\omega_0$ centered in the origin.
\end{statement}

\begin{proof}
The complex conjugate solutions of $\mathcal{D}=0$ are
\begin{align}
s&=-\sigma_p(A-1)\pm j\sqrt{\omega_0^2-\sigma^2_p(A-1)^2}~.
\end{align}
This is an alternate form of Eq.\ \req{eqn:le-reson-loop-poles-sol}, which makes the square root real when $\Delta\mathcal{D}<0$. The square distance $R^2$ of the poles from the origin is 
\begin{align}
R^2&=[\Re(s)]^2+[\Im(s)]^2\nonumber\\
   &=\sigma^2_p(A-1)^2+\omega_0^2-\sigma_p^2(A-1)^2~,\nonumber
\intertext{which simplifies as}
R^2 &=\omega_0^2.
\end{align}
\end{proof}

\begin{remark} 
That the poles of $H(s)$ are on a circle centered in the origin has an important consequence in metrology.  In the real world \emph{the gain fluctuates} around the value $A=1$. For small fluctuations of $A$, the poles fluctuate perpendicularly to the imaginary axis.  The effect on the oscillation frequency is of the second order only.
\end{remark}

\begin{remark} 
The exact condition $A=1$ can not be ensured without a gain control mechanism.  The latter can be interpreted as a control that stabilizes the oscillator poles onto the imaginary axis.

The gain control can be a feedback that sets $A$ for the output voltage to be constant.
This approach was followed in early times of electronics in the Wien bridge oscillator \cite{packard:thesis}.  Yet amplifier saturation proved to be an effective amplitude control, even in ultra-stable oscillators.  When the amplifier saturates, the output signal is clipped more or less smoothly.  Power leakage from the fundamental to the harmonics reduces the gain, and in turn stabilizes the output amplitude.  The resonator prevents the harmonics from being fed back to the amplifier input.
\end{remark}

\section{Resonator in the phase space}\label{sec:le-reson-phi-space}
%
\begin{figure}[t]
\namedgraphics{0.8}{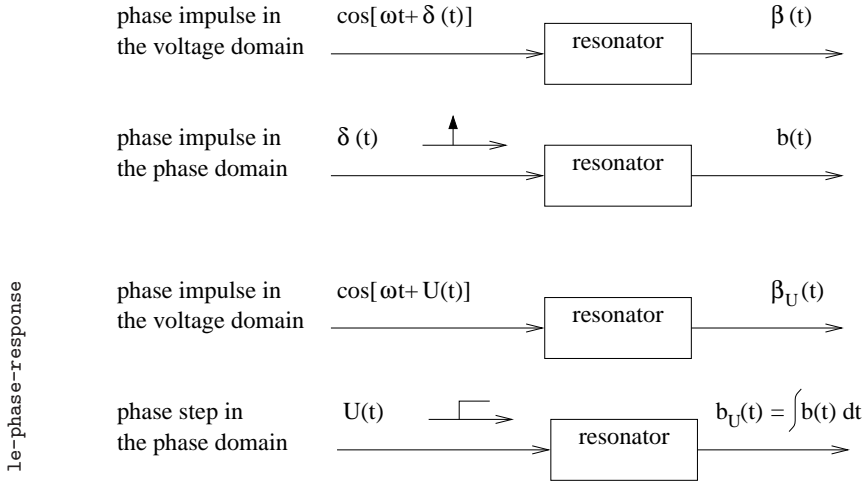}
\caption{Resonator response to a phase impulse $\delta(t)$.}
\label{fig:le-phase-response}
\end{figure}
We analyze the resonator phase response $b(t)$, and its Laplace transform $B(s)$, in quasi-stationary conditions.  The resonator is driven with a sinusoidal signal at the frequency $\omega$, which can be the natural frequency $\omega_0$ or any other frequency.
In the time domain, the phase transfer function $b(t)$ is the phase of the resonator response to a Dirac $\delta(t)$ function in the phase of the input.  More precisely, $b(t)$ is defined as follows (Fig.~\ref{fig:le-phase-response})
\begin{align}
v_i(t)&=\cos[\omega_0t+\delta(t)] &&\text{input with phase impulse} \nonumber\\
v_o(t)&=\cos[\omega_0t+b(t)]       &&\text{output response, defines $b(t)$}~. \nonumber
\end{align}
We find $b(t)$ from the response $b_U(t)$ to the Heaviside function (unit step)
\begin{align}
U(t)&=\int_{-\infty}^{\infty}\delta(t)dt=
  \begin{cases}0&t<0\\1&t>0\end{cases}\qquad\text{Heaviside}
\end{align}
using the property of linear systems that the impulse response is the derivative of the step response
$b_U(t)$
\begin{equation}
b(t)=\frac{d}{dt}b_U(t)~.
\label{eqn:le-deriv-prop}
\end{equation}
We calculate the response in small signal conditions, using a phase step $\kappa U(t)$, $\kappa\rightarrow0$, so that some functions can be linearized.  This is physically correct because phase noise in actual oscillators is a small signal.

\begin{figure}[t]
\namedgraphics{0.8}{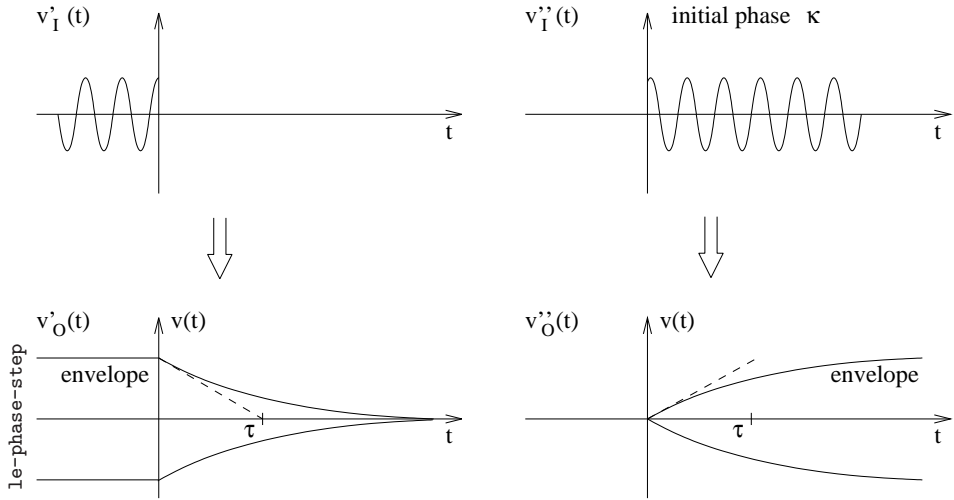}
\caption{A sinusoid with a phase step can be decomposed into a sinusoid switched off, plus a shifted sinusoid switched on.  The step response is the linear superposition of the two responses.}
\label{fig:le-phase-step}
\end{figure}

The method (Fig.~\ref{fig:le-phase-step}) consists of breaking the input sinusoid $v_i(t)$ at the time $t=0$
\begin{align}
v_i(t)&=v'_i(t)+v''_i(t)&  v_o(t)&=v'_o(t)+v''_o(t),\quad\text{output, for $t>0$}\nonumber\\
v'_i(t)&=v_i(t)U(-t)   &v'_o(t)&\quad\text{response to $v'_i(t)$, for $t>0$} \nonumber\\
v''_i(t)&=v_i(t)U(t)  &v''_o(t)&\quad\text{response to $v''_i(t)$, for $t>0$}~, \nonumber
\end{align}
so that the phase step $\kappa U(t)$ can be introduced in the phase of $v''_i(t)$.
The terms $v'_o(t)$ and $v''(t)$ are, respectively, the switch-on and the switch-off transient response for $t>0$.

\subsection{Input signal tuned at the exact natural frequency}
\begin{statement}{Property}
The impulse response of the resonator in the phase space is
\begin{align}
b(t)=\frac{1}{\tau}e^{-t/\tau}
\qquad\begin{array}{l}\text{impulse}\\[-0.5ex]\text{response}\end{array}
\label{eqn:le-tuned-timedom}
\end{align}
and its Laplace transform is
\begin{equation}
B(s)=\frac{1}{s\tau+1}
\qquad\begin{array}{l}\text{transfer}\\[-0.5ex]\text{function}\end{array}
\label{eqn:le-tuned-laplace}
\end{equation}
\end{statement}

\begin{proof}
Let
\begin{align}
v_i(t)&=\cos\bigl[\omega_0t + \kappa U(t)\bigr] \nonumber\\
	&=\cos(\omega_0t)U(-t) + \cos(\omega_0t+\kappa)U(t)\nonumber
\end{align}
the input signal.  The resonator response $v'_o(t)$ to the sinusoid switched off is the exponentially decaying sinusoid
\begin{align}
&v'_o(t)=\cos(\omega_0t)e^{-t/\tau}\qquad t>0~,\nonumber
\end{align}
where $\tau=\smash{\frac{2Q}{\omega_0}}$ is the resonator relaxation time 
(Appendix \ref{app:le-resonator}).
Similarly, the response $v''_o(t)$ to the switched-on sinusoid is the exponentially growing sinusoid
\begin{align}
&v''_o(t)=\cos(\omega_0t+\kappa)\bigl[1-e^{-t/\tau}\bigr]\qquad t>0~.\nonumber
\end{align}
The total output signal is
\begin{align}
v_o(t)
&=v'_o(t)+v''_o(t)\qquad\qquad t>0 \nonumber\\
&=\cos(\omega_0t)e^{-\frac{t}{\tau}}
     +\Bigl[\cos(\omega_0t)\cos(\kappa)
     -\sin(\omega_0t)\sin(\kappa) \Bigr]
     \Bigl[1-e^{-\frac{t}{\tau}}\Bigr] \nonumber\\
&=\cos(\omega_0t)
      \Bigl[e^{-\frac{t}{\tau}}+\cos(\kappa)
            -\cos(\kappa)e^{-\frac{t}{\tau}}\Bigr]
     -\sin(\omega_0t)\sin(\kappa)\Bigl[1-e^{-\frac{t}{\tau}}\Bigr]~. \nonumber
\end{align}
For $\kappa\rightarrow0$ we replace $\cos(\kappa)=1$ and $\sin(\kappa)=\kappa$, thus
\begin{align}
v_o(t)&=\cos(\omega_0t)-\kappa\sin(\omega_0t)\bigl[1-e^{-/\tau}\bigr]\nonumber
\end{align}
Deleting $\omega_0t$, the above can be seen as a slowly varying phasor
\begin{align}
V_o(t)& = 1+j\kappa\bigl[1-e^{-t/\tau}\bigr]\qquad\kappa\ll1~.\nonumber
\end{align}
The angle $\smash{\arctan\frac{\Im\{V_o(t)\}}{\Re\{V_o(t)\}}}$, normalized on $\kappa$, is the step response
\begin{align}
b_U(t)&=1-e^{-t/\tau}\qquad\text{step response}\nonumber
\end{align}
Taking the derivative, we obtain the impulse response
\begin{align}
b(t)=\frac{1}{\tau}e^{-t/\tau}~,
\end{align}
which is Eq.\ \req{eqn:le-tuned-timedom}.  
The Laplace transform [Eq.\ \req{eqn:le-tuned-laplace}]  is found in common mathematical tables.  The proof is omitted.
\end{proof}

\subsection{Detuned input signal}
\begin{figure}[t]
\namedgraphics{0.5}{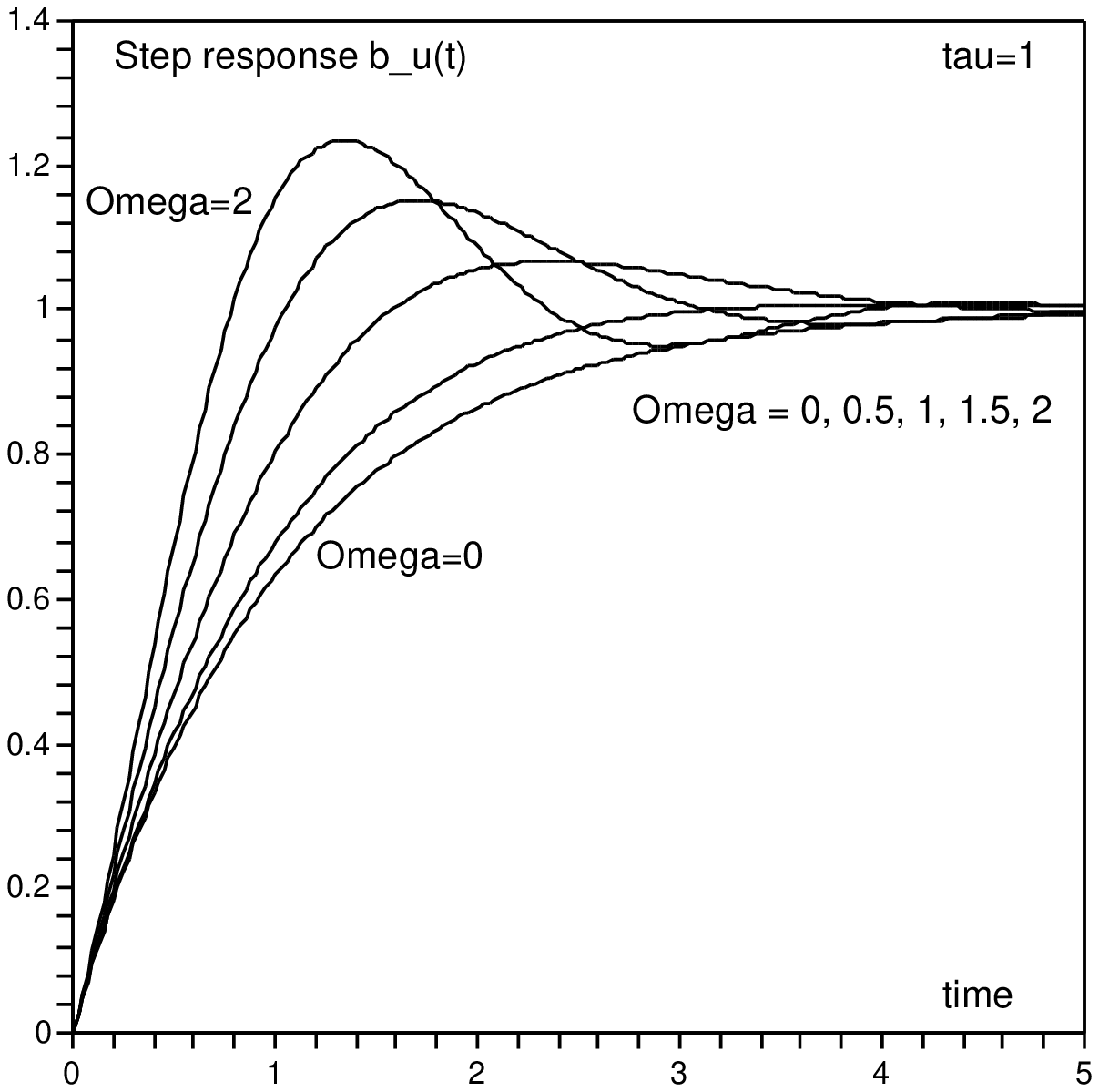}
\caption{Step response $b_U(t)$ evaluated for $\tau=1$.}
\label{fig:le-calc-bu}
\end{figure}
\begin{figure}[t]
\namedgraphics{0.5}{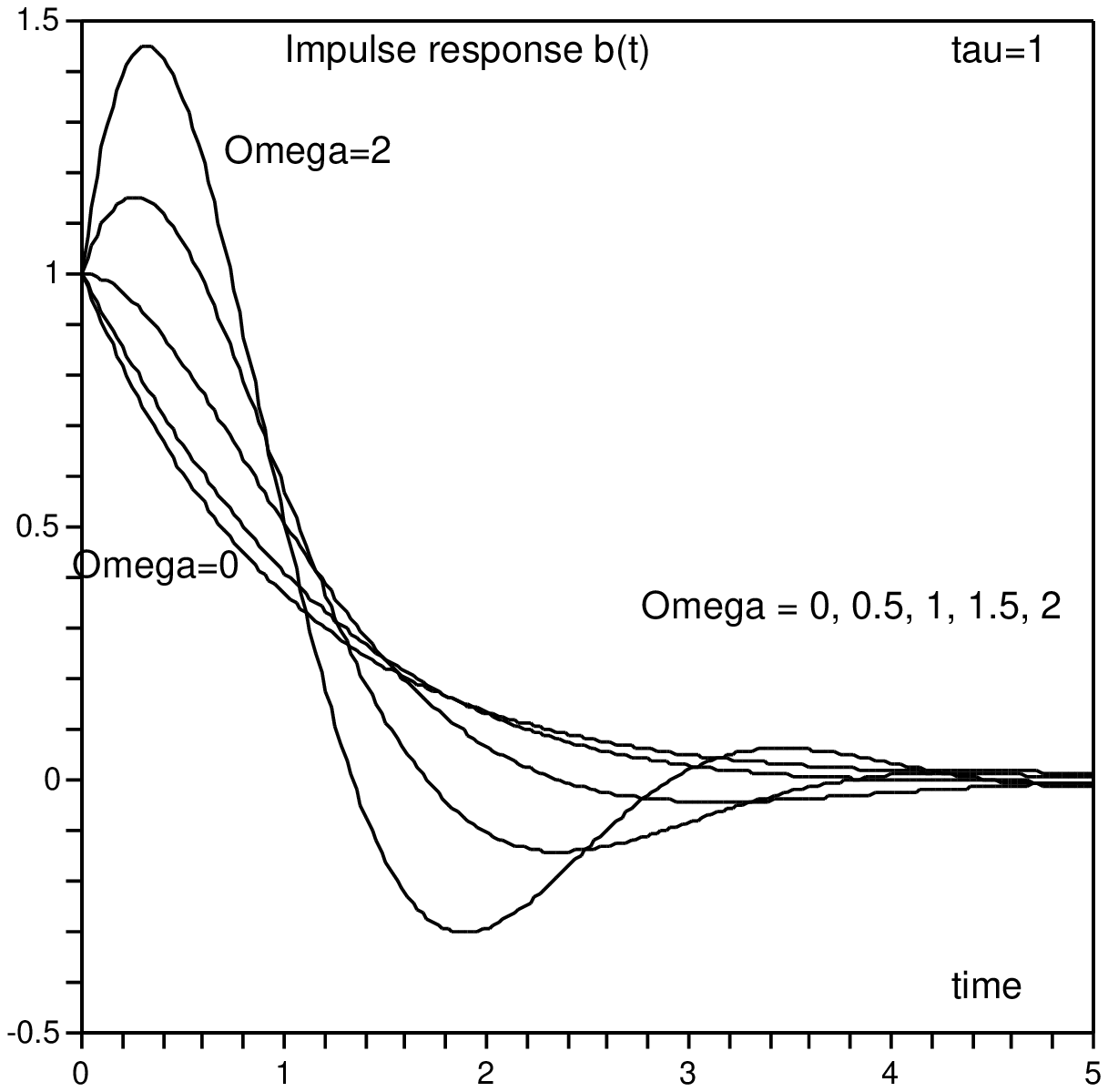}
\caption{Impulse response $b(t)$ evaluated for $\tau=1$.}
\label{fig:le-calc-b}
\end{figure}
\begin{statement}{Property}
The step response of the resonator in the phase space is
\begin{align}
b_U(t)&=1-\cos(\Omega t) e^{-t/\tau}
\qquad\begin{array}{l}\text{step response}\\[-0.5ex]
		\text{(Fig.\,\ref{fig:le-calc-bu})}\end{array}~,
\label{eqn:le-detun-bu}
\end{align}
and the impulse response impulse is
\begin{align}
b(t)&=\Bigl[\Omega\sin(\Omega t) + \frac1\tau \cos(\Omega t)\Bigr] e^{-t/\tau}
	\qquad\begin{array}{l}\text{impulse response}\\[-0.5ex]
	\text{(Fig.\ \ref{fig:le-calc-b})}\end{array}~,
\label{eqn:le-detun-b}
\end{align}
where 
\begin{align}
\Omega=\omega_1-\omega_0\qquad\text{def.\,of $\Omega$}~.
\end{align}
is the frequency offset (or detuning), i.e., the difference between the operating frequency $\omega_1$ and the natural frequency $\omega_0$ of the resonator.
\end{statement}

\begin{proof}
The resonator asymptotic response ($t\rightarrow\infty$) to the detuned input sinusoid $v_i(t)=\cos(\omega_1t)$, $\omega_1\neq\omega_0$, is $v_o(t)=V_\infty\cos(\omega_1t+\varphi)$, with $V_\infty$ and $\varphi$ constants (cfr.\ Appendix \ref{app:le-resonator}, p.\,\pageref{app:le-resonator}).
Without loss of generality, we can scale and shift the input signal
\begin{align}
v_i(t)&=\frac{1}{V_\infty}\cos(\omega_1t-\varphi)~,\nonumber
\intertext{in order to get the asymptotic response of the form}
v_o(t)&=\cos(\omega_1t) \qquad t\rightarrow\infty~.\nonumber
\end{align}
We use the method of the small phase step $\kappa U(t)$.  Yet we need some preliminary results before introducing the step.

\begin{subproof}{Input signal switched off at the time $t=0$.}
The input signal 
\begin{align}
v'_i(t)&=\frac{1}{V_\infty}\cos(\omega_1t-\varphi) U(-t)~,\nonumber
\intertext{switched off at $t=0$ by $U(-t)$, yields the output}
v'_o(t)&=\cos(\omega_0t) e^{-t/\tau}~\qquad t>0~.
\label{eqn:le-detun-decay}
\end{align}
\end{subproof}

\begin{subproof}{Input switched on at the time $t=0$.}
It is known from the differential equation theory that, in the presence of the input signal
\begin{align}
v''_i(t)&=\frac{1}{V_\infty}\cos(\omega_1t-\varphi) U(t) \nonumber
\end{align}
switched on at $t=0$ by $U(t)$, the output is of the form
\begin{align}
v''_o(t)&=a\cos(\omega_0t)e^{-\frac{t}{\tau}}+b\sin(\omega_0t)e^{-\frac{t}{\tau}}
	+c\cos(\omega_1t)+d\sin(\omega_1t)\qquad t>0\nonumber
\end{align}
with $a$, $b$, $c$, and $d$ constants to be determined.  As for $t\rightarrow\infty$ it holds that $v_o(t)=\cos(\omega_1t)$, we conclude that $c=1$ and $d=0$.
Then, $a$ and $b$ are found by assessing the continuity of the output at $t=0$ when the input signal is $v_i(t)=\smash{\frac{1}{V_\infty}}\cos(\omega_1t-\varphi)$.
The continuity condition is met for $a=-1$ and $b=0$. 

In summary, it holds the following input-output relationship
\begin{align}
\text{in/out pair}\quad&
\begin{array}{rl}
v''_i(t)&=\frac{1}{V_\infty}\cos(\omega_1t-\varphi) U(t)\\[1ex]
v''_o(t)&=\left[-\cos(\omega_0t)e^{-t/\tau}+\cos(\omega_1t)\right]U(t)
\end{array}
\label{eqn:le-detun-cos-pair}
\intertext{and, as an obvious extension,}
\text{in/out pair}\quad&
\begin{array}{rl}
v''_i(t)&=\frac{1}{V_\infty}\sin(\omega_1t-\varphi) U(t)\\[1ex]
v''_o(t)&=\left[-\sin(\omega_0t)e^{-t/\tau}+\sin(\omega_1t)\right]U(t)
\end{array}
\label{eqn:le-detun-sin-pair}
\end{align}
\end{subproof}

\begin{subproof}{Phase step at $t=0$.}
Let us define the input signal $v_i(t)=v'_i(t)+v''_i(t)$
\begin{align}
v'_i(t)&=\tfrac{1}{V_\infty}\cos(\omega_1t-\varphi) U(-t) \nonumber\\
v''_i(t)&=\tfrac{1}{V_\infty}\cos(\omega_1t-\varphi+\kappa) U(t) & \kappa\ll1~.\nonumber
\end{align}
The input $v''_i(t)$ can be rewritten as
\begin{align}
v''_i(t)&=\tfrac{1}{V_\infty}\bigl[\cos(\omega_1t-\varphi)\cos\kappa
	-\sin(\omega_1t-\varphi)\sin\kappa\bigr] U(t) \nonumber\\
&=\tfrac{1}{V_\infty}\bigl[\cos(\omega_1t-\varphi)
	-\kappa\sin(\omega_1t-\varphi)\bigr] U(t)  \qquad\kappa\ll1~.\nonumber
\end{align}
Using Eq.\ \req{eqn:le-detun-decay} and the pairs \req{eqn:le-detun-cos-pair} and \req{eqn:le-detun-sin-pair}, we get the output signal
\begin{align}
v_o(t)&=v'_o(t)+v''_o(t) &&t>0 \nonumber\\
&=\cos(\omega_0t) e^{-t/\tau}  &&\text{response to $v'_i$} \nonumber\\
&+\bigl[-\cos(\omega_0t) e^{-t/\tau} + \cos(\omega_1t)\bigr]
	&&\text{response to $v''_i$, 1st part} \nonumber\\
&+\kappa\bigl[\sin(\omega_0t) e^{-t/\tau} - \sin(\omega_1t)\bigr]
	&&\text{response to $v''_i$, 2nd part} \nonumber
\end{align}
hence
\begin{align}
v_o(t) &=\cos(\omega_1t) - \kappa\sin(\omega_1t)
	+ \kappa\cos(\omega_0t) e^{-t/\tau}\qquad t>0~.
\label{eqn:le-detun-vout-a}
\end{align}
Let us define the detuning frequency $\Omega=\omega_1-\omega_0$
Accordingly, it holds that $\omega_0=\omega_1-\Omega$, hence $\sin(\omega_0t)=\sin(\omega_1t-\Omega t)$, and consequently
\begin{align}
\sin(\omega_0t)
&=\sin(\omega_1t)\cos(\Omega t)-\cos(\omega_1t)\sin(\Omega t)~.\nonumber
\end{align}
The output signal \req{eqn:le-detun-vout-a}, written in terms of $\omega_1$ and $\Omega$, is
\begin{align*}
v_o(t)&=\cos(\omega_1t)-\kappa\sin(\omega_1t) + \\
	&{}+\kappa\sin(\omega_1t)\cos(\Omega t) e^{-\frac{t}{\tau}}
	-\kappa\cos(\omega_1t)\sin(\Omega t) e^{-\frac{t}{\tau}}
\end{align*}
which simplifies as
\begin{align}
v_o(t)&=\cos(\omega_1t)\Bigl[1-\kappa\sin(\Omega t) e^{-\frac{t}{\tau}}\Bigr]
	-\kappa\sin(\omega_1t)\Bigl[1-\cos(\Omega t) e^{-\frac{t}{\tau}}\Bigr]~.
\label{eqn:le-detun-vout}
\end{align}
\end{subproof}

\begin{subproof}{Output phase response.}
Freezing the oscillation $\omega_1t$, the output signal \req{eqn:le-detun-vout} turns into  a slow-varying phasor
\begin{align}
V_o(t)&=1+ j\kappa\bigl[1-\cos(\Omega t) e^{-t/\tau}\bigr]~\qquad\kappa\ll1.\nonumber
\end{align}
The angle $\smash{\arctan\frac{\Im\{V_o(t)\}}{\Re\{V_o(t)\}}}$, normalized on $\kappa$, is the phase-step response
\begin{align}
b_U(t)&=1-\cos(\Omega t) e^{-t/\tau}
\qquad\begin{array}{l}\text{step response}\\[-0.5ex]
		\text{(Fig.\,\ref{fig:le-calc-bu})}\end{array}~.\nonumber
\end{align}
Using the property $b(t)=\frac{d}{dt}b_U(t)$ [Eq.\,\req{eqn:le-deriv-prop}], we find the impulse response
\begin{align}
b(t)&=\Bigl[\Omega\sin(\Omega t) + \frac1\tau \cos(\Omega t)\Bigr] e^{-t/\tau}
	\qquad\begin{array}{l}\text{impulse response}\\[-0.5ex]
	\text{(Fig.\,\ref{fig:le-calc-bu})}\end{array}~.\nonumber
\end{align}
\end{subproof}
\end{proof}

\begin{remark}
In some unusual cases, a resonator may need a pump signal, for the measurement of the resonator parameter is difficult or impossible if the resonator is not in the oscillator loop.  This occurs in the domain of photonics with the coupled opto-electronic oscillator (COEO) \cite{yao97ol}.  In such case the phase step introduced in the oscillator loop proved to be a useful method for the measurement of the resonator parameters.  A microwave phase modulator (varactor) was inserted in series to the resonator and driven with a square-wave generator at a low frequency.
\end{remark}

\begin{figure}[t]
\namedgraphics{0.8}{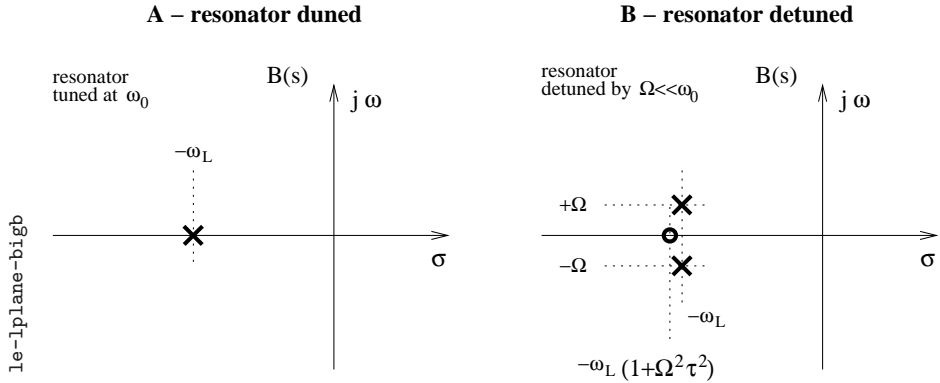}
\caption{Phase noise transfer function $B(s)$ on the Laplace plane.}
\label{fig:le-lplane-bigb}
\end{figure}

\begin{figure}[t]
\namedgraphics{0.5}{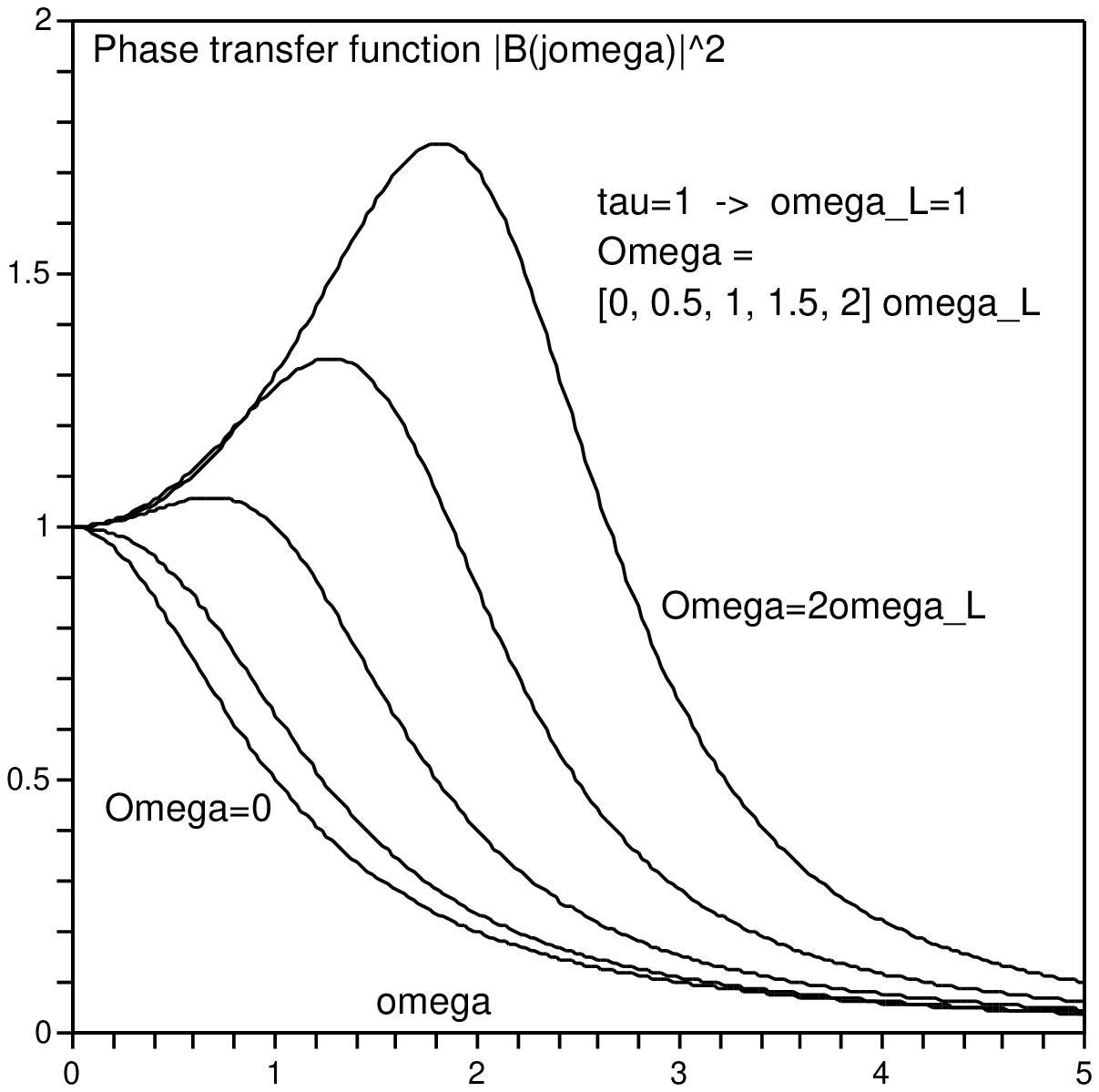}
\caption{Phase noise transfer function $|B(j\omega)|^2$ [Eq.\ \req{eqn:le-detun-laplace}] evaluated for $\tau=1$, thus $\omega_L=1$.  We observe that the transformation $\tau=\smash{\frac{1}{2\pi}}$ instead of $\tau=1$ (hence $f_L=1$), $\Omega\rightarrow\smash{\frac{\Omega}{2\pi}}=\nu_1-\nu_0$, and $\omega\rightarrow f$, lets the plot unchanged.  In other words, if the resonator Leeson frequency is the $f_L=1$ and one reads $f$ instead of $\omega$ on the horizontal axis, the curves are the phase response $\smash{|B(jf)|^2}$ of the resonator tuned, detuned by $\smash{\frac{1}{2}}f_L$, by $f_L$, etc.}
\label{fig:le-calc-bigb}
\end{figure}

\begin{statement}{Property}
The phase transfer function of the resonator in the presence of a detuned driving signal is
\begin{align}
B(s)&=\frac{1}{\tau}\:\frac{s+\frac1\tau+\Omega^2\tau}{%
    \left(s+\frac1\tau-j\Omega\right)\left(s+\frac1\tau+j\Omega\right)}
    \qquad\begin{array}{l}\text{transfer function}\\[-0ex]
    	\text{(Fig.\,\ref{fig:le-lplane-bigb} and \ref{fig:le-calc-bigb})}\end{array}~.
\label{eqn:le-detun-laplace}
\end{align}
\end{statement}


\begin{proof}
The Laplace transform $B(s)=\mathcal{L}\smash{\bigl\{b(t)\bigr\}}$ is found using the Euler identities
\begin{align*}
\cos(\Omega t)&=\tfrac12\left(e^{j\Omega t} + e^{-j\Omega t}\right)\\
\sin(\Omega t)&=\tfrac{1}{j2}\left(e^{j\Omega t} - e^{-j\Omega t}\right)
\end{align*}
and the properties of the Laplace transform
\begin{align}
\mathcal{L}\bigl\{e^{-t/\tau}\bigr\}&=\frac{1}{s+1/\tau} \nonumber\\
\mathcal{L}\bigl\{e^{at}f(t)\bigr\}&=F(s-a)~.\nonumber
\end{align}
Thus
\begin{align}
B(s)&=\mathcal{L}\Bigl\{\Bigl[
   \Omega\:\frac{1}{j2}\left(e^{j\Omega t} - e^{-j\Omega t}\right)
  +\frac1\tau\: \frac12\left(e^{j\Omega t} + e^{-j\Omega t}\right)
  \Bigr] e^{-t/\tau}\Bigr\}\nonumber
\end{align}
and finally
\begin{align}
B(s)&=\frac{1}{\tau}\:\frac{s+\frac1\tau+\Omega^2\tau}{%
    \left(s+\frac1\tau-j\Omega\right)\left(s+\frac1\tau+j\Omega\right)}
    \qquad\begin{array}{l}\text{transfer function}\\[-0ex]
    	\text{(Fig.\,\ref{fig:le-lplane-bigb} and \ref{fig:le-calc-bigb})}\end{array}~.\nonumber
\end{align}
\end{proof}

\begin{remark}
The phase noise bandwidth of the resonator increases when the resonator is detuned (Fig.\ \ref{fig:le-calc-bigb}).  This is related to the following two interesting facts.
\begin{enumerate}
\item At $\omega\neq\omega_0$, the slope $\smash{\left|\frac{d\,\arg[\beta(j\omega)]}{d\omega}\right|}$ is lower than at $\omega_0$ (Fig.\ \ref{fig:le-resonator}).  Hence, off the exact resonance, the resonator becomes less dispersive.
\item The step response is faster when the resonator is detuned, as it is seen in Fig.\ \ref{fig:le-calc-bu}.
\end{enumerate}
\end{remark}

\section{Another derivation of the resonator phase response}\label{sec:le-chuck}
The low-frequency phase transfer function of the resonator is
\begin{align}
\mathcal{H}(j\omega)=\frac{1}{2\left|H(j\omega_1)\right|}
	\Bigl[H(j(\omega-\omega_1)) + H(j(\omega+\omega_1))\Bigr]
	\label{eqn:le-chuck-phase}
\end{align}
The derivation\footnote{Thanks to Charles (Chuck) Greenhall, JPL, Pasadena, CA.} of the resonator phase response given here is more general than that
shown in Section \ref{sec:le-reson-phi-space}, which holds only for resonator. 

\begin{proof}
Let $h(t)$ the voltage impulse response of the resonator and $H(j\omega)$ its Fourier transform.  Let $v_i(t)=\cos[\omega_1t+\theta+\phi(t)]$ the input voltage.  The output voltage is 
\begin{align}
v_o(t)&=(h*v_i)(t)\qquad\qquad\text{convolution}\nonumber\\
&=\int_{-infty}^\infty h(u)v_i(u)\:du	\nonumber\\
&=\Re\biggl\{\int_{-\infty}^\infty h(u)e^{j[\omega_1(t-u)+\theta+\phi(t-u)]}\:du\biggr\}
	\nonumber\\
&=\Re\biggl\{e^{j(\omega_1t+\theta)}
	\int_{-\infty}^\infty h(u)e^{j[-\omega_1u+\phi(t-u)]}\:du\biggr\}
	\nonumber
\end{align}   
Linearize for $\phi\ll1$ 
\begin{align}
v_o(t)
&=\Re\biggl\{e^{j(\omega_1t+\theta)}
	\int_{-\infty}^\infty h(u)e^{-j\omega_1u}\left[1+j\phi(t-u)\right]\:du\biggr\}
	\nonumber\\
&=\Re\biggl\{e^{j(\omega_1t+\theta)}
	\left[H(j\omega_1) + j\int_{-\infty}^\infty h(u)e^{-j\omega_1u}\phi(t-u)\:du\right]\biggr\}
	\nonumber
\end{align}   
Replace $H(j\omega_1)=H_1e^{j\alpha_1}$
\begin{align*}
&v_o(t)=\Re\biggl\{e^{j(\omega_1t+\theta)}
	\left[H_1e^{j\alpha_1} + j\int_{-\infty}^\infty h(u)e^{-j\omega_1u}\phi(t-u)
	\:du\right]\biggr\}\\
&=\Re\biggl\{e^{j(\omega_1t+\theta)} H_1e^{j\alpha_1}
	\left[1 + \frac{1}{H_1e^{j\alpha_1}} 
	\int_{-\infty}^\infty h(u)\;je^{-j(\omega_1u+\alpha_1)}\phi(t-u)
	\:du\right]\biggr\}
\end{align*}
Observe that $je^{-j(\omega_1u+\alpha_1)}
=\sin(\omega_1u+\alpha_1)+j\cos(\omega_1u+\alpha_1)$.  Define
\begin{align*}
h_c(t)=h(t)\cos(\omega_1u+\alpha_1)\\
h_s(t)=h(t)\sin(\omega_1u+\alpha_1)
\end{align*}
Thus
\begin{multline*}
v_o(t)=\Re\biggl\{e^{j(\omega_1t+\theta)} H_1e^{j\alpha_1}\\
	\left[1 + \frac{1}{H_1} \int_{-\infty}^\infty h_s(u)\phi(t-u)\:du
	+j\frac{1}{H_1} \int_{-\infty}^\infty h_c(u)\phi(t-u)\:du
	\right]\biggr\}
\end{multline*}
Use $1+a+jb\approx(1+a)e^{jb}$ for small $a$ and $b$
\begin{align*}
v_o(t)=\Re\biggl\{e^{j(\omega_1t+\theta+\alpha_1)} H_1e^{j\alpha_1}
	H_1 \left[1 + \frac{(h_s*\phi)(t)}{H_1}\right] 
	\exp\left[j\frac{(h_c*\phi)(t)}{H_1}\right]\biggr\}
\end{align*}
Taking $\alpha_1=0$, the equivalent filters are
\begin{align*}
\text{AM:\qquad}&\frac{h_s(t)}{H_1}\:* &
\text{PM:\qquad}&\frac{h_c(t)}{H_1}\:*
\end{align*}
From $\cos(x)=\smash{\frac12}\left[e^{jx}+e^{-jx}\right]$ it follows that
\begin{align*}
h_c(t)&=\frac{1}{2}\Bigl[e^{jx}+e^{-jx}\Bigr]h(t)
\end{align*} 
Consequently
\begin{align*}
\mathcal{H}(j\omega)=\frac{1}{2H_1}
	\Bigl[H(j(\omega-\omega_1)) + H(j(\omega+\omega_1))\Bigr]
\end{align*} 
which is equivalent to Eq.\ \req{eqn:le-chuck-phase}.
\end{proof}

\begin{remark}
Eq.\ \req{eqn:le-chuck-phase} states that the poles (and the zeros, if any) of  $H(s)$ appear in $\mathcal{H}(s)$ translated by $\pm\omega_1$.  $H(s)$ has (at least) a pair of complex conjugate poles on the left-hand half-plane at about $\pm\omega_0$.   
A copy of them is present in $\mathcal{H}(s)$ translated close to the real axis.  Another copy translated in the opposite direction is present in $\mathcal{H}(s)$, about $\pm2\omega_1$.  This second copy should be discarded because it results from the approximations taken.  Anyway, questioning about the phase noise at a ``close-in'' frequency twice the carrier frequency is a nonsense.
\end{remark}

\begin{remark}
If $\omega_1\neq\omega_0$, the resonator natural frequency, there results a zero on the real axis, in the negative half plane.  This is still to be demonstrated with the formalism of this Section. 
\end{remark}

\begin{statement}{Property}
The resonator converts phase noise into amplitude noise.  The transfer function of this process is 
\begin{align}
\mathcal{H}_{X}(j\omega)=\frac{1}{2\left|H(j\omega_1)\right|}
	\Bigl[H(j(\omega-\omega_1) - H(j(\omega+\omega_1)\Bigr]
	\label{eqn:le-chuck-pm2am}
\end{align} 
\end{statement}

\begin{proof}
Proving Eq.\ \req{eqn:le-chuck-phase}, we come across the AM filter
$\smash{\frac{h_s(t)}{H_1}\,*}$.  This is the amplitude response to a phase modulation.
Using $\sin(x)=\smash{\frac{1}{2j}}\left[e^{jx}-e^{-jx}\right]$, we get
\begin{align*}
h_s(t)&=\frac{1}{j2}\Bigl[e^{jx}-e^{-jx}\Bigr]h(t)
\end{align*} 
Eq.\ \req{eqn:le-chuck-pm2am} follows immediately.
\end{proof}

\section{Phase noise in the oscillator loop}
\begin{figure}[t]
\namedgraphics{0.8}{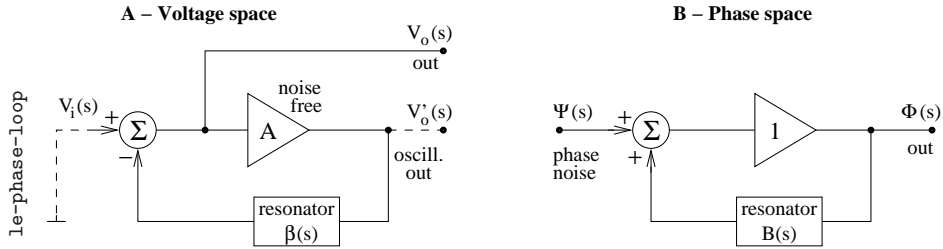}
\caption{Derivation of the oscillator phase-noise model (B) 
from the voltage-noise oscillator model (A).}
\label{fig:le-phase-loop}
\end{figure}
Whereas phase noise can be analyzed with the generic model of Fig.~\ref{fig:le-loop}, we turn our attention to the model of Fig.~\ref{fig:le-phase-loop}\,B, specific to phase noise.
The physical quantities of Fig.~\ref{fig:le-phase-loop}\,B are the Laplace
transform of the oscillator phase fluctuations.  Thus the input signal $\Psi(s)$ models the amplifier phase noise. Additionally $\Psi(s)$ can be used to introduce the resonator fluctuations, or the phase noise of an external signal to which the oscillator is locked by injection.  The amplifier gain is 1 because the amplifier repeats the input phase to the output. The filter transfer function $\beta(s)$ is replaced with $B(s)$, which is the 
Laplace transform of the resonator phase response.

The phase-noise transfer function is
\begin{align}
\mathcal{H}(s)&=\frac{\Phi(s)}{\Psi(s)}
	\qquad\text{def.\,of $\mathcal{H}(s)$}
	\label{eqn:le-pnoise-xfer-def}\\
\mathcal{H}(s)&=\frac{1}{1-B(s)}
	\qquad\text{oscillator}~.
	\label{eqn:le-pnoise-xfer}
\end{align}

\begin{statement}{Property}
The phase transfer function of the oscillator tuned at the exact resonant frequency $\omega_0$ is
\begin{align}
\mathcal{H}(s)&=\frac{1+s\tau}{s\tau}
&\begin{array}{l}\text{exact natural}\\\text{frequency}~\omega_0\end{array}
	\label{eqn:le-pnoise-h-tuned}\\
|\mathcal{H}(j\omega)|^2
&=\frac{\tau^2\omega^2+1}{\tau^2\omega^2}
	\label{eqn:le-pnoise-h2-tuned}
\end{align}
\end{statement}

\begin{proof}  
Replace $B(s)$ given by Eq.\ \req{eqn:le-tuned-laplace} in \req{eqn:le-pnoise-xfer}.
Using $\smash{\frac{1}{1-n/d}}=\smash{\frac{d}{d-n}}$, the result \req{eqn:le-pnoise-h-tuned} follows immediately.  The proof for $|\mathcal{H}(j\omega)|^2$ is omitted.
\end{proof} 

\begin{figure}[t]
\namedgraphics{0.8}{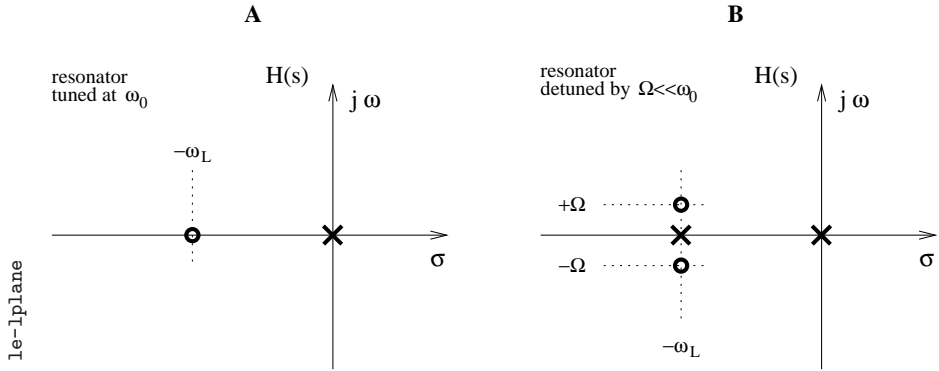}
\caption{Oscillator phase-noise transfer function $\mathcal{H}(s)$.}
\label{fig:le-lplane}
\end{figure}%

\begin{statement}{Leeson formula}%
\addcontentsline{toc}{section}{Formal derivation of the Leeson formula}
The oscillator phase noise spectrum
\begin{align} 
S_\phi(f)&=\left[1+\frac{1}{f^2}\left(\frac{\nu_0}{2Q}\right)^2\right]S_\psi(f)~.
	\qquad\text{[Eq.\ \req{eqn:le-leeson-heuristic}]}
\end{align}
where $S_\psi(f)$ is the spectrum of the phase fluctuation $\psi(t)$.
\end{statement}

\begin{proof}
From the definition of $\mathcal{H}(s)$ [Eq.\ \req{eqn:le-pnoise-xfer-def}], it follows that
\begin{align*} 
S_\phi(\omega)&=|\mathcal{H}(j\omega)|^2\:S_\psi(\omega)~.
\end{align*}
We replace Eq.\ \req{eqn:le-pnoise-h2-tuned} in the above
\begin{align*} 
S_\phi(\omega)&=\left[1+\frac{1}{\omega^2\tau^2}\right]\:S_\psi(\omega)~.
\end{align*}
Then, we replace $\omega=2\pi f$ and
$\tau=\smash{\frac{2Q}{\omega_0}}=\smash{\frac{Q}{\pi\nu_0}}$ 
[Eq.\ \req{eqn:le-reson-tau}]
\begin{align*} 
S_\phi(f)&=\left[1+\frac{1}{4\pi^2f^2}\:\frac{\pi^2\nu_0^2}{Q^2}\right]
	\:S_\psi(f)\\[1ex]
&=\left[1+\frac{1}{f^2}\left(\frac{\nu_0}{2Q}\right)^2\right]S_\psi(f)~.
\end{align*}
\end{proof}

\begin{statement}{Property}
The phase transfer function of the detuned oscillator,  oscillating at the frequency $\omega_1=\omega_0+\Omega$ is
\begin{align}
\mathcal{H}(s)&=\frac{(s\tau+1)^2+\Omega^2\tau^2}{s(s\tau+1)}
	\qquad\qquad\qquad\begin{array}{l}\text{detuned}\\\text{by}~\Omega\end{array}
	\label{eqn:le-pnoise-h-detuned}\\
&=\frac{\left(s+\frac1\tau-j\Omega\right)%
	\left(s+\frac1\tau+j\Omega\right)}{s\left(s+\frac1\tau\right)}
	\qquad\begin{array}{l}\text{alternate}\\[-0.4ex]\text{form}\end{array}\\[2ex] 
|\mathcal{H}(j\omega)|^2 &=\frac{\tau^4\omega^4+2(\tau^2-\Omega^2\tau^4)\omega^2
	+ (\Omega^4\tau^4+2\Omega^2\tau^2+1)}{%
	\tau^4\omega^4+\tau^2\omega^2}
	\label{eqn:le-pnoise-h2-detuned}
\end{align}
\end{statement}

\begin{proof}
Replace $B(s)$ given by Eq.\ \req{eqn:le-detun-laplace} in \req{eqn:le-pnoise-xfer}.
Use $\smash{\frac{1}{1-n/d}}=\smash{\frac{d}{d-n}}$.
\begin{align*}
\mathcal{H}(s)&=\frac{\tau\left(s+\frac1\tau-j\Omega\right)
		\left(s+\frac1\tau+j\Omega\right)}{%
	\tau\left(s+\frac1\tau-j\Omega\right)\left(s+\frac1\tau+j\Omega\right)
		- \left(s+\frac1\tau+\Omega^2\tau\right)}\\
&=\frac{(s\tau+1-j\Omega\tau)(s\tau+1+j\Omega\tau)}{%
	(s\tau+1-j\Omega\tau)(s\tau+1+j\Omega\tau)-(s\tau+1+\Omega^2\tau^2)}\\
&=\frac{(s\tau+1)^2+\Omega^2\tau^2}{%
	(s\tau+1)^2+\Omega^2\tau^2-(s\tau+1)-\Omega^2\tau^2}\\
&=\frac{(s\tau+1)^2+\Omega^2\tau^2}{s(s\tau+1)}
\end{align*}
The proof for $|\mathcal{H}(j\omega)|^2$ is omitted.
\end{proof}

\begin{remark}
Eq.\ \req{eqn:le-pnoise-h2-detuned} extends the Leeson formula to the case of the detuned oscillator.
\end{remark}

Figure~\ref{fig:le-lplane} shows $\mathcal{H}(s)$ on the Laplace plane.  When the oscillator is pulled out of $\omega_0$, the real zero at 
$s=-\omega_L$ splits into a pair of complex conjugate zeros at 
$s=-\omega_L\pm j\Omega$, leaving a real pole at
$s=-\omega_L$ in between.
The pole at $s=0$ in Fig.~\ref{fig:le-lplane} is an ideal integrator in the 
time domain.  This is the Leeson effect, which makes the phase diverge. 
At $\omega\gg\omega_L$, $\mathcal{H}(s)$ appears as a small cluster with equal 
number of poles and zeroes that null one another, for $\mathcal{H}(j\omega)$ is constant.
The resonator is a flywheel that blocks the phase fluctuations, for the oscillator is open loop.  Accordingly, the amplifier phase noise $\Psi(s)$ is repeated as the output.

\begin{figure}[t]
\namedgraphics{0.5}{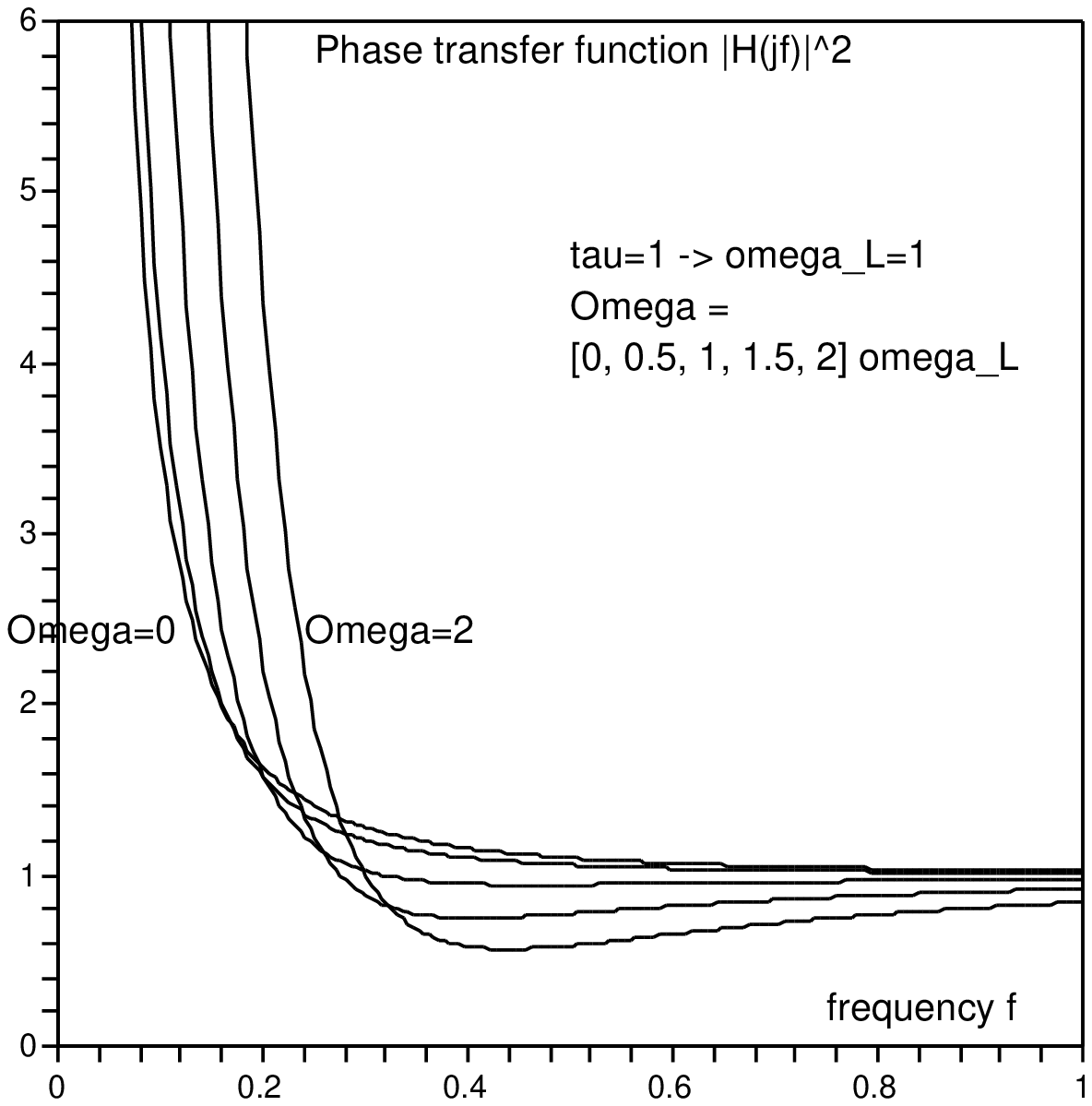}
\caption{Phase-noise transfer function $|\mathcal{H}(j\omega)|^2$  evaluated for $\tau=1$.}
\label{fig:le-calc-bigh}
\end{figure}

\begin{figure}[t]
\namedgraphics{0.5}{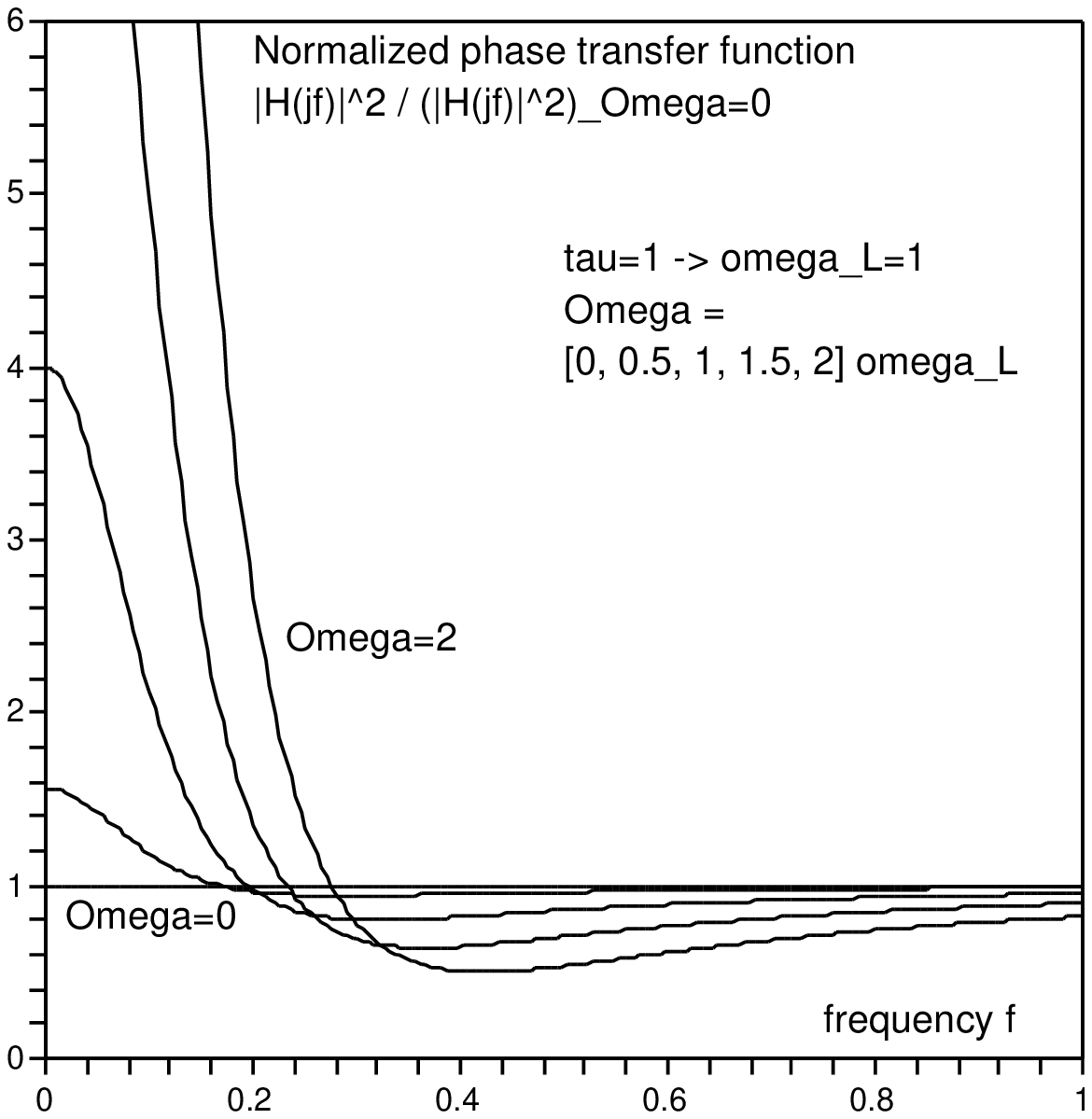}
\caption{Normalized phase noise transfer function $|\mathcal{H}(j\omega)|^2/|\mathcal{H}(j\omega)|^2_{\Omega=0}$  evaluated for $\tau=1$.}
\label{fig:le-calc-bighn}
\end{figure}

The plot of $|\mathcal{H}(j\omega)|^2$ (Fig.\,\ref{fig:le-calc-bigh}) reveals that the noise response is a function of $\Omega$, and increases as $\Omega$ increases.   
This is best shown in terms of the normalized noise transfer function $|\mathcal{H}_n(j\omega)|^2$, defined as
\begin{align}
|\mathcal{H}_n(j\omega)|^2&=\frac{\bigl|\mathcal{H}(j\omega)\bigr|^2}{
	\bigl|\mathcal{H}(j\omega)\bigr|^2_{\Omega=0}}
	\qquad\begin{array}{c}\text{def.\,of $|\mathcal{H}_n(j\omega)|^2$}\\[0.7ex]
		\text{Fig.\ \ref{fig:le-calc-bighn}~.}\end{array}
\end{align}
The denominator $\smash{|\mathcal{H}(j\omega)|^2_{\Omega=0}}$ is evaluated at $\Omega=0$.  $\smash{|\mathcal{H}_n(j\omega)|^2}$ has an horizontal asymptote at $\smash{|\mathcal{H}_n(j\omega)|^2}=1$, for can not be integrated to infinity.  
After translating the horizontal asymptote to zero, we define
\begin{align}
\mathcal{M}(\Omega)&=\int_0^\infty
	\left[|\mathcal{H}_n(j\omega)|^2-1\right]\:d\omega\\[1ex]
&=\int_0^\infty\left[
\frac{\bigl|\mathcal{H}(j\omega)\bigr|^2}{%
	\bigl|\mathcal{H}(j\omega)\bigr|^2_{\Omega=0}}-1
	\right]\:d\omega~,
\end{align}
$\mathcal{M}(\Omega)$ is a parameter that describes the noise increase due to the detuning $\Omega$, which converges with actual resonators.   
Interestingly, it holds that
\begin{align}
\mathcal{M}(\Omega)&=\frac\pi4\:\Omega^4\tau^3~.
\label{eqn:le-function-m}
\end{align}

\subsection{Pulling the oscillator frequency}\label{ssec:le-theory-tuning}
The oscillator can be pulled to a desired frequency in an interval around $\omega_0$, as shown in Section \ref{ssec:le-heur-tuning}.

If the oscillator is pulled by inserting a static phase in the loop, the oscillation frequency is not equal to the resonator natural frequency.  In this conditions it holds that $\Omega\neq0$.  Equation \req{eqn:le-function-m} indicate that the oscillator phase noise increases.  On the other hand, if $\omega_0$ is changed by making the resonator interact with an external reactance, there is no reason for $\Omega$ to be changed.  Thus, the phase noise does not increase. 
The following cases are of interest.

\paragraph{Quartz oscillator}
The quartz oscillator usually pulled by means of the external reactance, as in Figure \ref{fig:le-xtal-tuning}, for $\Omega$ is not affected.  Yet, the additional loss introduced by the external reactance may lower the merit factor, and in turn increase the resonator noise bandwidth.

\paragraph{Low-noise microwave oscillators}
In low-noise schemes, like the Galani oscillator \cite{galani84mtt}, it often happens that the resonator can not be tuned, for a phase shifter is used to pull the oscillation frequency.  There follows that $\Omega\neq0$, and additional noise is taken in.

\paragraph{Pound-stabilized oscillator}  In the Pound scheme \cite{pound46rsi}, the modulation and detection mechanism ensures that the resonator is tuned at the exact natural frequency $\omega_0$ by equating amplitude of the reflected sidebands.
In this oscillator it holds that  $\Omega=0$.

\section{Spectrum of frequency fluctuation and Allan variance}
If the phase noise spectrum inside the loop is $S_\psi(f)=b_0+b_{-1}f^{-1}+\ldots$, the oscillator fractional frequency spectrum is
\begin{align}
S_y(f)&=\frac{b_0}{\nu_0^2}f^2 + \frac{b_{-1}}{\nu_0^2}f
	+ \frac{b_0}{4Q^2} 
	+ \frac{b_{-1}}{4Q^2}\:\frac1f
	\label{eqn:le-oscill-sy}
\end{align}
and the Allan variance is
\begin{align}
\sigma_y^2(\tau)&=
	\left[\!\!\begin{array}{c}1/\tau^2\\[-1ex]\text{\small terms}\end{array}\!\!\right]
	+ \frac12\:\frac{1}{4Q^2}\:b_0\:\frac1\tau
	+ 2\ln(2)\:\frac{1}{4Q^2}\:b_{-1} 
	+ \ldots
	\label{eqn:le-oscill-avar}
\end{align} 
The proof extends this result to higher order terms, $b_{-2}f^{-2}$, etc. 

\begin{proof}[Proof (Spectrum)]
\begin{align*}
S_y(f)&=\frac{f^2}{\nu_0^2}S_\phi(f)
		\hspace{30ex}\text{derivative}\\
&=\frac{f^2}{\nu_0^2}\left[1+\frac{1}{f^2}\frac{\nu_0^2}{4Q^2}\right]S_\psi(f)
		\hspace{15ex}\text{Leeson}\\[1ex]
&=\left[\frac{f^2}{\nu_0^2}+\frac{1}{4Q^2}\right]S_\psi(f)\\[1ex]
&=\frac{b_0}{\nu_0^2}f^2 + \frac{b_{-1}}{\nu_0^2}f
	+ \frac{b_0}{\nu_0^2}\:\frac{\nu_0^2}{4Q^2} 
	+ \frac{b_{-1}}{\nu_0^2}\:\frac{\nu_0^2}{4Q^2}\:\frac1f\\[1ex] 
&=\frac{b_0}{\nu_0^2}f^2 + \frac{b_{-1}}{\nu_0^2}f
	+ \frac{b_0}{4Q^2} 
	+ \frac{b_{-1}}{4Q^2}\:\frac1f
\end{align*}
\end{proof}

\begin{proof}[Proof (Allan variance)]
Match Eq.\ \ref{eqn:le-oscill-sy} to $S_y(f)=\sum_ih_if^i$, and identify the terms
\begin{align*}
&h_0		=\frac{1}{2Q^2}\:b_0\\
&h_{-1}	=\frac{1}{2Q^2}\:b_{-1}~.
\end{align*}
Use Table \ref{tab:le-noise-conversion} (page \pageref{tab:le-noise-conversion}).
\end{proof}

\begin{remark}
For the reasons detailed in Section \ref{sec:le-ampli-noise}, terms steeper than $b_{-2}f^{-2}$ can not be present in the amplifier noise.  They can be added to for completeness, because $\psi(t)$ models all the phase fluctuations present in the loop.
\end{remark}

\begin{example} 
Calculate the Allan variance of a microwave DRO in which the resonator merit factor is $Q=2500$, and the amplifier noise is $S_\phi(f)=10^{-15}+10^{-11}/f$
(white $-150$ \unit{dBrad^2/Hz}, which results from $f=4$ dB and $P_0=-20$ dBm, and flicker $-150$ \unit{dBrad^2/Hz} at 1 Hz).  Using Eq.\ \req{eqn:le-oscill-avar}
\begin{align*}
\sigma_y^2(\tau)&=\frac{2{\times}10^{-23}}{\tau}+5.55{\times}10^{-19}&
\sigma_y(\tau)&=\frac{4.47{\times}10^{-12}}{\sqrt{\tau}}+7.45{\times}10^{-10}
\end{align*} 
\end{example}

\chapter{Noise in the delay-line oscillator}\label{chap:le-delayline}
The delay-line oscillator (Fig.\ \ref{fig:le-dly-loop}\,A) is an oscillator in which the feedback path is a mere delay $\tau$, for if the gain is $A=1$ oscillation can take place at any frequency $\omega_n$ multiple of $\frac{2\pi}{\tau}$ in a range.  A selector filter selects a given mode $\omega_m=\frac{2\pi}{\tau}m$ by introducing attenuation at all other frequencies but $\omega_m$.  The delay-line oscillator has the relevant property that the oscillation frequency can be switched between modes by tuning the selector, like a synthesizer.  Of course, the pitch is $\Delta\omega=\frac{2\pi}{\tau}$.

We analyze the phase noise mechanism in the delay-line oscillator following the same approach of previous chapter, based on the elementary theory of linear feedback systems.

\section{Basic delay-line oscillator}
\begin{figure}[t]
\namedgraphics{0.8}{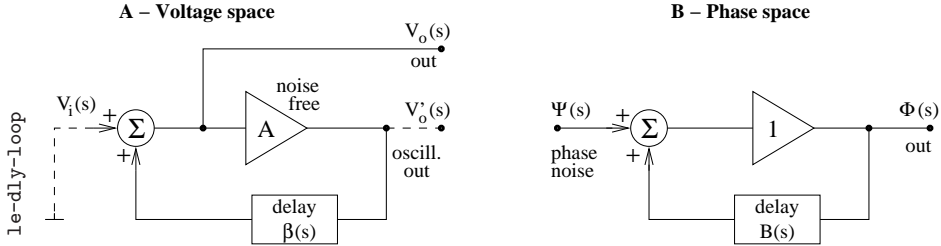}
\caption{Voltage-noise (A) and phase-noise (B) oscillator model.}
\label{fig:le-dly-loop}
\end{figure}

\begin{figure}[t]
\namedgraphics{0.79}{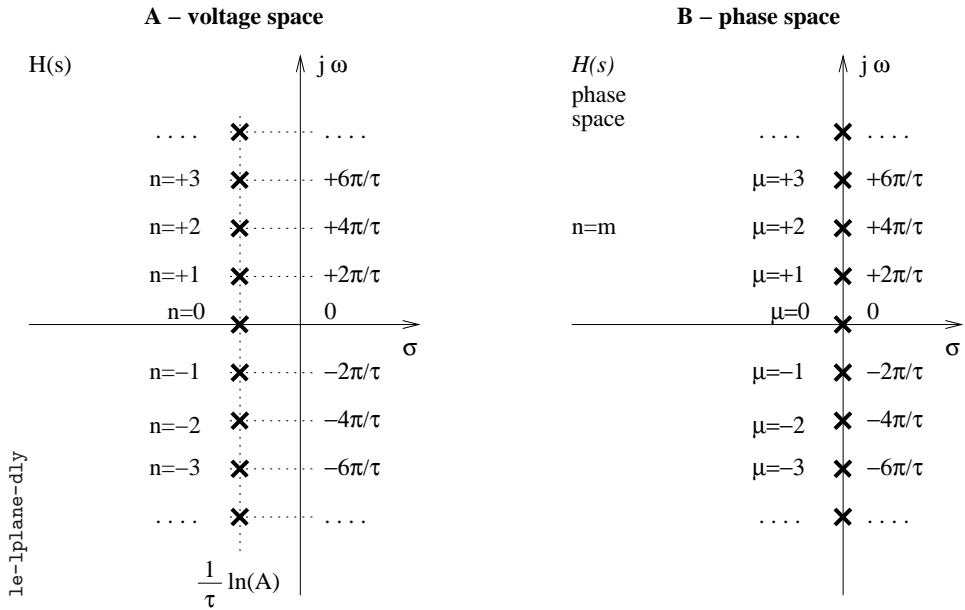}
\caption{Noise transfer function $H(s)$, and phase noise transfer function $\mathcal{H}(s)$ on the Laplace plane.}
\label{fig:le-lplane-dly}
\end{figure}

The delay-line oscillator can be represented as the feedback model of Fig.\ \ref{fig:le-dly-loop}\,A\@. This is similar to Fig.\ \ref{fig:le-loop} but for the resonator replaced with a delay line of delay $\tau$.  The signal $V_i(s)$ allows initial conditions and noise to be introduced in the system, as well as the driving signal if the oscillator is injection locked.  We first study the voltage-noise transfer function, defined as  $H(s)=\smash{\frac{V_o(s)}{V_i(s)}}$ [Eq.\ \req{eqn:le-def-hvolt}].  
In our system, it holds that $H(s)=\smash{\frac{1}{1-A\beta(s)}}$ [Eq.\ \req{eqn:le-closed-loop}].
In the Laplace transforms, a delay $\tau$ maps into $e^{-s\tau}$. Thus
\begin{align}
H(s)&=\frac{1}{1-Ae^{-s\tau}}\qquad\text{(Fig.\,\ref{fig:le-lplane-dly}\,A)}~.
\label{eqn:le-dly-hv}
\end{align}
$H(s)$ has an infinite array of poles that are solutions of $\mathcal{D}(s)=0$, with
\begin{align}
\mathcal{D}(s)&=1-Ae^{-s\tau}=1-Ae^{-\sigma}e^{-j\omega\tau}~.\nonumber
\end{align}
The poles are located at $s=s_n$,
\begin{align}
s_n&=\frac1\tau \ln(A) + j\frac{2\pi}{\tau}\,n\qquad\text{integer $n$}~.
\end{align}
Figure \ref{fig:le-lplane-dly}\,A shows the poles of $H(s)$ on the complex plane.
The poles are in the left half plane for $A<1$, move rightwards as $A$ increases, and reach the imaginary axis for $A=1$.  The derivative $\smash{\frac{ds_n}{dA}}=\smash{\frac{1}{\tau A}}$
indicates that a gain change $\delta A$ cause the poles to move by
$\delta s_n=\smash{\frac{1}{\tau}\,\frac{\delta A}{A}}$. Thus, the poles move horizontally by $\delta\sigma_n=\frac{1}{\tau}\:\frac{\delta A}{A}$ if $\delta A$ is real, and vertically by $j\delta\omega_n=\frac{1}{\tau}\:\frac{\delta A}{A}$ if $\delta A$ is imaginary. Yet for notation consistency, we will keep $A$ a real constant, and introduce a separate function $\rho e^{j\theta}$ (Sec.\ \ref{sec:le-mode-selection}).

A pair of imaginary conjugate poles represents a loss-free resonator that sustains a pure sinusoidal oscillation in the time domain if excited by an appropriate initial condition.  
Consequently, the delay-line oscillator can oscillate at any frequency $\omega_n=\smash{\frac{2\pi}{\tau}n}$. 
If more than one pole pair is excited, the oscillation is a linear superposition of sinusoids at the frequencies $\omega_n$.  This is the Fourier series expansion of an arbitrary periodic waveform.  The property of completeness of the Fourier series tells us that the delay-line oscillator can sustain any periodic waveform.  

Let us now choose one oscillation frequency, $\omega_m$.  This means that only
one pair of poles of $H(s)$ is excited by appropriate initial conditions, at the frequencies $\omega_{\pm m}=\smash{\pm\frac{2\pi}{\tau}}m$.  Having defined the carrier frequency $\omega_m$, we are now able to analyze the phase-noise transfer function, defined as $\mathcal{H}(s)=\smash{\frac{\Phi(s)}{\Psi(s)}}$ [Eq.\ \req{eqn:le-pnoise-xfer-def}].  
The phase noise model of the oscillator is shown in Fig.\ \ref{fig:le-dly-loop}\,B\@.  Once again, the amplifier gain is 1 because the amplifier just ``copies'' the input phase to the output.  Thus it holds that
\begin{align}
\mathcal{H}(s)&=\frac{1}{1-B(s)}~,\nonumber
\end{align}
which is the same as Eq.\ \req{eqn:le-pnoise-xfer}.  Yet, the phase transfer function is now $B(s)=e^{-s\tau}$, independent of the carrier frequency. A proof can be obtained following the approach of Section \ref{sec:le-reson-phi-space}. Nonetheless, we observe that the delay line has infinite bandwidth, for the group delay is the same at all frequencies, and equal to the line delay $\tau$.  Hence the phase, as well as any information-related parameter of the signal, take a time $\tau$ to propagate through the delay line.  Consequently it holds that   
\begin{align}
B(s)=e^{-s\tau}
\label{eqn:le-dlyl-hphase}~.
\end{align}
\begin{figure}[t]
\namedgraphics{0.5}{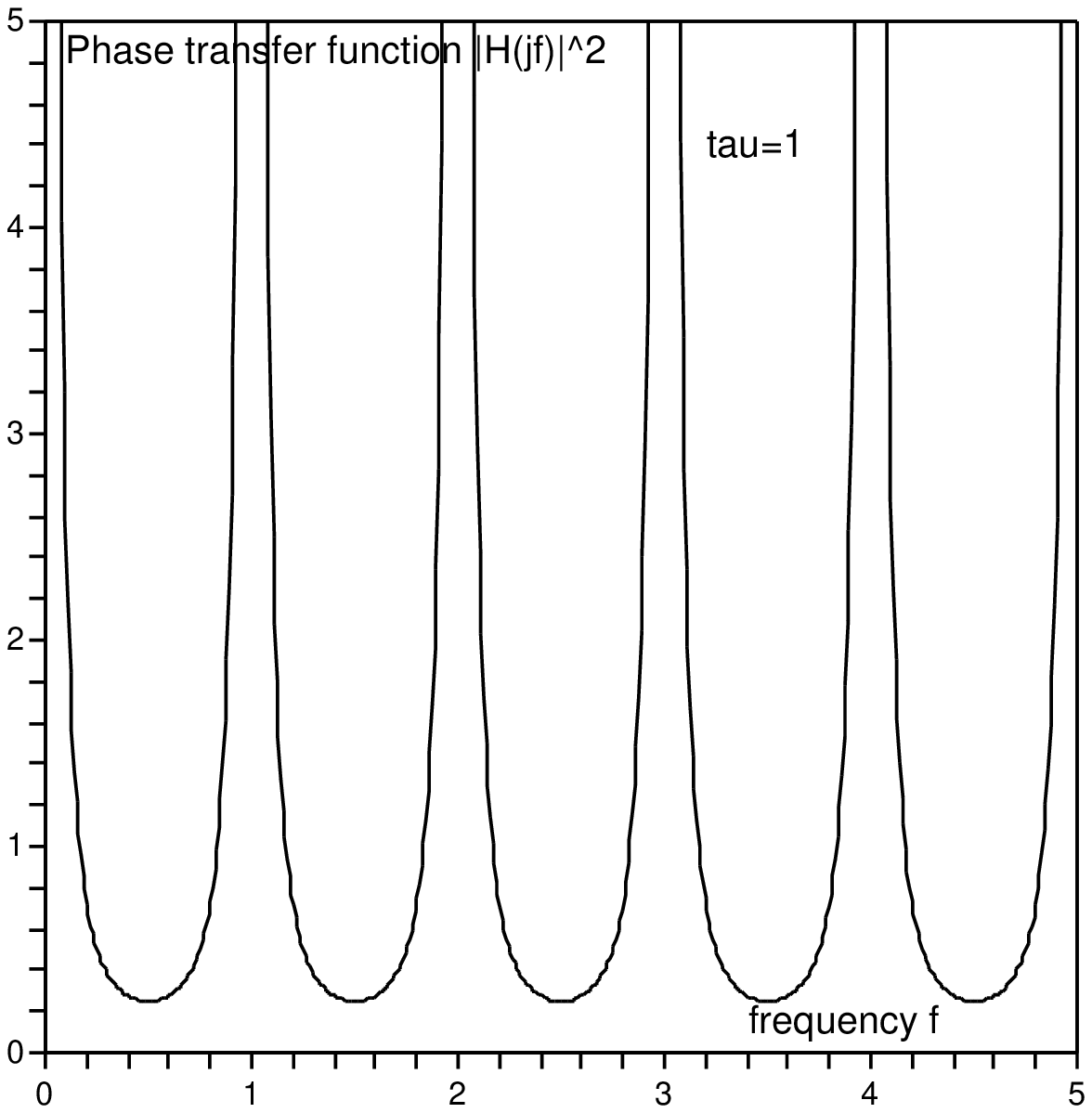}
\caption{Phase noise transfer function $|\mathcal{H}(jf)|^2$ evaluated for $\tau=1$.}
\label{fig:le-calc-hdly}
\end{figure}%
By substituting $B(s)=e^{-s\tau}$ in $\mathcal{H}(s)$, we get
\begin{align}
\mathcal{H}(s)&=\frac{1}{1-e^{-s\tau}}\qquad
\begin{array}{l}\text{delay-line}\\[-0.5ex]\text{oscillator}\end{array}~.
\end{align}
$\mathcal{H}(s)$ has an infinite array of poles that are solutions of
$1-e^{-s\tau}=0$.  Introducing the mode offset
\begin{align}
\mu &= n-m \qquad\text{mode offset}~,
\label{eqn:le-dly-mu-def} 
\end{align}
i.e., the integer frequency offset from the carrier, the poles (Fig.\ \ref{fig:le-lplane-dly}\,B) are located at  $s=s_\mu$,
\begin{align}
s_\mu&=j\frac{2\pi}{\tau}\,\mu\qquad\text{integer $\mu$}~.
\end{align}
In the spectral domain, the square modulus of the phase noise transfer function is 
\begin{align}
\left|\mathcal{H}(j\omega)\right|^2&=\frac{1}{2\left(1-\cos\omega\tau\right)}~.
\qquad\text{(Fig.\ \ref{fig:le-calc-hdly})}
\end{align}
Interestingly, it holds that $|\mathcal{H}(jf)|^2_\mathrm{min}=\frac14$.

\section{Mode selection}\label{sec:le-mode-selection}
A pure delay-line oscillator, with an infinite array of poles on the imaginary axis, is a nice exercise of mathematics with scarce practical usefulness.  A mode selector is necessary, which ensures that that the Barkhausen condition $A\beta=1$ is met at the privileged $m$ only, thus at the frequencies $\omega_{\pm m}=\smash{\pm\frac{2\pi}{\tau}m}$, and that all the other poles $s_n$, $n\neq\pm m$, are on the left-hand half plane ($\Re\{s_n\}<1$).  Thus we replace 
\begin{align}
\beta(s)&\rightarrow\beta_d(s)\beta_f(s) \label{eqn:le-beta-d-f}
\end{align}
in Eq.\ \req{eqn:le-dly-hv}, with 
\begin{align}
\beta_d(s)&=e^{-s\tau} &&\text{delay line}\\
\beta_f(s)&=\rho(s) e^{-j\theta(s)} &&\text{selector filter}
\end{align}
The function $\beta_f(s)$, as any network function, has the following properties
\begin{align}
&\beta_f(s)=\beta_f^*(s^*) &&\left\{\begin{array}{ll}
		\Re\{\beta_f(s)\}=\Re\{\beta_f(s^*)\}&\text{even in $\omega$}\\
		\Im\{\beta_f(s)\}=-\Im\{\beta_f(s^*)\}&\text{odd in $\omega$}
		\end{array}\right.\\[1ex]
&\rho(s)=\rho(s^*) &&\text{the modulus $\rho(s)$ is even in $\omega$}\\
&\theta(s)=-\theta(s^*) &&\text{the argument $\theta(s)$ is odd in $\omega$}
\end{align}
These conditions are necessary for $\beta_f(s)$ to be a real-coefficient function, or by extension a function whose series expansion has real coefficients of $s$, and ultimately for the inverse transform (i.e., the time-domain response) to be a real function of time.  

Introducing the selector filter, the transfer function [Eq.\ \req{eqn:le-dly-hv}] becomes 
\begin{align}
H(s)&=\frac{1}{1-A\rho(s) e^{j\theta(s)}e^{-s\tau}}~.
\end{align}
The poles of $H(s)$ are the solutions of $\mathcal{D}(s)=0$, with 
\begin{align}
\mathcal{D}(s)&=1-A\rho(s)e^{j\theta(s)}e^{-s\tau}
\end{align}
Splitting $\mathcal{D}(s)$ into real and imaginary parts, we get 
\begin{align}
&A\rho e^{-\sigma\tau}=1\nonumber\\
&e^{j\theta}e^{-j\omega\tau}=1 \qquad\Longrightarrow\qquad
	\theta-\omega\tau=0\quad\mod2\pi\nonumber
\end{align}
and finally
\begin{align}
\begin{array}{l} \sigma=\dfrac1\tau\ln(A\rho)\\[2ex]
	\omega=\dfrac\theta\tau \quad \mod\dfrac{2\pi}{\tau}\end{array}
	\qquad\text{poles of $H(s)$}~.
\end{align}
As a consequence, the modulus $\rho$ affects only the real part $\sigma$ of the pole, thus the damping; the argument $\theta$ affects only the imaginary part, thus the frequency.  The poles of $H(s)$ are located at $s=s_n$
\begin{align}
s_n=\frac1\tau\ln(A\rho)+j\frac{2\pi}{\tau}\:n+j\frac\theta\tau
\qquad\text{poles of $H(s)$}
\label{eqn:le-poles-of-bigh}
\end{align}
The above Eq.\ \req{eqn:le-poles-of-bigh} will be used to analyze some relevant  types of selector filter. 

\begin{figure}[t]
\namedgraphics{0.8}{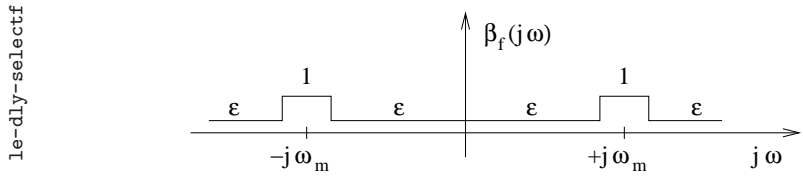}
\caption{Amplitude-only selector filter.}
\label{fig:le-dly-selectf}
\end{figure}

\subsection{Amplitude-only filter}
This selector filter is defined as
\begin{align}
\beta_f(s)&=\begin{cases}
	1	& \begin{array}{l}\omega\approx\pm\omega_m\\[-0.5ex]
			(n=\pm m)\end{array}\\[1ex]
	\epsilon\ll1& \begin{array}{l}\text{elsewhere}\\[-0.5ex]
			(n\neq\pm m)\end{array}	
	\end{cases}
\qquad\text{(Fig.\ \ref{fig:le-dly-selectf})~.}
\end{align}
Such filter attenuates by a factor $\epsilon\ll1$ all the signals but those in the vicinity of $\pm\omega_m$.  Needless to say, this filter is an abstraction, which can not be implemented with real-world devices.

\begin{figure}[t]
\namedgraphics{0.8}{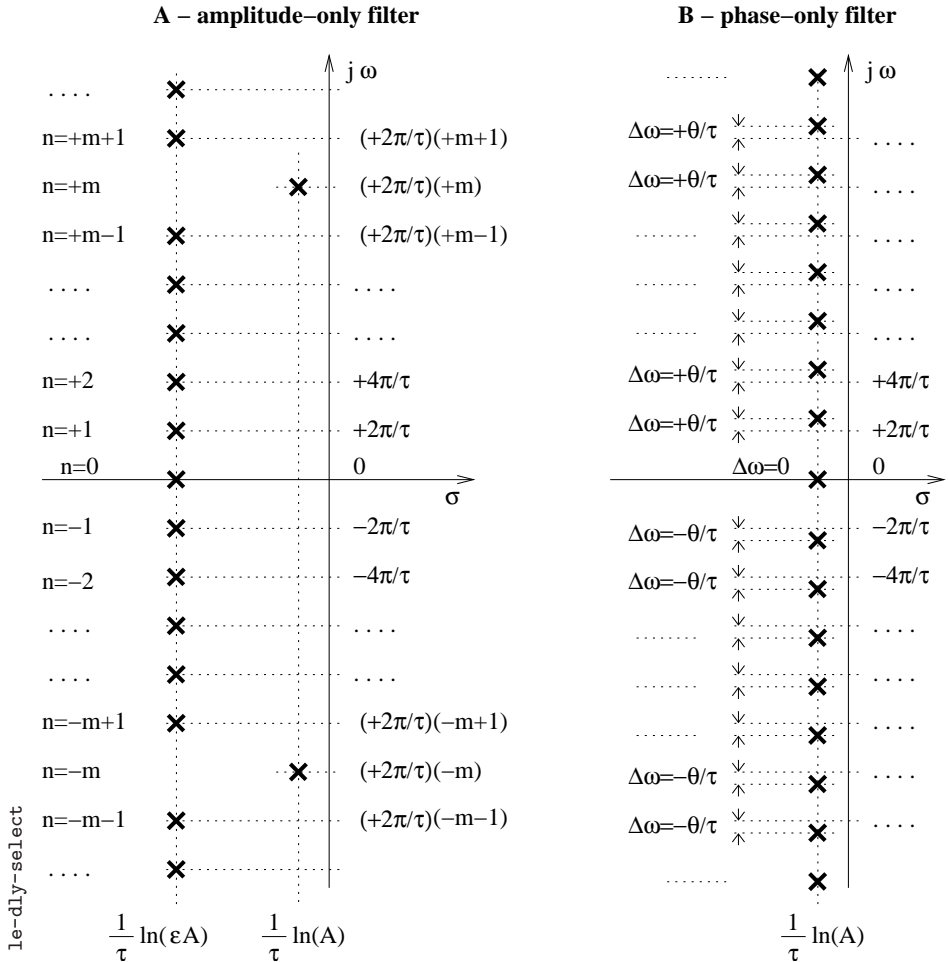}
\caption{Transfer function $H(s)$ of the delay-line oscillator with a selector filter $\beta_f(s)$ inserted in the oscillator loop.}
\label{fig:le-dly-select}
\end{figure}
The poles of $H(s)$ are the solution of $\mathcal{D}(s)=0$, that is, 
\begin{align}
&1-Ae^{-s\tau}=0 
	&\begin{array}{c}\omega\approx\pm\omega_m\\[-0.5ex]
		(n=\pm m)\end{array}\\[1ex]
&1-\epsilon Ae^{-s\tau}=0
	&\begin{array}{c}\text{elsewhere}\\[-0.5ex]
		(n\neq\pm m)\end{array}
\end{align}
Hence the poles (Fig.\ \ref{fig:le-dly-select}\,A) are located at $s=s_n$,
\begin{align}
s_n&=\begin{cases}
	\frac{1}{\tau}\ln(A)+j\frac{2\pi}{\tau}n	
			&n=\pm m~~(\omega=\pm\omega_m)\\[1ex]
	\frac{1}{\tau}\ln(\epsilon A)+j\frac{2\pi}{\tau}n	&n\neq\pm m
	\end{cases}
\end{align}

\subsection{Phase-only filter}
The phase-only filter is defined by
\begin{align}
\beta_f(s)&=e^{j\,\mathrm{sgn}(\omega)\,\theta}~.
\end{align}
This filter shifts the signals by $\theta$ for $\omega>0$, and by $-\theta$ for $\omega<0$.  The signum function $\mathrm{sgn}(\omega)$ is necessary for the condition $\beta_f(s)=\beta_f^*(s^*)$ to be satisfied.  Once again, this filter is an abstraction.

By replacing the above $\beta_f(s)$ in $H(s)$, we get
\begin{align}
H(s)&=\frac{1}{1-Ae^{j\,\mathrm{sgn}(\omega)\,\theta}e^{-s\tau}}~.
\end{align}
The poles are the solution of $\mathcal{D}=0$, where
\begin{align}
\mathcal{D}(s)&=1-Ae^{j\,\mathrm{sgn}(\omega)\,\theta} e^{-s\tau}\nonumber
\end{align}
By splitting the modulus and the argument of $\mathcal{D}$, we find
\begin{align}
&Ae^{-\sigma\tau}=1 \nonumber \\
&\mathrm{sgn}(\omega)\,\theta - \omega\tau = 0 \mod 2\pi~. \nonumber
\end{align}
There follows that the poles (Fig.\ \ref{fig:le-dly-select}\,B) are located at $s=s_n$
\begin{align}
s_n=\frac{1}{\tau}\ln(A)+j\left[\frac{2\pi}{\tau}n+\frac\theta\tau\mathrm{sgn}(n)\right]
\qquad\text{integer $n$}
\end{align}
In conclusion, the real part of the poles is not affected, for the gain condition is still $A=1$. On the other hand, the phase $\theta$ results in a frequency shift
\begin{align}
\Delta\omega=\frac\theta\tau\quad\text{for $\omega>0$}
\qquad\text{and}\qquad
\Delta\omega=-\frac\theta\tau\quad\text{for $\omega>0$}~.
\end{align}

\section{The use of a resonator as the selector filter}\label{sec:le-dly-resonator}
The practical reason to  choose a resonator as the selector filter is that it is simpler and easier to tune, as compared to other types of filter.  Let us assume that the resonator is tuned at the exact frequency $\omega_m=\frac{2\pi}{\tau}m$ of our interest, among the permitted frequencies of the delay-line oscillator. Thus, the feedback function is 
\begin{align}
\beta(s)&=\beta_d(s)\beta_f(s)
	&&[Eq.\ \req{eqn:le-beta-d-f}]~, 
\intertext{with} 
\beta_d(s) &=e^{-s\tau}
	&&\begin{tabular}{c}\text{delay line}\\
	\text{[Eq.\ \req{eqn:le-dlyl-hphase}]}\end{tabular}\\
\beta_f(s)&=\frac{\omega_m}{Q}\frac{s}{s^2+\frac{\omega_m}{Q}s+\omega_m^2}
	&&\begin{tabular}{c}\text{filter}\\\text{[Eq. \req{eqn:le-app-def-h}]}\end{tabular}
\end{align}
It is important to understand the role of delay line and of the filter.  As a consequence of the Leeson effect, stable oscillation frequency requires high slope $\frac{d}{d\omega}\arg\beta(j\omega)$, i.e., long group delay. Nonetheless, the oscillator tracks the natural frequency of the feedback network, which is sensitive to the environment parameters, chiefly the ambient temperature. When $\beta_d$ and $\beta_f$ are cascaded, the frequency fluctuations are weighted by the phase slope $\frac{d}{d\omega}\arg\beta(j\omega)$ because the Barkhausen condition requires $\arg\beta(j\omega)=0$.
The appropriate choice for the delay-line oscillator is therefore
\begin{align}
\left.\frac{d}{d\omega}\arg\beta_d(j\omega)\right|_{\omega=\omega_m} \gg \;\;\left.\frac{d}{d\omega}\arg\beta_f(j\omega)\right|_{\omega=\omega_m}~,
\label{eqn:le-delay-gg-q}
\end{align}
which means that the oscillator frequency is stabilized to the delay line, while the resonator serves only as the mode selector.  The main technical reason to choose a delay-line oscillator is the availability of a delay line that exhibit much higher stability than the resonator.  Accordingly, the resonator merit factor must be sufficiently low to prevent the resonator from impairing the oscillator stability.  
Another reason for the condition \req{eqn:le-delay-gg-q} is that it makes easy to switch between the delay-line modes, allowing the resonator to be tuned in a narrow range around the exact frequency.

By replacing $\beta(s)=\beta_d(s)\beta_f(s)$ in $H(s)=\smash{\frac{1}{1-A\beta(s)}}$, under the assumption that the amplifier gain is $A=1$, we find the oscillator transfer function
\begin{align}
H(s)&=\frac{s^2+\frac{\omega_m}{Q}s+\omega_m^2}{%
	s^2+\frac{\omega_m}{Q}s+\omega_m^2-\frac{\omega_m}{Q}se^{-s\tau}}~.
	\label{eqn:le-hrf-rational}
\end{align}
This function has a pair of complex conjugate zeros at $s=s_z$ and $s=s^*_z$
\begin{align}
s_z, s_z^*&=-\frac{\omega_m}{2Q} \pm j\omega_m~,\nonumber
\end{align}
and a series of complex conjugate poles.
As a consequence of the condition \req{eqn:le-delay-gg-q}, the resonator bandwidth $\smash{\frac{\omega_m}{Q}}$ is large as compared to the mode pitch $\smash{\frac{2\pi}{\tau}}$.  This means that in a frequency range $\mathcal{F}$ around $\omega_m$, and also around $-\omega_m$, the resonator dissonance $\chi$ is such that $Q\chi\ll1$.  The range $\mathcal{F}$ contains a few modes $\omega_n$ around $\omega_m$.  In $\mathcal{F}$ the resonator function $\beta_f(j\omega)$ is about constant and close to 1.  Thee follows that in $\mathcal{F}$
\begin{enumerate}
\item The poles of $H(s)$ are chiefly determined by the oscillation of $\beta_d(s)$; they are close to $j\frac{2\pi}{\tau}n$, which is the result already obtained without the selector filter.
\item An approximate solution for the poles is found by replacing $\beta_f(s)$ with $\beta_f(j\omega)=\rho(\omega)\smash{e^{j\theta(\omega)}}$
\begin{align}
\rho(\omega)&=\frac{1}{\sqrt{1+Q^2\chi^2}}
	&&\text{modulus}\nonumber\\
\theta(\omega)&=-\arctan\left(Q\chi\right)
	&&\text{phase}\nonumber\\
\chi&=\Bigl(\frac{\omega}{\omega_m}-\frac{\omega_m}{\omega}\Bigr)^2
	&&\text{dissonance}\nonumber
\end{align}
\item From $Q\chi\ll1$ it follows that $\chi\simeq2\frac{\omega-\omega_m}{\omega_m}$.  Consequently $\rho(\omega)$ and $\theta(\omega)$ can be further approximated as
\begin{align}
\rho(\omega)&=1-2Q^2\Bigl(\frac{\omega-\omega_m}{\omega_m}\Bigr)^2
	\nonumber\\
\theta(\omega)&=-2Q\,\frac{\omega-\omega_m}{\omega_m}\nonumber
\end{align}
\end{enumerate}
Thus, the poles of $H(s)$ are found by inserting the above $\rho(\omega)|_{\omega=\omega_n}$ and $\theta(\omega)|_{\omega=\omega_n}$ in Eq.\ \req{eqn:le-poles-of-bigh}, with $A=1$.  The poles are at $s=s_n=\sigma_n+j\omega_n$
\begin{align}
\sigma_n
&=\frac1\tau\ln\left[1-2Q^2\Bigl(\frac{\omega_n-\omega_m}{\omega_m}\Bigr)^2\right]
	\simeq-\frac{2Q^2}{\tau}\Bigl(\frac{\omega_n-\omega_m}{\omega_m}\Bigr)^2\\
\omega_n&=\frac{2\pi}{\tau}\:n+\frac{\theta}{\tau}
	\simeq\frac{2\pi}{\tau}\:n-\frac{2Q}{\tau}\:
	\frac{\omega_n-\omega_m}{\omega_m}~,
\end{align}
hence
\begin{align}
s_n&=-\frac{2Q^2}{\tau}\Bigl(\frac{\omega_n-\omega_m}{\omega_m}\Bigr)^2
	+j\frac{2\pi}{\tau}\:n-j\frac{2Q}{\tau}\:\frac{\omega_n-\omega_m}{\omega_m}~,
\end{align}
Additionally, the pols can be given in terms of the mode offset $\mu=n-m$ [Eq.\ \req{eqn:le-dly-mu-def}] through the simple property that 
$\frac{\omega_n-\omega_m}{\omega_m} = \frac{\mu}{m}$.  Hence
\begin{align}
s_n&=-\frac{2Q^2}{\tau}\left(\frac{\mu}{m}\right)^2
	+j\frac{2\pi}{\tau}\:n-j\frac{2Q}{\tau}\:\frac{\mu}{m}~.
\end{align}
\begin{figure}[t]
\namedgraphics{0.8}{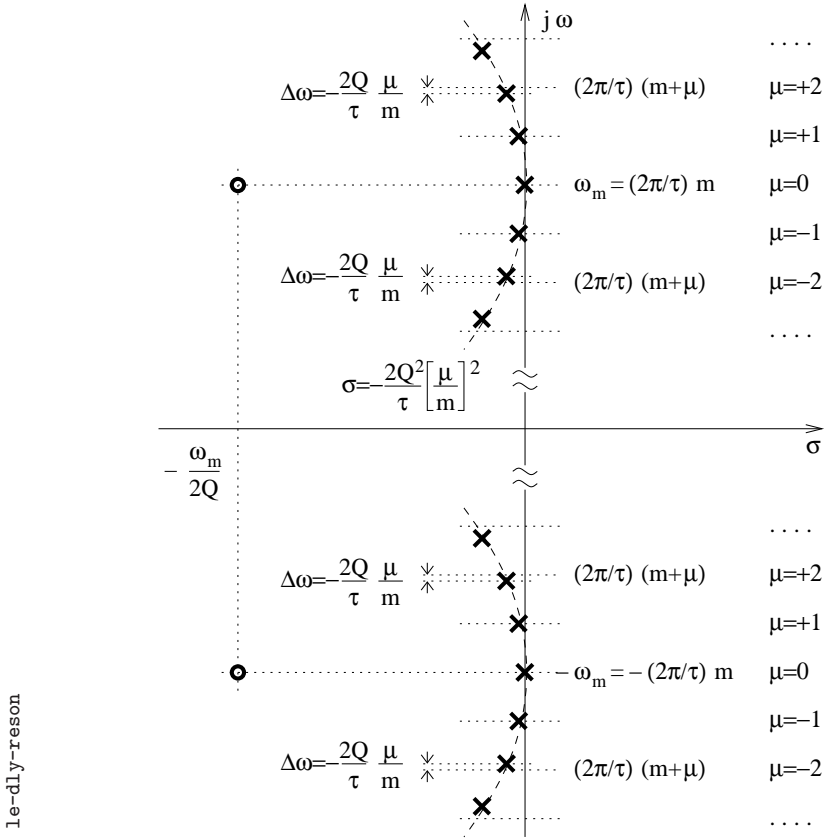}
\caption{Transfer function $H(s)$ of the delay-line oscillator with a resonator as the selector filter.}
\label{fig:le-dly-reson}
\end{figure}%
Figure \ref{fig:le-dly-reson} shows the poles of $H(s)$, located on a horizontal parabola centered at $\omega_m$.  The pole frequencies are shifted by $-\frac{2Q}{\tau}\frac{\mu}{m}$ with respect to the exact mode $\frac{2\pi}{\tau}n$ of the delay line.  This frequency shift results from the off-resonance phase of the filter.  The negative-frequency part of the Laplace plane follows by symmetry.

Interestingly, every pair of complex conjugate poles of $H(s)$ can be seen as a separate resonator [Fig.\ \ref{fig:le-resonator}, and Eq.\ \req{eqn:le-app-def-h}], whose merit factor is enhanced by the positive feedback.  The pole pair selected by the resonator, denoted with the subscript $m$, is onto the imaginary axis, for the equivalent merit factor is $Q_m=\infty$. The other pole pairs, denoted with the subscript $n$, yield a merit factor
\begin{align}
Q_n&=\frac{\tau\omega_n}{4Q^2}\Bigl(\frac{m}{\mu}\Bigr)^2
\intertext{or equivalently}
Q_n&=\frac{\pi n}{2Q^2}\Bigl(\frac{m}{\mu}\Bigr)^2~.\nonumber
\end{align}
In the presence of noise, these poles are excited by random signal.  Thus, the radio-frequency spectrum shows a series of sharp lines at frequency $\omega_\mu$ off the carrier frequency $\omega_m$, on both sides because $\mu$ takes positive and negative values.  These lines are easily mistaken for competitor oscillation modes, which they can not be\footnote{Mode competition and multimode oscillation do exist, yet under different hypotheses.  This behavior is commonly found in lasers.}.  The relevant differences are that random excitation causes incoherent oscillation, and that such oscillation is damped.  With the $\mu=0$ pair, coherent and incoherent oscillation coexist. This is the Leeson effect. 

With the operating parameters of actual or conceivable oscillators, the equivalent merit factor $Q_n$ is so high that the noise transfer function $|H(j\omega)|^2$ [Eq.\ \ref{eqn:le-hrf-rational}] can hardly be plotted.  Examples are provided in Section 
\ref{sec:le-dly-osc-examp}.

\begin{remark}\textbf{(Multi-pole filters)}
In the domain of telecommunications, multi-pole passband filters are widely used, which exhibit nearly flat response in a bandwidth. The experimentalist may be inclined to use such filters as the mode selector because of their easy commercial availability, and out of the common belief that the flat response is the most desirable one for a bandpass filter.
But the maximally flat filter is the \emph{worst} choice for an oscillator.  In fact, with a flat filters the $\mu\neq0$ poles are maximally close to the imaginary axis.  There follow that
\begin{enumerate}
\item selecting the desired oscillation frequency is maximally difficult,
\item mode jump can occur,
\item close-in noise peaks are maximally high.
\end{enumerate} 
\end{remark}

\section{Phase response}\label{sec:le-dly-phase-response}
Let the delay-line oscillator oscillate at the frequency $\omega_m=\smash{\frac{2\pi}{\tau}}$, thanks to a resonator in the feedback loop tuned at the exact frequency $\omega_m$. Under this condition, we analyze the phase-noise transfer function $\mathcal{H}(s)=\smash{\frac{\Psi(s)}{\Phi(s}}$ [Eq.\ \req{eqn:le-pnoise-xfer-def}], i.e., 
\begin{align}
\mathcal{H}(s)&=\frac{1}{1-B(s)}~.
\end{align}
Introducing the resonator as the selector filter, the feedback function is
\begin{align}
B(s)&=B_d(s)B_f(s)
\intertext{with}
B_d(s)&=e^{-s\tau_d}		&&\text{delay}\\
B_f(s)&=\frac{1}{1+s\tau_f}	&&\text{filter}
\end{align}
The subscript `$d$' and `$f$' are introduced in order to prevent confusion between the delay of the line ($\tau_d$) and the relaxation time $\tau_f=\smash{\frac{2Q}{\omega_m}}$ [Eq.\ \req{eqn:le-reson-tau}] of the resonator. 

\begin{figure}[t]
\namedgraphics{0.8}{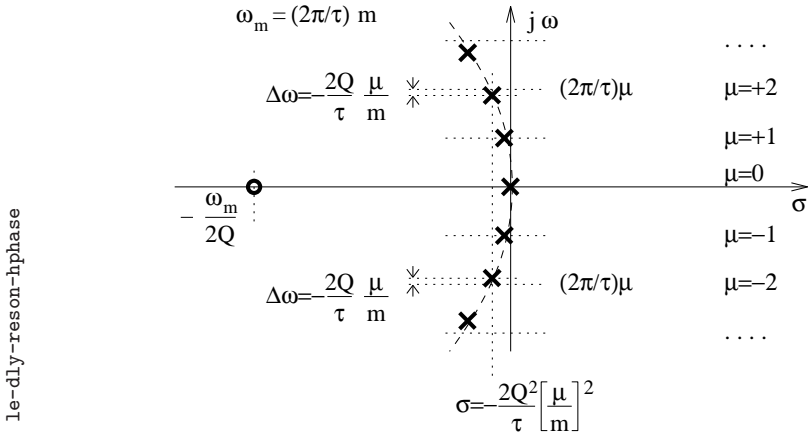}
\caption{Phase-noise transfer function $\mathcal{H}(s)$ of the delay-line oscillator with a resonator as the selector filter.}
\label{fig:le-dly-reson-hphase}
\end{figure}%
Expanding the phase-noise transfer function, we find 
\begin{align}
\mathcal{H}(s)&=\frac{1+s\tau_f}{1+s\tau_f-e^{-s\tau_d}}
\label{eqn:le-dly-hphase-exp}
\end{align}
$\mathcal{H}(s)$ has a real zero at $s=\smash{-\frac{1}{\tau_f}}=\smash{-\frac{\omega_m}{2Q}}$, a pole at $s=0$, and a series of complex conjugate poles in the left half-plane. The location of these poles is best found by searching for the zeros of $1-B(s)$, without using the simplified form \req{eqn:le-dly-hphase-exp}.   
As a consequence of the design choice $\tau_d\gg\tau_f=\frac{2Q}{\omega_m}$, the poles are expected to be close to the position already found in the absence of the selector, that is, $\smash{j\frac{2\pi}{\tau_d}\mu}$.  In other words, in a region around the origin that contains a few pole pairs, the selector has a small effect on the modulus and on the phase of $B(s)$.  This is exactly the same situation already found searching for the poles of $H(s)$, transposed from $\pm\omega_m$ to the origin.

In order to the find poles, solution of $1-B(s)=0$, we first focus on the low-pass phase filter $B_f(s)$.  The poles are expected to be close to the imaginary axis. Thus we can approximate $B_f(s)\simeq B_f(j\omega)$, and write in polar form
\begin{align}
B_f(s)&=\rho(\omega)e^{j\theta(\omega)}		\nonumber\\
\rho(\omega)&=\sqrt{\frac{1}{1+\omega^2\tau^2_f}}	\nonumber\\
\theta(\omega)&=-\arctan\left(\omega\tau_f\right)~.	\nonumber
\end{align}
Then we replace the low-pass time constant $\tau_f$ as $\tau_f=\smash{\frac{2Q}{\omega_m}}$, and we expand for $\omega\tau_f\ll1$ 
\begin{align}
\rho(\omega)&=1-2Q^2\Bigl(\frac{\omega}{\omega_m}\Bigr)^2		\nonumber\\
\theta(\omega)&=-2Q\,\frac{\omega}{\omega_m}			\nonumber
\end{align}
The equation $1-B(s)=0$ can be rewritten as 
\begin{align}
\rho(\omega)e^{j\theta(\omega)}\;e^{-\sigma\tau_d}e^{-j\omega\tau_d}=1~,\nonumber
\end{align}
and further split into modulus and phase.  The poles are the solutions of
\begin{align}
&\rho(\omega)e^{-\sigma\tau_d}=1		&&\text{modulus}	\nonumber\\
&\theta(\omega)-\omega\tau_d=0 \mod 2\pi	&&\text{phase}~.	\nonumber
\end{align}
The modulus condition yields
\begin{align}
&e^{-\sigma\tau_d}\Bigl(1-2Q^2\frac{\omega^2}{\omega_m^2}\Bigr)=1	\nonumber\\
&e^{\sigma\tau_d}=1-2Q^2\frac{\omega^2}{\omega_m^2}		\nonumber\\
&\sigma\tau_d=\ln\Bigl(1-2Q^2\frac{\omega^2}{\omega_m^2}\Bigr)	\nonumber\\
&\sigma=-\frac{2Q^2}{\tau_d}\:\frac{\omega^2}{\omega_m^2}
	\qquad\qquad\text{for $2Q^2\frac{\omega^2}{\omega_m^2}\ll1$}~.
	\label{eqn:le-dly-lp-sigma}
\end{align}
The phase condition yields
\begin{align}
&\omega\tau_d + 2Q\frac{\omega}{\omega_m} = 0 \mod 2\pi	\nonumber\\
&\omega = \frac{2\pi}{\tau_d}\mu - \frac{2Q}{\tau_d}\:\frac{\omega}{\omega_m}
	\qquad\text{integer $\mu$}~.
	\label{eqn:le-dly-lp-omega}
\end{align}
The poles are located at $s=s_\mu=\sigma_\mu+j\omega_\mu$, which results from Eq.\ \req{eqn:le-dly-lp-sigma} and \req{eqn:le-dly-lp-omega}
\begin{align}
s_\mu&=-\frac{2Q^2}{\tau_d}\:\frac{\omega^2_\mu}{\omega_m^2}
	+j\frac{2\pi}{\tau_d}\mu - j\frac{2Q}{\tau_d}\:\frac{\omega_\mu}{\omega_m}
\end{align}
Additionally, by replacing $\smash{\frac{\omega_\mu}{\omega_m}}=\mu$ we find
\begin{align}
s_\mu&=-\frac{2Q^2}{\tau_d}\:\Bigl(\frac{\mu}{m}\Bigr)^2 
	+j\frac{2\pi}{\tau_d}\:\mu - \frac{2Q}{\tau_d}\:\frac{\mu}{m}~.
\end{align}
Figure \ref{fig:le-dly-reson-hphase} shows the phase-noise transfer function
$\mathcal{H}(s)$ on the Laplace plane.  The pole in the origin represents a pure integrator in the time domain, which causes the Leeson effect. The other poles are on a horizontal parabola centered in the origin, at a small negative distance $\sigma\propto-\mu^2$ from the imaginary axis.  
It is easily seen by comparing Fig.\ \ref{fig:le-dly-reson-hphase} to Fig.\ \ref{fig:le-resonator} that each complex conjugate pair is equivalent to a resonator of merit factor
\begin{align}
Q_\mu&=\frac{\omega_\mu\tau_d}{4Q^2}\:\Bigl(\frac{m}{\mu}\bigr)^2~,
\end{align}
which can be rewritten as 
\begin{align}
Q_\mu&=\frac{\pi}{2Q^2}\:\frac{m^2}{\mu}
\end{align}
because $\omega_\mu=\smash{\frac{2\pi}{\tau_d}}\mu$.

\begin{figure}[t]
\namedgraphics{0.5}{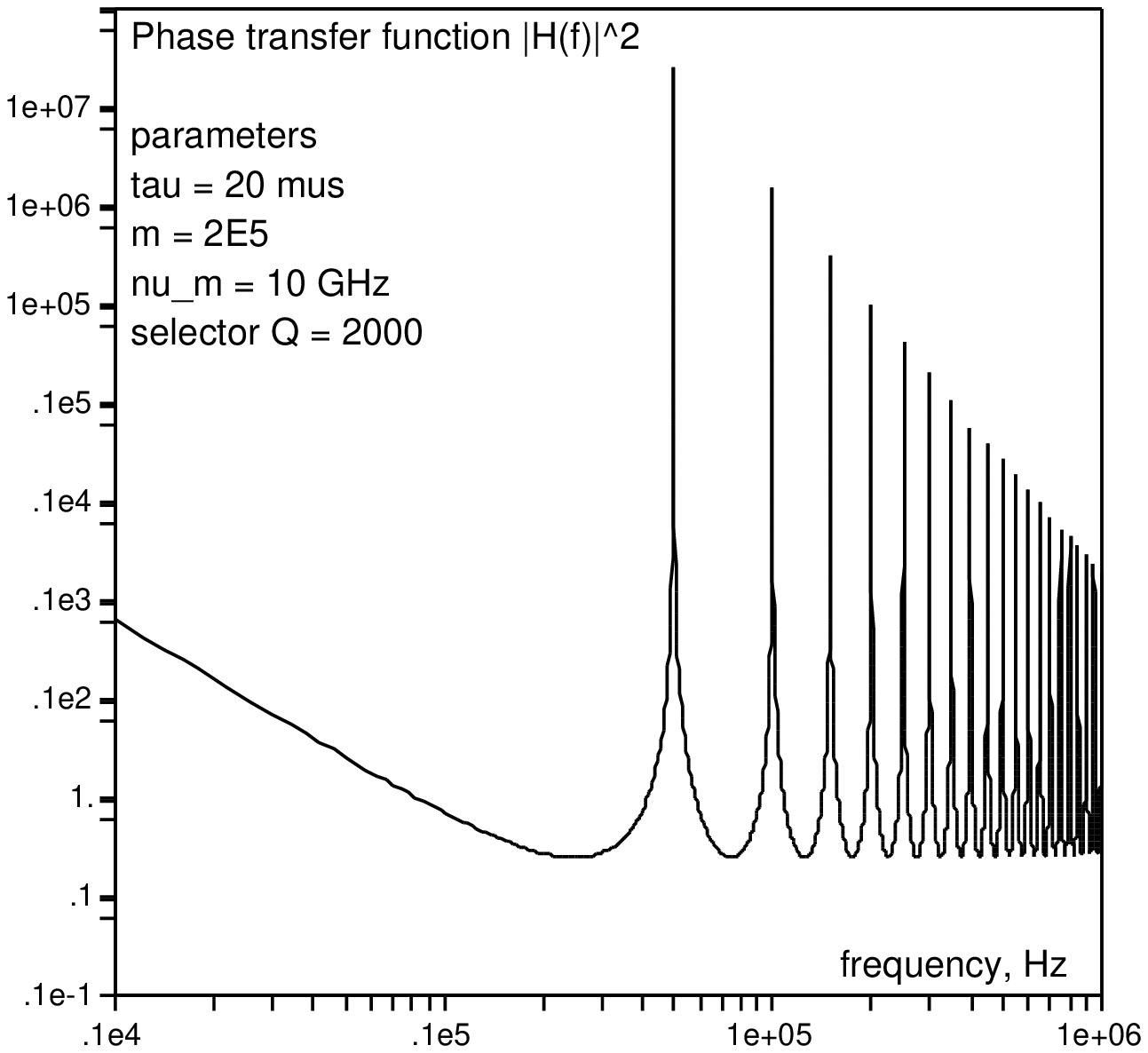}
\caption{Phase transfer function $|\mathcal{H}(jf)|^2$ evaluated for $\tau_d=20$ $\mu$s, $m=2{\times}10^5$ (thus $\nu_m=10$ GHz), and $Q=2000$ (data of Example \ref{ssec:le-photon-osc-examp}).}
\label{fig:le-calc-dly-hphase}
\end{figure}

From Eq.\ \req{eqn:le-dly-hphase-exp}, after some tedious algebra it is proved that
\begin{align}
\left|\mathcal{H}(j\omega)\right|^2&=
\frac{1+\omega^2\tau_f}{2-2\cos(\omega\tau_d)+\omega^2\tau_f^2+
	2\omega\tau_f\sin(\omega\tau_d)}
	\qquad\text{(Fig.\ \ref{fig:le-calc-dly-hphase})}.
\label{eqn:le-dly-h2phase-exp}
\end{align}
$|\mathcal{H}(j\omega)|^2$ shows a series of sharp peaks at $\omega=\smash{\frac{2\pi}{\tau_d}}\mu$, or $f=\smash{\frac{1}{\tau_d}}\mu$, integer $\mu$, around which phase noise taken in is enhanced. The peaks derive from the poles of $\mathcal{H}(s)^2$ close to the imaginary axis, but not on the axis, for they are not the signature of competing oscillation modes.
The minima of $|\mathcal{H}(j\omega)|^2$ are located at
$\omega\simeq\frac{2\pi}{\tau_d}\mu$, or 
$f\simeq\frac{1}{\tau_d}\mu$, $\mu=\frac12,\frac32,\frac52,\ldots$.
Interestingly, at these minima it holds that $|\mathcal{H}(j\omega)|^2\simeq\frac14$.
Finally, at low frequencies ($\omega\ll\frac{2\pi}{\tau_d}\ll\frac{2\pi}{\tau_f}$) 
$|\mathcal{H}(j\omega)|^2$ is approximated as
\begin{align}
\left|\mathcal{H}(j\omega)\right|^2
&\simeq\frac{1}{\tau^2_d}\:\frac{1}{\omega^2}+\frac{1}{12}~.
\end{align}

\section{Close-in noise spectra and Allan variance}\label{sec:le-dly-noise}
The phase noise spectrum is 
\begin{align}
S_\phi(f)&=\left|\mathcal{H}(jf)\right|^2 S_\psi(f)
\end{align}
At low frequencies, it holds that
\begin{align}
S_\phi(f)&\simeq\frac{1}{4\pi^2\tau^2_d}\:\frac{1}{f^2}\:S_\psi(f)
	&&\omega\ll\frac{2\pi}{\tau_d}\\
&=\frac{\nu_m^2}{4\pi^2m^2}\:\frac{1}{f^2}\:S_\psi(f)
	&&\text{use~}\nu_m=\tfrac{m}{\tau_d}\\
&=\frac{\nu_m^2b_0}{4\pi^2m^2}\:\frac{1}{f^2} +
	\frac{\nu_m^2b_{-1}}{4\pi^2m^2}\:\frac{1}{f^3}
	&&\!\!\begin{array}{c}\text{amplifier}\\[-1ex]\text{noise}\end{array}
\end{align}
The spectrum of the fractional frequency fluctuation $y$ is found using 
$S_y(f)=\smash{\frac{f^2}{\nu_m^2}}S_\phi(f)$
\begin{align}
S_y(f)&=\frac{1}{4\pi^2m^2}\:S_\psi(f)
\end{align}
By replacing $S_\psi(f)=b_0+b_{-1}f^{-1}$ and matching the above to $S_y(f)=\sum_ih_if^i$, we find 
\begin{align*}
h_0=\frac{b_0}{4\pi^2m^2} \qquad\text{and}\qquad
h_{-1}=\frac{b_{-1}}{4\pi^2m^2}\\
\end{align*}
Using Table \ref{tab:le-noise-conversion}, the Allan variance is
\begin{align*}
\sigma^2_y(\tau)=
	\left[\!\!\begin{array}{c}1/\tau^2\\[-1ex]\text{\small terms}\end{array}\!\!\right] +
	\frac{b_0}{4\pi^2m^2}\:\frac{1}{2\tau} +
	\frac{b_{-1}}{4\pi^2m^2}\:2\ln(2)
\end{align*}

\section{Examples}\label{sec:le-dly-osc-examp}

\begin{example} \textbf{Photonic delay-line oscillator}\label{ssec:le-photon-osc-examp}
We analyze a photonic delay-line oscillator based on the following parameters
\begin{center}\begin{tabular}{ll}
$\tau_d=20$ $\mu$s & 4-km optical fiber, refraction index of 1.5\\
$1/\tau_d=50$ kHz & mode spacing\\
$\nu_m=10$ GHz & selected frequency\\
$m=2{\times}10^5$ & mode order, equal to $\nu_m\tau_d$\\
$Q=2{\times}10^3$ & selector filter (tunable microwave cavity)\\
$b_0=3.2{\times}10^{-14}$ & white phase noise, $-135$ \unit{rad^2/Hz}\\
$b_{-1}=10^{-10}$ & flicker phase noise, $-100$ \unit{dBrad^2/Hz} at $f=1$ Hz.
\end{tabular}\end{center} 
Figure \ref{fig:le-calc-dly-hphase} shows the phase noise transfer function $|\mathcal{H}(jf)|^2$.  A series of spectral lines due to noise is evident at $f=\smash{\frac1\tau_d}=50$ kHz and multiples.  The following table shows the relevant parameters of the resonance due to the neighboring modes of the delay line.  
\begin{center}
\begin{tabular}{cccccc}\hline
$\mu$ & $|\nu{-}\nu_m|$& $Q\chi$ & $\sigma_\mu$ & 
	$Q_n$ & $Q_\mu$\\\hline
    0     &     0  &    0   & $  0   $ &        $\infty$             & $\infty$\\  
$\pm1$&$ 5{\times}10^4$& 0.02 & $ -10$&$3.14{\times}10^9$ &$15.7{\times}10^3$\\  
$\pm2$&$10{\times}10^4$& 0.04 & $  -40$ & $785{\times}10^6$& $7.85{\times}10^3$\\  
$\pm3$&$15{\times}10^4$& 0.06 & $  -90$ & $349{\times}10^6$& $5.24{\times}10^3$\\  
$\pm4$&$20{\times}10^4$& 0.08 & $-160$ & $196{\times}10^6$& $3.93{\times}10^3$\\  
$\pm5$&$25{\times}10^4$& 0.10 & $-250$ & $126{\times}10^6$&$3.14{\times}10^3$\\
\hline
\end{tabular}\end{center}
The reader should observe the high equivalent merit factor $Q_\mu$ of the phase-noise response, which explains the sharp resonances of Fig.\ \ref{fig:le-calc-dly-hphase}.

\begin{figure}[t]
\namedgraphics{0.5}{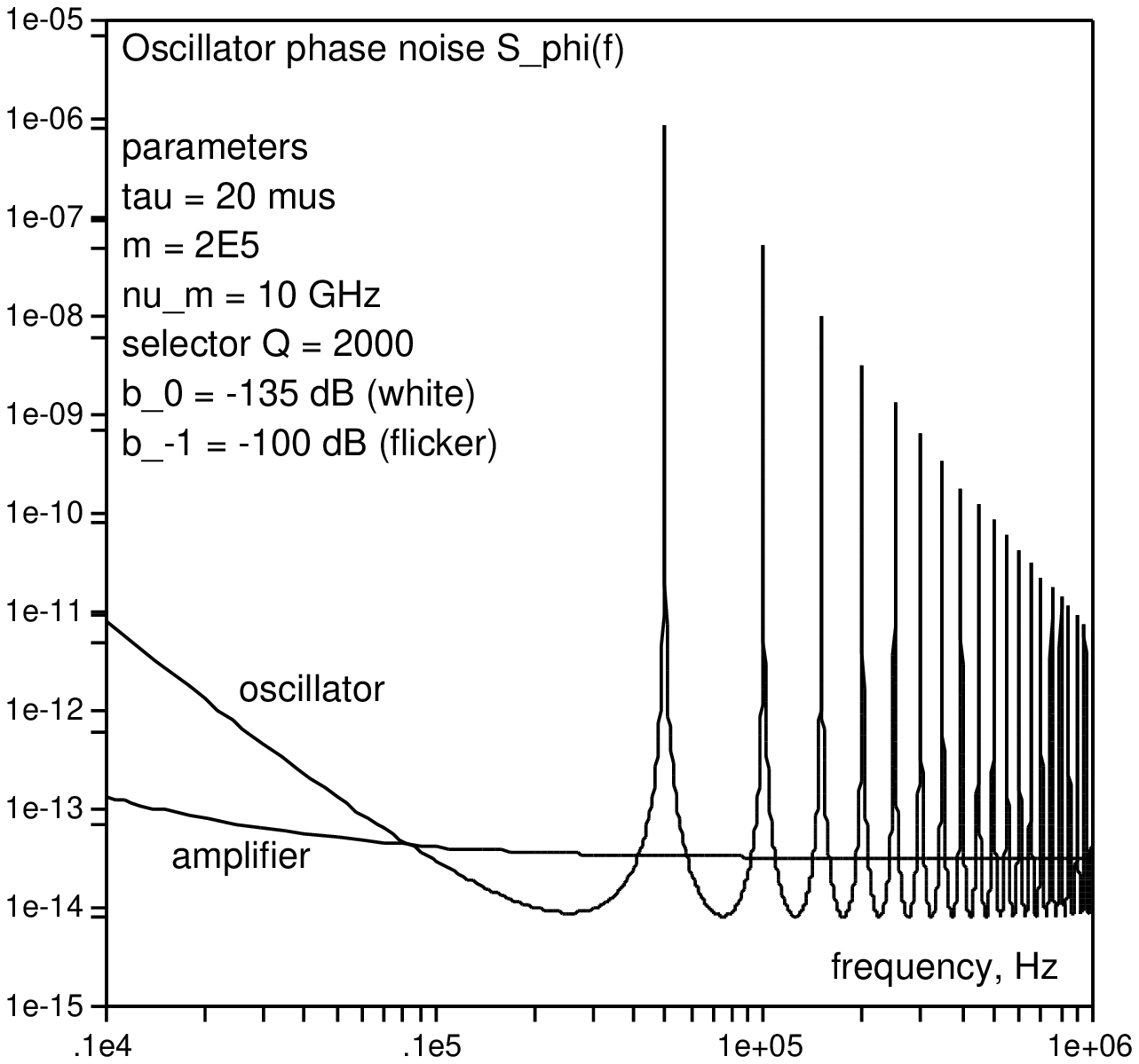}
\caption{Expected phase noise spectrum $S_\phi(f)$ for a photonic delay-line oscillator.  Parameters: $\tau=20$ $\mu$s, $m=2{\times}10^5$ (thus $\nu_m=10$ GHz), and $Q=2000$. Data refer to Example \ref{ssec:le-photon-osc-examp}.}
\label{fig:le-calc-dly-sphi}
\end{figure}%
The amplifier corner frequency is $f_c=3.16$ kHz, as results from $b_{-1}f^{-1}=b_0$.  
The estimated phase noise spectrum is shown in Fig.\ \ref{fig:le-calc-dly-sphi}.
Restricting our attention to the white and flicker frequency noise, the oscillator phase noise is
\begin{align*}
S_\phi(f)&=\frac{2{\times}10^{-6}}{f^2}+\frac{6.3{\times}10^{-3}}{f^3}~.
\end{align*}
The spectrum of the fractional frequency fluctuation is 
\begin{align*}
S_y(f)&=2{\times}10^{-26} + \frac{6.3{\times}10^{-23}}{f}~,
\end{align*}
and the Allan variance
\begin{align*}
\sigma^2_y(\tau)=\frac{10^{-26}}{\tau} + 8.8{\times}10^{-23}
\qquad\text{thus}\qquad
\sigma_y(\tau)=\frac{10^{-13}}{\sqrt{\tau}} + 9.4{\times}10^{-12}
\end{align*}
\end{example}

\begin{remark}
The numerical values of Example \ref{ssec:le-photon-osc-examp} are 
inspired to the references \cite{eliyahu03fcs} and \cite{yao96josab}. 
\end{remark}

\begin{example} \textbf{SAW delay-line oscillator}\label{ssec:le-saw-osc-examp}
We analyze a SAW delay-line oscillator based on the following parameters
\begin{center}\begin{tabular}{ll}
$\tau=5$ $\mu$s & 15 mm  SAW, sound speed 3 km/s\\
$1/\tau=200$ kHz & mode spacing\\
$\nu_m=900$ MHz & selected frequency\\
$m=4500$ & mode order, equal to $\nu_m\tau$\\
$Q=80$ & selector filter ($LC$ filter)\\
$b_0=3.2{\times}10^{-15}$ & white phase noise, $-145$ \unit{rad^2/Hz}\\
$b_{-1}=10^{-13}$ & flicker phase noise, $-100$ \unit{dBrad^2/Hz} at $f=1$ Hz.
\end{tabular}\end{center} 
The amplifier corner frequency is $f_c=31.6$ Hz, as results from $b_{-1}f^{-1}=b_0$.  Figure \ref{fig:le-calc-dly-hphase} shows the expected phase noise $S_\phi(f)$.  The neighboring resonances show up at $f=\smash{\frac1\tau}=200$ kHz and multiples.  The following table shows the relevant resonance parameters.
\begin{center}
\begin{tabular}{cccccc}\hline
$\mu$ & $|\nu{-}\nu_m|$& $Q\chi$ & $\sigma_\mu$ & 
	$Q_n$ & $Q_\mu$\\\hline
    0     &     0  &    0   & $  0   $ &        $\infty$             & $\infty$\\  
$\pm1$&$ 2{\times}10^5$& 0.035 & $ -126$&$2.24{\times}10^7$ &$4970$\\  
$\pm2$&$4{\times}10^5$& 0.071 & $  -506$ & $5.59{\times}10^6$&$2485$\\  
$\pm3$&$8{\times}10^5$& 0.107 & $-1138$ & $2.49{\times}10^6$&$1657$\\  
$\pm4$&$10{\times}10^5$& 0.142 &$-2023$ &$1.40{\times}10^6$&$1243$\\  
$\pm5$&$12{\times}10^5$& 0.178 &$-3160$ &$8.95{\times}10^6$&$994$\\
\hline
\end{tabular}\end{center}

\begin{figure}
\namedgraphics{0.5}{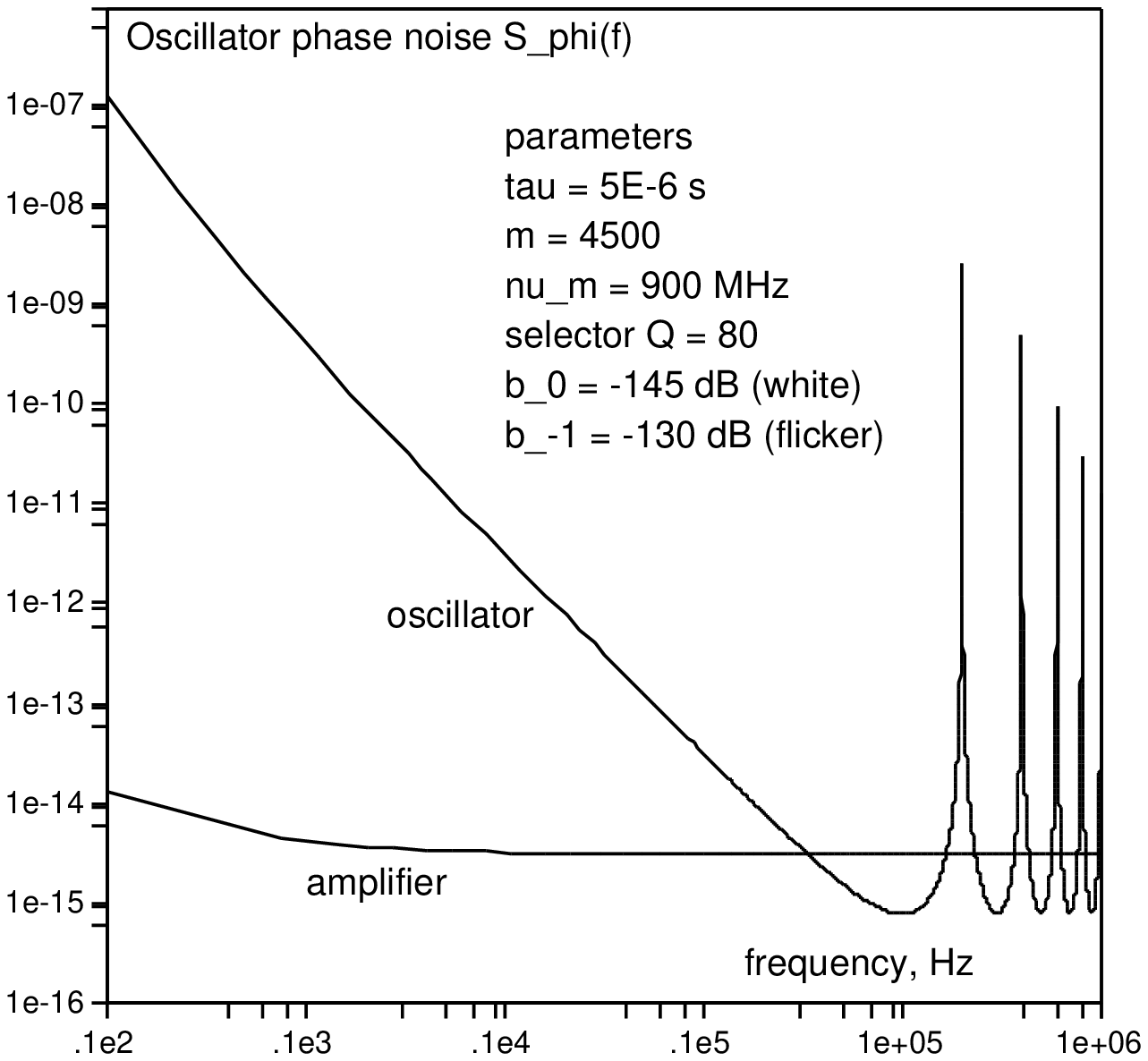}
\caption{Expected phase noise spectrum $S_\phi(f)$ for a SAW delay-line oscillator.  Parameters: $\tau=20$ $\mu$s, $m=2{\times}10^5$ (thus $\nu_m=10$ GHz), and $Q=2000$. Data refer to Example \ref{ssec:le-saw-osc-examp}.}
\label{fig:le-calc-saw-sphi}
\end{figure}%
The estimated phase noise spectrum is shown in Fig.\ \ref{fig:le-calc-saw-sphi}.
Focusing our attention to the white and flicker frequency noise, the oscillator phase noise is
\begin{align*}
S_\phi(f)&=\frac{2.5{\times}10^{-6}}{f^2}+\frac{8{\times}10^{-5}}{f^3}~.
\end{align*}
The spectrum of the fractional frequency fluctuation is 
\begin{align*}
S_y(f)&=3.96{\times}10^{-24} + \frac{1.25{\times}10^{-22}}{f}~,
\end{align*}
and the Allan variance
\begin{align*}
\sigma^2_y(\tau)=\frac{2{\times}10^{-24}}{\tau} + 1.7{\times}10^{-22}
\quad\text{thus}\quad
\sigma_y(\tau)=\frac{1.4{\times}10^{-13}}{\sqrt{\tau}} + 1.3{\times}10^{-12}
\end{align*}
\end{example}

\begin{remark}
The numerical values of Example \ref{ssec:le-saw-osc-examp} are inspired to the european GSM standard for mobile phones, which operates in the 900 MHz band with a channel spacing of 200 kHz.  By tuning the selector filter, the SAW oscillator can be used as a frequency synthesizer that works at the appropriate frequency, with the appropriate step.  
\end{remark}

\section{Phase noise in lasers}
\begin{figure}[t]
\namedgraphics{0.8}{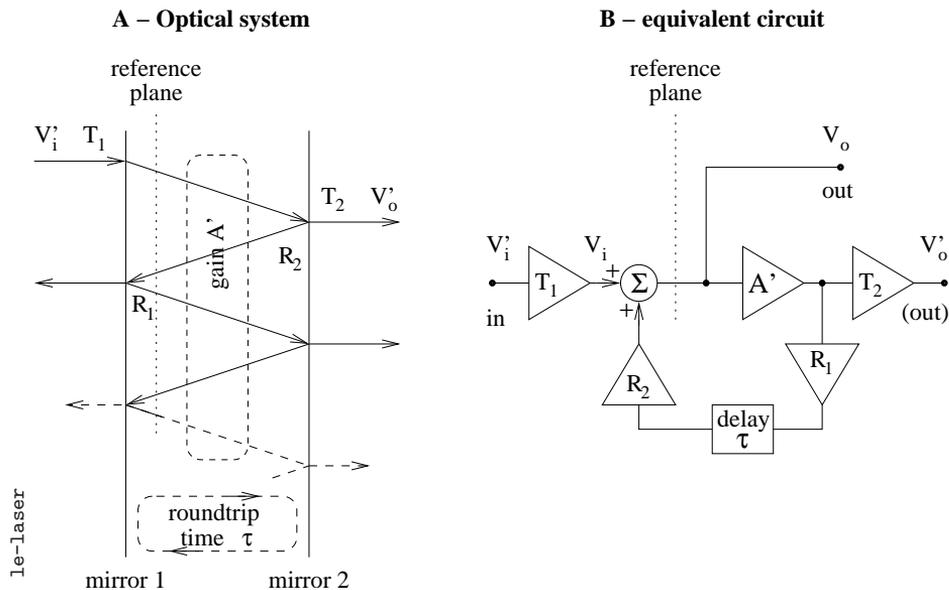}
\caption{A laser can be regarded as a feedback system.}
\label{fig:le-laser}
\end{figure}%
Figure \ref{fig:le-laser} sketches the optical system and the equivalent electric circuit of a laser.  $T$ and $R$ are the mirror transmission and reflection coefficients, $A$ is the round-trip gain, and $\tau$ is the round-trip transit time.  Gain and transmission coefficients refer to electric field, thus to voltage in the equivalent circuit of Fig.\  \ref{fig:le-laser}\,B\@. 
We have to prove that the laser is equivalent to the feedback scheme of Fig.~\ref{fig:le-dly-loop}, for all the framework of Sections \ref{sec:le-dly-resonator}, \ref{sec:le-dly-phase-response} and \ref{sec:le-dly-noise} applies.

At the reference plane, the input signal (electric field) is added to the signal amplified and fed back after delay $\tau$.  Without loss of generality, we drop the trivial factor $T_1$.  This means that the true input signal is $V_i$ on the reference plane, instead of the external signal $V'_i$.  Then, we take the output signal  $V_o$ at the reference plane, instead of $V'_o$ after the mirror 2. The transfer function is defined as
\begin{align}
H(s)&=\frac{V_o(s)}{V_i(s)}	\qquad\text{[def.\ of $H(s)$]}~.
\end{align}
By breaking the circuit at the reference plane and equating the left-hand signal to the right-hand signal, we find
\begin{align*}
V_o(s)=V_i(s)+V_o(s)\left[1-A'(s)\beta_d(s)R_1R_2)\right]
\end{align*}
where $\beta_d(s)=e^{-s\tau_d}$ is the delay.  Hence
\begin{align*}
H(s)&=\frac{1}{1-A'(s)\beta_d(s)R_1R_2}
\end{align*}
We observe that in actual lasers the gain $A'$ of the active medium is a smooth function of frequency, wide as compared to the mode spacing of the optical cavity, which shows a maximum centered at the laser frequency.  This is the same analysis/design criterion that we have adopted in Section \ref{sec:le-dly-resonator}, introducing the mode-selector filter.  
Thus 
\begin{enumerate}
\item We move the frequency dependence from $A'(s)$ to a new filter function $\beta_f(s)$. The latter is normalized for the maximum to be equal one, at the laser frequency.
\item We approximate $\beta_f(s)$, wide and smooth around the laser frequency, with a resonator. 
\item We introduce the gain $A$ that accounts for the maximum $|A(j\omega)|$, and for the two reflection coefficients $R_1$ and $R_2$.
\end{enumerate}
After these manipulations, the laser transfer function takes the form
\begin{align*}
H(s)&=\frac{1}{1-A\beta_f(s)\beta_d(s)}~,
\end{align*}
which is equivalent to Eq.\ \req{eqn:le-hrf-rational}.
Needless to say, the Barkhausen condition still holds, for the loop gain must be equal to one as a result of some saturation phenomena.

\begin{figure}[t]
\namedgraphics{0.8}{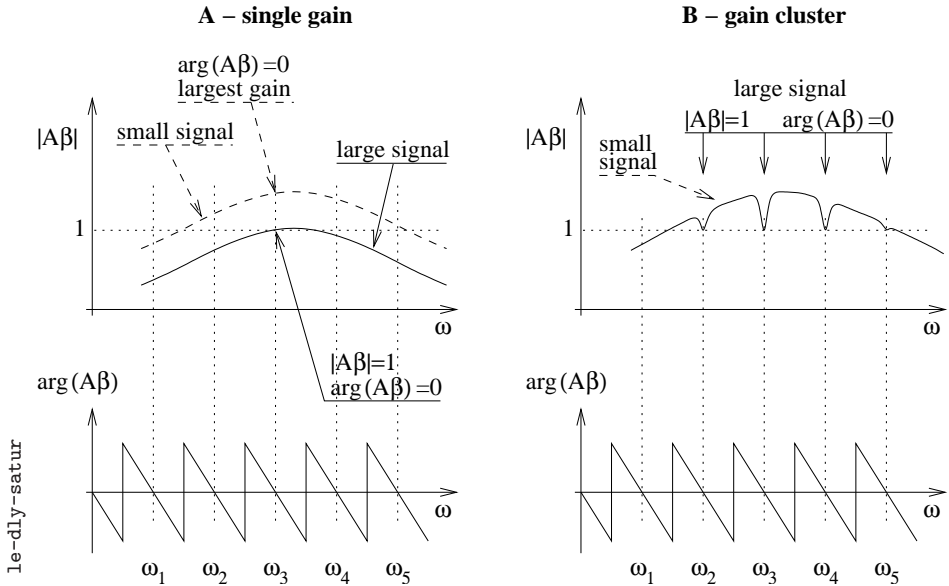}
\caption{Different types of saturation in amplifier.}
\label{fig:le-dly-satur}
\end{figure}%
\section{Saturation and multimode oscillation}
Multimode oscillation requires that the Barkhausen condition $A\beta=1$ is met at more than one frequency.  This is of course possible.

\paragraph{Single-mode oscillation} In the amplifiers of the first type, when the signal exceeds a threshold the gain drops more or less smoothly, yet uniformly in the pass band (Fig.\ \ref{fig:le-dly-satur}\,A).  
This is always the case of the oscillators in which the power is controlled in closed loop by measuring the output power and by acting on a variable-gain element.
This behavior is also typical of saturated electronic amplifiers, either vacuum or solid state.  In this case, distortion reduces the gain and squeezes power into higher harmonics, at frequencies multiple of $\omega_0$.   
Additionally, this is found in the quantum amplifiers (masers and lasers) in which amplification is due to a single pumped level, or to a cluster of levels that act as a single one because of the linewidth.
Oscillation sinks atoms from the pumped level, for the active population is reduced.
With these amplifiers, the gain condition can only be met at one frequency.  Generally, the oscillation frequency is that at which the small-signal gain is the highest among those at which the loop phase is zero.  In the example of Fig.\ \ref{fig:le-dly-satur}\,A, $\omega_1$ is not in the mode competition because of the insufficient the small-signal gain, and $\omega_3$ wins.

\paragraph{Multimode oscillation} In the amplifier of the second type, found in some lasers, the bandwidth is due to a cluster of energy levels narrow enough for power to sink population selectively (Fig.\ \ref{fig:le-dly-satur}\,B).   
When oscillation takes place at a given frequency, the active population decreases only in the neighbors.  Thus oscillation can also rise at the other frequencies at which the phase condition is met and the small-signal gain is greater than one.  In the example of Fig.\ \ref{fig:le-dly-satur}\,B, the small-signal gain is insufficient at $\omega_1$; oscillation takes place at $\omega_2$, $\omega_3$, $\omega_4$, and $\omega_5$.%
\clearpage

\appendix%
\chapter{Resonator model and parameters}\label{app:le-resonator}
\section{Laplace plane}
\begin{figure}[t]
\namedgraphics{0.8}{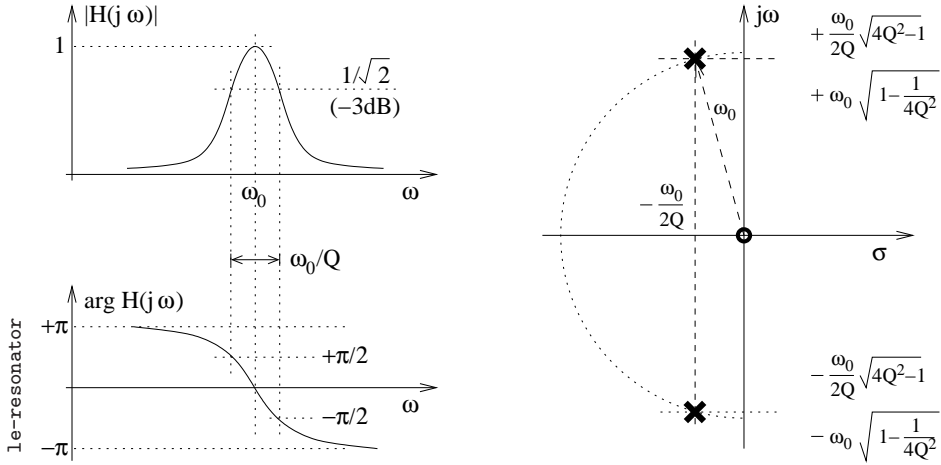}
\caption{Resonator transfer function in the frequency domain and on the Laplace plane.}
\label{fig:le-resonator}
\end{figure}

Around the resonant frequency, real world resonators are well described by a linear second-order differential equation with constant coefficients.  Accordingly The Laplace transform of the impulse response is a rational function of the form
\begin{align}
H_r(s) &= \frac{\omega_0}{Q}\;\frac{s}{s^2+\frac{\omega_0}{Q}\:s+\omega_0^2}~,
\label{eqn:le-app-def-h}
\end{align} 
which can be rewritten as 
\begin{align}
H_r(s) &= \frac{\omega_0}{Q}\:\frac{s}{(s-s_p)(s-s_p^*)}\qquad
\begin{array}{l}s_p=\sigma_p+j\omega_p\\s_p^*=\sigma_p-j\omega_p\end{array}
\intertext{with}
&\sigma_p =-\frac{\omega_0}{2Q}\label{eqn:le-app-sigma-p}\\[1ex]
&\omega_p=\frac{\omega_0}{2Q}\sqrt{4Q^2-1} 
	= \omega_0 \sqrt{1-\frac{1}{\smash{4Q^2}}}\label{eqn:le-app-omega-p}\\[1ex]
&\sigma_p^2+\omega_p^2=\omega_0^2~.
\end{align}
Figure \ref{fig:le-resonator} shows the resonator transfer function in the frequency domain and on the Laplace plane. 

\section{Frequency domain}
The frequency response is easily found by replacing $s=j\omega$ in the Laplace transform.  Defining the dissonance $\chi$ as
\begin{align}
\chi&=\frac{\omega}{\omega_0}-\frac{\omega_0}{\omega}
	&&\text{dissonance}\label{eqn:le-app-disson}\\
\chi&\simeq2\:\frac{\omega-\omega_0}{\omega_0}
	&&\text{for}~\left|\omega-\omega_0\right|\ll\tfrac{\omega_0}{Q}
	\label{eqn:le-app-disson-approx}
\end{align}
we find the following relevant relationships
\begin{align}
H_r(j\omega)&=\frac{1}{1+jQ\chi}   &&\text{frequency response}\\
R(j\omega)&=\frac{1}{1+Q^2\chi^2} &&\text{real part}~~\Re{\{H_r(j\omega)\}}\\
I(j\omega)&=-\frac{Q\chi}{1+Q^2\chi^2} &&\text{imag. part}~~\Im{\{H_r(j\omega)\}}\\
\rho(j\omega)&=\frac{1}{\sqrt{1+Q^2\chi^2}} &&\text{modulus}~~|H_r(j\omega)|
	\label{eqn:le-app-reson-mod}\\
\theta(j\omega)&=-\arctan Q\chi &&\text{argument}~~\arg\left[H_r(j\omega)\right]
	\label{eqn:le-app-reson-arg}
\end{align}

\section{Time domain}\label{sec*:le-resonator-td}
In the time domain, the impulse response of the resonator an exponentially decaying sinusoid of the form
\begin{align}
v_o(t)&=\cos\left(\omega_0t\right)e^{-t/\tau}~,
\end{align}
where the relaxation time $\tau$ is related to the other resonator parameters by
\begin{align}
\tau=\frac{Q}{\pi}T_0=\frac{Q}{\pi\nu_0}=\frac{2Q}{\omega_0}=
	\frac{1}{\omega_L}=\frac{1}{2\pi f_L}
\qquad\begin{array}{l}\text{relaxation}\\[-0.5ex]\text{time}\end{array}~.
\label{eqn:le-reson-tau}
\end{align}

\def\bibfile#1{/Users/rubiola/Documents/work/bib/#1}
\addcontentsline{toc}{chapter}{References}
\bibliographystyle{amsalpha}
\bibliography{\bibfile{ref-short},%
              \bibfile{references},%
              \bibfile{rubiola}}

\end{document}